\documentclass[a4paper, 11 pt]{article}
\usepackage[hmargin=1.25in,vmargin=1.25in]{geometry}
\usepackage{amsmath,amssymb,amsthm,mathrsfs,amsfonts,dsfont} 
\usepackage[dvipsnames,svgnames, x11names]{xcolor}
\usepackage[round]{natbib}
\usepackage{bibunits}
\usepackage{grffile}
\usepackage{pdfpages,selectp}
\usepackage{fancyhdr}
\usepackage[titletoc]{appendix}
\usepackage{setspace, titlesec,rotating,titling} 
\usepackage{booktabs,threeparttable, multirow,caption,makecell} 
\usepackage{tikz,pgfplots} 
\pgfplotsset{compat=1.11}
\usetikzlibrary{arrows,intersections}
\usetikzlibrary{backgrounds,positioning,fit,petri}
\usetikzlibrary{mindmap,trees,calendar,external,shapes}
\usetikzlibrary{decorations.fractals,decorations.pathmorphing,shadows,shadings,patterns}
\usetikzlibrary{spy,calc} 

\usepackage[open,openlevel=2,numbered,atend]{bookmark}
\usepackage{hyperref}

\newtheorem{thm}{Theorem}[section]

\newtheorem{lem}{Lemma}[section]
\newtheorem{pro}{Proposition}[section]
\newtheorem{ass}{Assumption}[section]

\theoremstyle{definition}
\newtheorem{defn}{Definition}[section]
\newtheorem{rem}{Remark}[section]

\newtheoremstyle{varass}
{\topsep} {\topsep}%
{\upshape}
{}
{\bfseries\scshape}
{.}
{1em}
{\thmname{#1} \thmnumber{ #2 \prime} }
\theoremstyle{varass}

\theoremstyle{definition}
\newtheorem{ex}{Example}[section]

\newtheoremstyle{exctd}
{\topsep} {\topsep}%
{\upshape}
{}
{\bfseries\scshape}
{.}
{1em}
{\thmname{#1} \thmnumber{ #2}\thmnote{#3} (Continued)}
\theoremstyle{exctd}
\newtheorem*{exctd}{Example}

\DeclareMathOperator{\cov}{Cov}
\DeclareMathOperator{\var}{Var}

\newcommand{\amin}{\operatornamewithlimits{arg\,min}}

\DeclareMathOperator{\bl}{BL}
\newcommand{\obar}[1]{\mkern 1.5mu\overline{\mkern-1.5mu#1\mkern-1.5mu}\mkern 1.5mu}

\newcommand{\transpose}{\text{\scalebox{0.7}{$\intercal$}}}
\newcommand{\comment}[1]{}

\newcommand{\convl}{\xrightarrow{L}}
\newcommand{\convln}{\xrightarrow{L_n}}

\newcommand{\convp}{\xrightarrow{p}}

\newcommand{\darrow}{\downarrow}
\newcommand{\uarrow}{\uparrow}

\DeclareMathOperator{\vecop}{vec}
\defaultbibliographystyle{ims}
\defaultbibliography{mybibliography}

\begin{document}

\begin{bibunit}
\pdfbookmark[1]{Title}{title}

\title{Inference on Functionals under First Order Degeneracy}
\author{Qihui Chen\\ School of Management and Economics\\ The \hspace{-0.03cm}Chinese\hspace{-0.03cm} University\hspace{-0.03cm} of\hspace{-0.03cm} Hong\hspace{-0.03cm} Kong, \hspace{-0.03cm}Shenzhen\\ qihuichen@cuhk.edu.cn
\and
Zheng Fang\thanks{We would like to thank Brendan Beare, Andres Santos, Yixiao Sun and anonymous referees for valuable suggestions that have helped greatly improve this paper. We are also grateful for Xiaohong Chen, Qi Li and seminar participants for helpful discussions and comments.} \\ Department of Economics \\ Texas A\&M University \\ zfang@tamu.edu}
\date{\today}
\maketitle

\begin{abstract}
This paper presents a unified second order asymptotic framework for conducting inference on parameters of the form $\phi(\theta_0)$, where $\theta_0$ is unknown but can be estimated by $\hat\theta_n$, and $\phi$ is a known map that admits null first order derivative at $\theta_0$. For a large number of examples in the literature, the second order Delta method reveals a nondegenerate weak limit for the plug-in estimator $\phi(\hat\theta_n)$. We show, however, that the ``standard'' bootstrap is consistent if and only if the second order derivative $\phi_{\theta_0}''=0$ under regularity conditions, i.e., the standard bootstrap is inconsistent if $\phi_{\theta_0}''\neq 0$, and provides degenerate limits unhelpful for inference otherwise. We thus identify a source of bootstrap failures distinct from that in \citet{Fang_Santos2014HDD} because the problem (of consistently bootstrapping a {\it nondegenerate} limit) persists even if $\phi$ is differentiable. We show that the correction procedure in \citet{Babu1984bootstrap} can be extended to our general setup. Alternatively, a modified bootstrap is proposed when the map is {\it in addition} second order nondifferentiable. Both are shown to provide local size control under some conditions. As an illustration, we develop a test of common conditional heteroskedastic (CH) features, a setting with both degeneracy and nondifferentiability -- the latter is because the Jacobian matrix is degenerate at zero and we allow the existence of multiple common CH features.
\end{abstract}

\begin{center}
\textsc{Keywords:} First order degeneracy, Second order Delta method, Bootstrap consistency, Babu correction, Common CH features, $J$-test.
\end{center}

\vspace{-0.15in}

\hspace{0.07in} {JEL Classification: C12, C15}

\newpage
\section{Introduction}

There is a large number of inference problems in economics and statistics in which the parameter of interest is of the form $\phi(\theta_0)$, where $\theta_0$ is an unknown parameter depending on the underlying distribution of the data and $\phi$ is a known map. In these settings, it is common practice to employ the plug-in estimator $\phi(\hat\theta_n)$, where $\hat\theta_n$ is an estimator for $\theta_0$, as a building block for conducting inference on $\phi(\theta_0)$. The Delta method asserts that if $r_n\{\hat\theta_n-\theta_0\}\convl\mathbb G$ for some sequence $r_n\uarrow\infty$, then
\begin{align}\label{Eqn: delta method, 1st}
r_n\{\phi(\hat\theta_n)-\phi(\theta_0)\}\convl\phi_{\theta_0}'(\mathbb G)~,
\end{align}
provided $\phi$ is at least Hadamard directionally differentiable at $\theta_0$, where $\phi_{\theta_0}'$ is the derivative of $\phi$ at $\theta_0$ \citep{Shapiro1991, Dumbgen1993}. As powerful as the Delta method has proven to be \citep{Vaart1998,Fang_Santos2014HDD}, an implicit and yet crucial assumption for the convergence \eqref{Eqn: delta method, 1st} to be useful for inferential purposes is that $\phi_{\theta_0}'(\mathbb G)$ or $\phi_{\theta_0}'$ is nondegenerate, i.e., $\phi_{\theta_0}'\neq 0$. Unfortunately, such {\it first order degeneracy} arises frequently in asymptotic analysis, with applications including Wald tests or Wald type functionals \citep{Wald1943tests,Engle1984Handbook}, unconditional and conditional moment inequality models \citep{AndrewsandSoares2010,Andrews_Shi2013CMI}, Cram\'{e}r-von Mises functionals \citep{Darling1957KSCvM}, the study of stochastic dominance \citep{Linton2010}, and the $J$-test for overidentification in GMM settings \citep{Hall_Horowitz1996bootstrap}.

In the presence of first order degeneracy, one may resort to a higher order analysis for the sake of a nondegenerate limiting distribution. \citet{Shapiro2000inference} established that if $\phi$ is second order Hadamard directionally differentiable (see Definition \ref{Defn: HDD 2nd}) -- a feature shared by aforementioned examples, then
\begin{align}\label{Eqn: delta method, 2nd, intro}
r_n^2\{\phi(\hat\theta_n)-\phi(\theta_0)-\phi_{\theta_0}'(\hat\theta_n-\theta_0)\}\convl\phi_{\theta_0}''(\mathbb G)~,
\end{align}
where $\phi_{\theta_0}''$ denotes the second order derivative of $\phi$ at $\theta_0$. Thus, when first order degeneracy occurs, \eqref{Eqn: delta method, 2nd, intro} suggests that we may base our asymptotic analysis on
\begin{align}\label{Eqn: delta method, 2nd, intro2}
r_n^2\{\phi(\hat\theta_n)-\phi(\theta_0)\}\convl\phi_{\theta_0}''(\mathbb G)~.
\end{align}
Usefulness of the limiting distribution in \eqref{Eqn: delta method, 2nd, intro2}, however, relies on our ability to consistently estimate it. In this regard, \citet{Efron1979}'s bootstrap seems to be a potential option. Specifically, if $\hat\theta_n^*$ is a bootstrap analog of $\hat\theta_n$ that works for estimating the law of $\mathbb G$, then in view of \eqref{Eqn: delta method, 2nd, intro2} one may hope that
\begin{align}\label{Eqn: standard bootstrap, intro}
r_n^2\{\phi(\hat\theta_n^*)-\phi(\hat\theta_n)\}
\end{align}
can be employed as an estimator for the law of $\phi_{\theta_0}''(\mathbb G)$, at least when $\phi$ is smooth. Unfortunately, there are simple examples where the law of \eqref{Eqn: standard bootstrap, intro} conditional on the data, referred to as the standard bootstrap, fails to provide consistent estimates \citep{Babu1984bootstrap}.

As the first contribution of this paper, we show that the standard bootstrap \eqref{Eqn: standard bootstrap, intro} is consistent if and only if $\phi_{\theta_0}''=0$ under mild conditions. Thus, the standard bootstrap is necessarily inconsistent when $\phi_{\theta_0}''$ is nondegenerate, while when $\phi_{\theta_0}''$ is degenerate, the resulting asymptotic distribution is degenerate and hence not useful for inference. Therefore, the failure of the standard bootstrap is an inherent implication of first order degeneracy. It is worth noting that the failure of the standard bootstrap persists even when $\phi$ is differentiable. Hence, we identify a source of bootstrap inconsistency distinct from that in \citet{Fang_Santos2014HDD}, i.e., nondifferentiability of the map $\phi$, as explained further towards the end of this section.

Heuristically, the reason why the standard bootstrap fails is that even though $r_n^2\phi_{\theta_0}'(\hat\theta_n-\theta_0)=0$ in the ``real world'', its bootstrap counterpart is nondegenerate, i.e., $r_n^2\phi_{\hat\theta_n}'(\hat\theta_n^*-\hat\theta_n)=O_p(1)$, echoing \citet{Efron1979}'s point that the bootstrap provides approximate frequency statements rather than approximate likelihood statements. This observation was picked up by \citet{Babu1984bootstrap} who provided a consistent resampling procedure by including the first order correction term:
\begin{align}\label{Eqn: Babu extension, intro}
r_n^2\{\phi(\hat\theta_n^*)-\phi(\hat\theta_n)-\phi_{\hat\theta_n}'(\hat\theta_n^*-\hat\theta_n)\}~.
\end{align}
As the second contribution, we generalize the above modified bootstrap \eqref{Eqn: Babu extension, intro}, referred to as the Babu correction, to settings that accommodate infinite dimensional models and a wide range of bootstrap schemes for $\hat\theta_n^*$. However, we stress that the Babu correction is inappropriate when $\phi$ is only Hadamard directionally differentiable.

As the third contribution, we follow \citet{Fang_Santos2014HDD} and provide a modified bootstrap which is consistent regardless of the presence of first order degeneracy and nondifferentiability of $\phi$. The insight we exploit is that the weak limit $\phi_{\theta_0}''(\mathbb G)$ in \eqref{Eqn: delta method, 2nd, intro2} is a composition of the limit $\mathbb G$ and the derivative $\phi_{\theta_0}''$. Therefore, we may estimate the law of $\phi_{\theta_0}''(\mathbb G)$ by composing a suitable estimator $\hat\phi_n''$ for $\phi_{\theta_0}''$ with a bootstrap approximation $r_{n}\{\hat\theta_n^*-\hat\theta_n\}$ for $\mathbb G$. Since the conditions on $\hat\phi_n''$ proposed by \citet{Fang_Santos2014HDD} in order for this approach to work are either demanding or hard to check in our setup, we provide a high level condition that is easy to verify. We further demonstrate that numerical differentiation provides a desirable estimator $\hat\phi_n''$ in general; alternatively, we show how to estimate $\phi_{\theta_0}''$  by exploiting its structure in particular examples. Our inference procedures are also shown to enjoy the local size control property under a key condition that is algebraically simple.

Finally, to further demonstrate the applicability of our framework, we develop a test of common conditional heteroskedastic (CH) features studied by \citet{Dovonon_Renault2013testing} but under weaker assumptions that allow more than one common CH features. Thus, in addition to the first order identification failure they focused on, we further allow second order (and hence global) identification failures, which renders the functional involved highly (second-order) nondifferentiable as well as first order degenerate. Such a generalization is important because it is unknown {\it a priori} how many common features there are and in the context of asset pricing the number can be large \citep{Engle_Ng_Rothschild1990asset}. Moreover, the linear normalization in \citet{Dovonon_Renault2013testing} can falsely exclude the existence of common features even when there does exist a unique common CH feature, a deficiency which we avoid by the unit-length normalization. Monte Carlo simulations indicate our tests substantially alleviate size distortion and have good power performance. We stress that first order degeneracy is of a nature different from that of the degeneracy of Jacobian matrices which is the focus of \citet{Dovonon_Renault2013testing}; see Section \ref{Sec: Application} for details. Our approach may also be used to develop tests for other common features \citep{Engle_Kozicki1993CF}.

There have been extensive studies on the bootstrap consistency \citep{Hall1992bootstrap,HorowitzBoot}. It was realized soon after \citet{Efron1979} that the bootstrap is not always successful \citep{BickelandFreedman1981bootstrap}; see also \citet{Andrews2000Bootstrap} for a summary. \citet{Babu1984bootstrap} provided a simple example of bootstrap failure due to first order degeneracy, and established the validity of the Babu correction for the special case studied there. \citet{Shao1994bootstrap} and \citet{Bertail_Politis_Romano1999subsampling} showed that $m$ out of $n$ resampling and subsampling can serve as alternative remedies. There are, however, three reasons we choose not to use these methods. First, they entail the choice of tuning parameters while our proposal can work without such nuisances when $\phi$ is differentiable. Second, when $\phi$ is nondifferentiable, both can lead to invalid tests due to lack of uniform approximations \citep{AndrewsandGuggen2010ET}. We provide a simple algebraic condition which, together with regularity of $\hat\theta_n$,  delivers local uniformity of our inferential procedure. Third, they have been shown to be dominated by other inferential methods, for example, in moment inequality models \citep{AndrewsandSoares2010} which our framework includes as special cases. \citet{Datta1995bootstrap} revisited Babu's example and offered a bias correction procedure that depends on a first stage shrinkage type estimator. Somewhat similar methods were later proposed in \citet{Andrews2000Bootstrap} and \citet{Giurcanu2012bootsrtap}. These methods are not easily extendable to more general settings.

Bootstrap inconsistency due to nondifferentiability of $\phi$ was studied in \citet{Dumbgen1993} and recently in \citet{Fang_Santos2014HDD} who formally established that (first order) differentiability of $\phi$ is a necessary as well as sufficient condition for the standard bootstrap to work under regularity conditions. Our work complements theirs by identifying a different source of bootstrap failure. Specifically, given bootstrap consistency of $\hat\theta_n^*$ and if $\phi$ is first order degenerate (and hence fully differentiable!), then \citet{Fang_Santos2014HDD} implies that the standard bootstrap $r_n\{\phi(\hat\theta_n^*)-\phi(\hat\theta_n)\}$ is consistent for the law of $\phi_{\theta_0}'(\mathbb G)$ which is degenerate (and unhelpful for inference). We further show that the law of the second order limit $\phi_{\theta_0}''(\mathbb G)$ cannot be consistently estimated by the second order standard bootstrap \eqref{Eqn: standard bootstrap, intro} unless $\phi_{\theta_0}''$ itself is degenerate -- {\it this remains true regardless of whether $\phi$ is (second order) differentiable or not!} Moreover, extra work is needed in order to show our bootstrap inferential procedures work well in the local uniformity sense. In applications, first order degeneracy and second order nondifferentiability are often mixed together, for example, in \citet{Romano_Shaikh2010}, \citet{AndrewsandSoares2010}, \citet{Linton2010}, and \citet{Andrews_Shi2013CMI}. The numerical differentiation approach of estimating derivatives was somewhat implicit in \citet{Dumbgen1993}'s rescaled bootstrap, recently employed by \citet{Song2014minimax} and studied by \citet{Hong_Li2015numericaldelta}. We provide a more general condition that may be used to verify ``consistency'' of derivative estimators (not necessarily constructed via numerical differentiation). Our theory has been utilized in \citet{ChenFang2016Rank} to develop a rank test where, unlike previous studies, the true rank is potentially strictly less than the hypothesized value, a longstanding problem in the literature.

We now introduce some notation. For a set $T$, we let $\ell^\infty(T)$ denote the space of bounded real-valued functions defined on $T$ and $C(T)$ the space of real-valued continuous functions on a compact set $T$ (endowed with some topology). Both $\ell^\infty(T)$ and $C(T)$ are equipped with the uniform norm, i.e., $\|f\|_\infty\equiv\sup_{t\in T}|f(t)|$. For a normed space $\mathbb D$ endowed with norm $\|\cdot\|_{\mathbb D}$ and $m\in\mathbf N$, we equip the product space $\prod_{j=1}^m \mathbb D$ with the product norm $\max_{j=1}^m\|\theta^{(j)}-\vartheta^{(j)}\|_{\mathbb D}$, denoted $\|\cdot\|_{\mathbb D}$ with some abuse of notation, for $\theta,\vartheta\in \prod_{j=1}^m \mathbb D$, where $\theta^{(j)}$ and $\vartheta^{(j)}$ are the $j$th coordinates of $\theta$ and $\vartheta$ respectively. For a subset $A\subset T$, we write $1\{A\}$ for the indicator function of $A$.

The remainder of the paper is structured as follows. Section \ref{Sec: setup and backgroud} formalizes the general setup, shows the wide applicability of our framework by introducing related examples, and establishes the asymptotic framework by presenting a mild extension of the second order Delta method. Section \ref{Sec: Bootsrtap} characterizes the inherent difficulties caused by first order degeneracy, extends the Babu correction to our general setup, and offers a flexible modified bootstrap procedure. Section \ref{Sec: Application} develops a test for common CH features that allows multiple common CH features, while Section \ref{Sec: conclusion} concludes. Appendix \ref{Sec: Testing} demonstrates that our inferential procedure is robust to local perturbations of the distribution of the data under regularity conditions. The remaining appendices collect all the proofs and additional discussions.

\section{Setup and Background}\label{Sec: setup and backgroud}

In this section, we formalize the general setup, introduce related examples, and review notions of differentiability based on which we present the second order Delta method.

\subsection{General Setup}

The treatment in this paper is general in the sense that we allow both the parameter $\theta_0$ and the map $\phi$ to take values in infinite dimensional spaces, though attention is confined to real-valued $\phi$ when studying tests. In particular, we assume $\theta_0\in\mathbb D_\phi\subset\mathbb D$ and $\phi: \mathbb D_\phi\to\mathbb E$, where $\mathbb D$ and $\mathbb E$ are normed spaces with norms $\|\cdot\|_{\mathbb D}$ and $\|\cdot\|_{\mathbb E}$ respectively. Moreover, the data generating process is general as well in that the model can be parametric, semiparametric and nonparametric and that the data $\{X_i\}_{i=1}^n$ need not be i.i.d.. However, we do impose i.i.d.\ assumption in our local analysis, but only for simplicity. The results there can presumably be extended to general asymptotically normal experiments \citep{Vaart_Wellner1990prohorov}.

The common probability space on which all (random) maps are defined is the canonical one. For example, in the simplest i.i.d.\ setup, we think of the data  $\{X_i\}_{i=1}^n$ as the coordinate projections on the first $n$ coordinates in the product probability space $(\prod_{i=1}^\infty\mathscr X, \bigotimes_{i=1}^\infty \mathcal A, \prod_{i=1}^\infty P)$ where $(\mathscr X,\mathcal A)$ is the sample space each $X_i$ lives in and $P$ is the common Borel probability measure that governs each $X_i$. In the presence of bootstrap weights, we further think of the product space as the ``first $\infty$'' coordinates of the even ``larger'' product space $\big((\prod_{i=1}^\infty\mathscr X)\times \mathscr W, (\bigotimes_{i=1}^\infty \mathcal A) \otimes\mathcal W, (\prod_{i=1}^\infty P)\times Q\big)$, where $(\mathscr W,\mathcal W,Q)$ governs the infinite sequence of bootstrap weights.

Given the generality of our setup, weak convergence throughout the paper is meant in the Hoffmann-J{\o}rgensen sense \citep{Vaart1996}. Expectations and probabilities should therefore be interpreted as outer expectations and outer probabilities respectively defined relative to the canonical probability space, though we obviate the distinction in the notation. The notation is made explicit in the appendices whenever differentiating between inner and outer expectations is necessary.

\subsection{Related Examples}\label{Sec: Examples}

To fix ideas, we now turn to related examples that serve to illustrate the wide applicability of our framework. The first example is taken from \citet{Babu1984bootstrap}, which provides an easy illustration of bootstrap inconsistency in the presence of first order degeneracy even if the transformation $\phi$ is smooth.

\begin{ex}[Wald Functional: Squared Mean]\label{Ex: mean squared}
Let $X\in\mathbf R$ be a random variable, and suppose that we are interested in conducting inference on
\begin{align}\label{Eqn: example, squared}
\phi(\theta_0)=(E[X])^2~.
\end{align}
Here, $\theta_0=E[X]$, $\mathbb D=\mathbb E=\mathbf R$, and $\phi: \mathbf R\to\mathbf R$ is defined by $\phi(\theta)=\theta^2$. In fact, $\phi$ is a special case of the more general quadratic functionals of the form $\|W\theta\|^2$ for $\theta\in\mathbf R^k$ and $W$ a $k\times k$ weighting matrix. This seemingly toy example also arises in VAR models for inference on impulse responses \citep{Benkwitz_Neumann_Lutekpohl2000} and in some nonseparable models with structural measurement errors \citep{Hoderlein_Winter2010}. \qed
\end{ex}

The second example is a special case of the unconditional moment inequality models studied in \citet{CHT2007}, \citet{Romano_Shaikh2008,Romano_Shaikh2010}, \citet{AndrewsandGuggen2009ET}, and \citet{AndrewsandSoares2010}.

\begin{ex}[Unconditional Moment Inequalities]\label{Ex: moment inequalities}
Let $X\in\mathbf R$ be a scalar random variable and suppose we want to test the moment inequality $E[X]\le 0$. The modified method of moments approach is based on estimating the functional
\begin{align}
\phi(\theta_0)=(\max\{\theta_0,0\})^2~,
\end{align}
where $\theta_0=E[X]$, $\mathbb D=\mathbb E=\mathbf R$, and $\phi:\mathbf R\to\mathbf R$ is defined by $\phi(\theta)=(\max\{\theta,0\})^2$. The functional $\phi$ can be easily adapted to handle general moment inequality models.\qed
\end{ex}

The third example concerns the classical Cram\'{e}r-von Mises functional employed to test goodness of fit \citep{Darling1957KSCvM,Vaart1998}.

\begin{ex}[Cram\'{e}r-von Mises Functional]\label{Ex: Cramer-von Mises}
Suppose that we are interested in testing if the distribution function of a random vector $X\in\mathbf R^{d_x}$ is a given function $F_0$. The Cram\'{e}r-von Mises approach considers the functional
\[
\phi(\theta_0)=\int(F-F_0)^2\,dF_0~.
\]
Here, $\theta_0=F$, $\mathbb D=\ell^\infty(\mathbf R^{d_x})$, $\mathbb E=\mathbf R$, and $\phi: \ell^\infty(\mathbf R^{d_x})\to\mathbf R$ is defined to be $\phi(\theta)=\int(\theta-F_0)^2\,dF_0$. More generally, it is possible to test if $F$ belongs to a parametric family $\{F_\gamma: \gamma\in\Gamma\}$ by studying $\phi(\theta_0)=\inf_{\gamma\in\Gamma}\int(\theta_0-F_\gamma)^2\,dF_\gamma$. \qed
\end{ex}

The fourth example, closely related to but significantly different from Example \ref{Ex: Cramer-von Mises}, is based on \citet{Linton2010} for testing stochastic dominance.

\begin{ex}[Stochastic Dominance]\label{Ex: stochastic dominance}
Let $X = (X^{(1)},X^{(2)})^\intercal \in \mathbf R^2$ be continuously distributed, and define the marginal cdfs $F^{(j)}(u) \equiv P(X^{(j)} \leq u)$ for $j \in \{1,2\}$. For a weighting function $w:\mathbf R\rightarrow \mathbf R^+\equiv \{x\in\mathbf R: x\ge 0\}$, \citet{Linton2010} estimate
\begin{equation}\label{Eqn: example, stochastic dominance}
\phi(\theta_0) = \int_{\mathbf R} \max\{F^{(1)}(u) - F^{(2)}(u),0\}^2w(u)du ~,
\end{equation}
to construct a test of whether $X^{(1)}$ first order stochastically dominates $X^{(2)}$. In this example, we set $\theta_0 = (F^{(1)},F^{(2)})$, $\mathbb D = \ell^\infty(\mathbf R)\times \ell^\infty(\mathbf R)$, $\mathbb E = \mathbf R$ and $\phi(\theta) = \int \max\{\theta^{(1)}(u) - \theta^{(2)}(u),0\}^2w(u)du$ for any $\theta\equiv(\theta^{(1)},\theta^{(2)}) \in \ell^\infty(\mathbf R)\times \ell^\infty(\mathbf R)$. We note that the Cram\'{e}r-von Mises type functionals in \citet{Andrews_Shi2013CMI,Andrews_Shi2014CMI} shares the common structure of the functional $\phi$ in \eqref{Eqn: example, stochastic dominance} and hence can be taken care of by our framework as well.\qed
\end{ex}

The fifth example is a special case of the Kolmogorov-Smirnov type functionals for inference on conditional moment inequalities studied by \citet{Andrews_Shi2013CMI}.
\begin{ex}[Conditional Moment Inequalities]\label{Ex: conditional moment inequalities, KS}
Let $Z\in\mathbf R^2$ and $W\in\mathbf R^{d_w}$ be random vectors satisfying $E[Z^{(1)}|W]\le 0$ and $E[Z^{(2)}|W]=0$. For a suitably chosen class of nonnegative functions $\mathcal F$ on $\mathbf  R^{d_w}$, the above conditional moment inequality is equivalent to $E[Z^{(1)}f(W)]\le 0$ and $E[Z^{(2)}f(W)]=0$ for all $f\in\mathcal F$. \citet{Andrews_Shi2013CMI} propose testing the above restriction by estimating the functional
\begin{align}
\phi(\theta_0)=\sup_{f\in \mathcal F}\{[\max(E[Z^{(1)}f(W)],0)]^2+(E[Z^{(2)}f(W)])^2\} ~.
\end{align}
Here, $\theta_0\in\ell^\infty(\mathcal F)\times \ell^\infty(\mathcal F)$ satisfies $\theta_0(f)=E[Zf(W)]$ for all $f\in\mathcal F$, $\mathbb D=\ell^\infty(\mathcal F)\times \ell^\infty(\mathcal F)$, $\mathbb E=\mathbf R$, and $\phi: \mathbb D\to\mathbb E$ is given by $\phi(\theta)=\sup_{f\in\mathcal F}\{[\max(\theta^{(1)}(f),0)]^2+[\theta^{(2)}(f)]^2\}$. \qed
\end{ex}

Our final example is concerned with the $J$-test of overidentification in GMM settings proposed by \citet{Sargan1958iv,Sargan1959IV} and further developed in \citet{Hansen1982}.

\begin{ex}[Overidentification Test]\label{Ex: Overidentification Test}
Let $X\in\mathbf R^{d_x}$ be a random vector and consider the model defined by the moment restriction $E[g(X,\gamma_0)]=0$ for some $\gamma_0\in\Gamma\subset \mathbf R^k$ where $g: \mathbf R^{d_x}\times\Gamma\to\mathbf R^m$ is a known function with $m>k$. The conventional $J$-test can be recast by estimating the functional $\phi$ defined as: for some known $m\times m$ symmetric positive definite matrix $W$,
\begin{align}\label{Eqn: J test}
\phi(\theta_0)=\inf_{\gamma\in\Gamma}E[g(X,\gamma)]^\intercal WE[g(X,\gamma)]~.
\end{align}
Here, $\theta_0\in \prod_{j=1}^m \ell^\infty(\Gamma)$ is defined by $\theta_0(\gamma)=E[g(X,\gamma)]$, $\mathbb D=\prod_{j=1}^m\ell^\infty(\Gamma)$, $\mathbb E=\mathbf R$, and $\phi: \prod_{j=1}^m\ell^\infty(\Gamma)\to\mathbf R$ is defined by $\phi(\theta)=\inf_{\gamma\in\Gamma}\theta(\gamma)^\intercal W\theta(\gamma)$. The bootstrap for the $J$ statistic has been studied by \citet{Hall_Horowitz1996bootstrap} and \citet{Andrews2002higher}. Note that $\theta_0$ is always identified even though $\gamma_0$ is potentially partially identified, which makes $\phi$ second order nondifferentiable as will be shown below. \qed
\end{ex}

\subsection{Concepts of Differentiability}

All examples in the previous subsection exhibit first order degeneracy, i.e., there exist points $\theta$ in $\mathbb D$ such that the first order derivative $\phi_\theta'$ is $0$ and in some cases $\phi$ is not even differentiable at $\theta$, which can be seen from Examples \ref{Ex: mean squared} and \ref{Ex: moment inequalities} respectively. As such, we resort to a second order expansion that handles first order degeneracy and meanwhile accommodates potential nondifferentiability of $\phi$. Let us proceed by recalling notions of first order differentiability \citep{Shapiro1990,Fang_Santos2014HDD}

\begin{defn}\label{Defn: HDD}
Let $\mathbb D$ and $\mathbb E$ be normed spaces equipped with norms $\|\cdot\|_{\mathbb D}$ and $\|\cdot\|_{\mathbb E}$ respectively, and $\phi:\mathbb D_\phi\subseteq \mathbb D\to\mathbb E$.
\begin{itemize}
    \item[(i)] The map $\phi$ is said to be {\it Hadamard differentiable} at $\theta \in\mathbb D_\phi$ {\it tangentially} to a set $\mathbb D_0\subseteq\mathbb D$, if there is a continuous linear map $\phi_\theta':\mathbb D_0\to\mathbb E$ such that:
    \begin{equation}
    \lim_{n\rightarrow \infty}\| \frac{\phi(\theta +t_nh_n)-\phi(\theta)}{t_n} - \phi_\theta'(h)\|_{\mathbb E} = 0 ~,
    \end{equation}
    for all sequences $\{h_n\}\subset\mathbb D$ and $\{t_n\}\subset\mathbf R$ such that $t_n\to 0$, $h_n\to h\in\mathbb D_0$ as $n\to\infty$ and $\theta+t_nh_n\in\mathbb D_\phi$ for all $n$.
    \item[(ii)] The map $\phi$ is said to be {\it Hadamard directionally differentiable} at $\theta \in\mathbb D_\phi$ {\it tangentially} to a set $\mathbb D_0\subseteq\mathbb D$, if there is a continuous map $\phi_\theta':\mathbb D\to\mathbb E$ such that:\footnote{We note that the ``tangential set'' in \citet{Shapiro1991} refers to the domain of $\phi$ (i.e., $\mathbb D_\phi$ in our context), whereas here it refers to the domain $\mathbb D_0$ of the derivative $\phi_\theta'$.}
    \begin{equation}\label{Eqn: HDD 1st}
    \lim_{n\rightarrow \infty}\|\frac{\phi(\theta +t_n h_n)-\phi(\theta)}{t_n} -\phi_\theta'(h)\|_{\mathbb E} = 0 ~,
    \end{equation}
    for all sequences $\{h_n\}\subset\mathbb D$ and $\{t_n\}\subset\mathbf R_+$ such that $t_n\downarrow 0$, $h_n\to h\in\mathbb D_0$ as $n\to\infty$ and $\theta+t_nh_n\in\mathbb D_\phi$ for all $n$.
\end{itemize}
\end{defn}

Inspecting Definition \ref{Defn: HDD}, we see that the main difference between Hadamard differentiability and directional differentiability lies in the linearity of the derivative. This turns out to be the exact gap between these two notions of differentiability. In particular, \eqref{Eqn: HDD 1st} ensures that the directional derivative $\phi_\theta'$ is necessarily continuous and positively homogeneous of degree one, though potentially nonlinear \citep{Shapiro1990}.

Given the introduced notions of differentiability and in view of the remarkable fact that Delta method is valid under even Hadamard directional differentiability in terms of deriving asymptotic distributions \citep{Shapiro1991,Dumbgen1993}, it seems a natural next step to invoke the Delta method. However, in the presence of first order degeneracy, the resulting limiting distribution is degenerate at zero, rendering substantial challenges for inferential purposes. In essence, the Delta method is a stochastic version of Taylor expansion. Therefore, one could go one step further to explore the quadratic term when the linear term is degenerate. We thus follow \citet{Shapiro2000inference} and define

\begin{defn}\label{Defn: HDD 2nd}
Let $\phi:\mathbb D_\phi\subseteq \mathbb D\to\mathbb E$ be a map as in Definition \ref{Defn: HDD}.
\begin{itemize}
\item[(i)] Suppose that $\phi:\mathbb D_\phi\to\mathbb E$ is Hadamard differentiable tangentially to $\mathbb D_0\subset\mathbb D$ such that the derivative $\phi_{\theta}': \mathbb D_0\to\mathbb E$ is well defined on $\mathbb D$. We say that $\phi$ is {\it second order Hadamard differentiable} at $\theta\in\mathbb D_\phi$ {\it tangentially} to $\mathbb D_0$ if there is a bilinear map $\Phi_\theta'':\mathbb D_0\times \mathbb D_0\to\mathbb E$ such that: for $\phi_\theta''(h)\equiv\Phi_\theta''(h,h)$,
\begin{align}\label{Eqn: HD 2nd}
\lim_{n\rightarrow \infty}\|\frac{\phi(\theta +t_n h_n)-\phi(\theta)-t_n\phi_\theta'(h_n)}{t_n^2} -\phi_\theta''(h)\|_{\mathbb E} = 0 ~,
\end{align}
for all sequences $\{h_n\}\subset\mathbb D$ and $\{t_n\}\subset\mathbf R^+$ such that $t_n\to 0$, $h_n\to h\in\mathbb D_0$ as $n\to\infty$ and $\theta+t_nh_n\in\mathbb D_\phi$ for all $n$.
\item[(ii)] Suppose that $\phi:\mathbb D_\phi\to\mathbb E$ is Hadamard directionally differentiable tangentially to $\mathbb D_0\subset\mathbb D$ such that the derivative $\phi_{\theta}': \mathbb D_0\to\mathbb E$ is well defined on $\mathbb D$. We say that $\phi$ is {\it second order Hadamard directionally differentiable} at $\theta\in\mathbb D_\phi$ {\it tangentially} to $\mathbb D_0$ if there is a map $\phi_\theta'':\mathbb D_0\to\mathbb E$ such that:\footnote{Compared with \citet{Shapiro2000inference}, we omitted $\frac{1}{2}$ in the denominator for notational compactness.}
\begin{align}\label{Eqn: HDD 2nd}
\lim_{n\rightarrow \infty}\|\frac{\phi(\theta +t_n h_n)-\phi(\theta)-t_n\phi_\theta'(h_n)}{t_n^2} -\phi_\theta''(h)\|_{\mathbb E} = 0 ~,
\end{align}
for all sequences $\{h_n\}\subset\mathbb D$ and $\{t_n\}\subset\mathbf R^+$ such that $t_n\downarrow 0$, $h_n\to h\in\mathbb D_0$ as $n\to\infty$ and $\theta+t_nh_n\in\mathbb D_\phi$ for all $n$.
\end{itemize}
\end{defn}

The second order derivative $\phi_\theta''$ in both cases is necessarily continuous on $\mathbb D_0$, which can be shown in a straightforward manner as in the proof of Proposition 3.1 in \citet{Shapiro1990}. Similar in spirit to Definition \ref{Defn: HDD},
the key difference between the above two notions of second order differentiability is that the former is a quadratic form corresponding to a bilinear map while the latter is in general only positively homogeneous of degree two, i.e., $\phi_\theta''(th)=t^2\phi_\theta''(h)$ for all $t\ge 0$ and all $h\in\mathbb D_0$. Note that it is possible that $\phi$ is first order Hadamard differentiable but only second order Hadamard directionally differentiable (see Example \ref{Ex: moment inequalities}). In all our examples, $\phi$ is first order Hadamard differentiable though $\phi_{\theta}'$ may be degenerate; see Subsection \ref{Sec: Examples, 2nd HDD}. We stress that requiring $\phi_{\theta}'$ to be well defined on the entirety of $\mathbb D$ does not demand differentiability on $\mathbb D$. Instead, it just means that $\phi_{\theta}'$ can take elements potentially not in $\mathbb D_0$ as arguments. Finally, we note that first and second order (directional) derivatives share the same domain $\mathbb D_0$.

If $\phi_\theta''$ in turn is degenerate, one can go beyond the second order, a possibility we do not pursue at length in this paper; see Remark \ref{Rem: Higher Order}.

\begin{rem}\label{Rem: Higher Order}
Suppose that $\phi:\mathbb D_\phi\subseteq \mathbb D\to\mathbb E$ is $(p-1)$-th order Hadamard directionally differentiable tangentially to $\mathbb D_0\subset\mathbb D$ such that the derivative $\phi_{\theta}^{(j)}: \mathbb D_0\to\mathbb E$ is well defined on $\mathbb D$ for all $j=1,\ldots,p-1$, where $p\ge 2$. Then we say that $\phi$ is $p$th order Hadamard directionally differentiable at $\theta\in\mathbb D_\phi$ tangentially to $\mathbb D_0$ if there is a map $\phi_\theta^{(p)}:\mathbb D_0\to\mathbb E$ such that:
\begin{align}
\phi(\theta +t_n h_n)=\phi(\theta)+\sum_{j=1}^{p-1}t_n^j\phi_\theta^{(j)}(h_n)+t_n^p\phi_\theta^{(p)}(h)+o(t_n^p) ~,
\end{align}
for all sequences $\{h_n\}\subset\mathbb D$ and $\{t_n\}\subset\mathbf R^+$ such that $t_n\downarrow 0$, $h_n\to h\in\mathbb D_0$ as $n\to\infty$ and $\theta+t_nh_n\in\mathbb D_\phi$ for all $n$. Note that, similar to the treatment of $\phi_\theta''$, the factors $1/j!$ are incorporated in the definition of the derivatives $\phi_\theta^{(j)}$ to reflect the nature of them as approximating maps. \citet{Demyanov1974Minimax} established the above high order expansion for $\mathbb D=\mathbf R^k$ with $k\in\mathbf N$ and $\mathbb E=\mathbf E$;\footnote{We thank an anonymous referee for bringing this reference to our attention.} see also \citet{Demyanov2009Minimax}. \qed
\end{rem}

\subsubsection{Examples Revisited}\label{Sec: Examples, 2nd HDD}

From now on, we shall focus on Examples \ref{Ex: mean squared} and \ref{Ex: Overidentification Test} exclusively for conciseness; Examples \ref{Ex: moment inequalities}, \ref{Ex: Cramer-von Mises}, \ref{Ex: stochastic dominance} and \ref{Ex: conditional moment inequalities, KS} will be treated in Appendix \ref{Sec: Results for Exs}.

\begin{exctd}[\ref{Ex: mean squared}]
In this example, the functional involved is second order Hadamard differentiable. Trivially we have
\begin{align}
\phi_\theta'(h)=2\theta h~,\,\phi_\theta''(h)=h^2~.
\end{align}
Note that the first order derivative $\phi_\theta'$ is degenerate when $\theta=0$, whereas $\phi_\theta''$ is everywhere nondegenerate. The bilinear map $\Phi_\theta'': \mathbf R^2\to\mathbf R$ here is given by $\Phi_\theta''(h,g)=hg$. \qed
\end{exctd}

In Example \ref{Ex: Overidentification Test}, the domain $\mathbb D_0$ of the derivative $\phi_{\theta_0}''$ is a strict subset of $\mathbb D$.

\begin{exctd}[\ref{Ex: Overidentification Test}]
Consider $\theta\in\prod_{j=1}^{m}\ell^\infty(\Gamma)$ such that $\theta(\gamma_0)=0$ for some $\gamma_0\in\Gamma$. Then $\phi$ is Hadamard differentiable at $\theta$ and $\phi_{\theta}'(h)=0$ for all $h\in\prod_{j=1}^m\ell^\infty(\Gamma)$. Suppose further that $\Gamma$ is compact and that $\Gamma_0(\theta)\equiv\{\gamma_0\in\Gamma: \theta(\gamma_0)=0\}$ is in the interior of $\Gamma$. For $C^1(\Gamma)$ the space of continuously differentiable functions on $\Gamma$, if $\theta\in \prod_{j=1}^m C^1(\Gamma)$, then by Lemma \ref{Lem: over ID, HDD}, under additional regularity conditions, $\phi$ is second order Hadamard directionally differentiable at $\theta$ tangentially to $\prod_{j=1}^m C(\Gamma)$ with the derivative given by: for any $h\in \prod_{j=1}^m C(\Gamma)$,
\[
\phi_{\theta}''(h)=\min_{\gamma_0\in\Gamma_0(\theta)}h(\gamma_0)^\intercal W^{1/2}M(\gamma_0) W^{1/2} h(\gamma_0)~,
\]
where $M(\gamma_0)=I_m-W^{1/2}J(\gamma_0)[J(\gamma_0)^\intercal W J(\gamma_0)]^{-1}J(\gamma_0)^\intercal W^{1/2}$ with $J(\gamma_0)\equiv\frac{d\theta(\gamma)}{d\gamma^\intercal}\big|_{\gamma=\gamma_0}$ the Jacobian matrix and $I_m$ the identity matrix of size $m$. Here, invertibility of $J(\gamma_0)$ is an implied requirement in Lemma \ref{Lem: over ID, HDD}; see Remark \ref{Rem: over ID, HDD}. Note that if $\gamma_0$ is point identified, then $\phi$ becomes second order Hadamard differentiable with
\[
\phi_{\theta}''(h)=h(\gamma_0)^\intercal W^{1/2}M(\gamma_0) W^{1/2} h(\gamma_0)~,
\]
which in turn yields $\chi^2(m-k)$ as the asymptotic distribution of the $J$-statistic under optimal weighting. We emphasize that the regularity conditions in Lemma \ref{Lem: over ID, HDD} are sufficient for applying our framework but by no means necessary -- as explained in Section \ref{Sec: Application}, those sufficient conditions exclude the setup of \citet{Dovonon_Renault2013testing}, and so we shall provide an alternative set of sufficient conditions there. \qed
\end{exctd}

\subsection{Second Order Delta Method}\label{Sec: Delta method 2nd}

The Delta method for potentially directionally differentiable maps as well as differentiable ones has proven powerful in asymptotic analysis \citep{Vaart1998,Shapiro1991,Fang_Santos2014HDD,Hansen2015regression}. Unfortunately, it is insufficient to handle substantial challenges for inference arising from first order degeneracy. Heuristically, if $r_n\{\hat\theta_n-\theta_0\}\convl\mathbb G$ and $\phi_{\theta_0}'=0$, then the Delta method implies that
\[
r_n\{\phi(\hat\theta_n)-\phi(\theta_0)\}\convl \phi_{\theta_0}'(\mathbb G)\equiv 0~.
\]
For real-valued $\phi$, the usual confidence interval for $\phi(\theta_0)$ at asymptotic level $1-\alpha$ is
\begin{align}\label{Eqn: CI given 1st order degeneracy}
[\phi(\hat\theta_n)-\frac{c_{1-\alpha/2}}{r_n},\phi(\hat\theta_n)-\frac{c_{\alpha/2}}{r_n}]=\{\phi(\hat\theta_n)\}~,
\end{align}
where the $c_\alpha$ is the $\alpha$-th quantile of $\phi_{\theta_0}'(\mathbb G)\equiv 0$ and is zero for all $\alpha\in(0,1)$. Clearly, $P(\phi(\theta_0)\in \{\phi(\hat\theta_n)\})=0$ if, for example, $\phi(\hat\theta_n)$ is a continuous random variable.

To circumvent the above difficulty, we impose the following conditions in order to obtain a suitable second order Delta method.

\begin{ass}\label{Ass: HDD 2nd}
(i) $\mathbb D$ and $\mathbb E$ are normed spaces with norms $\|\cdot\|_{\mathbb D}$ and $\|\cdot\|_{\mathbb E}$ respectively; (ii) $\phi:\mathbb D_\phi\subset\mathbb D\to\mathbb E$ is second order Hadamard directionally differentiable at $\theta_0\in\mathbb D_\phi$ tangentially to $\mathbb D_0\subset\mathbb D$; (iii) $\phi_{\theta_0}'(h)=0$ for all $h\in\mathbb D_0$.
\end{ass}

\begin{ass}\label{Ass: weak convergence for theta}
(i) There is $\hat\theta_n:\{X_i\}_{i=1}^n\to\mathbb D_\phi$ such that $r_n\{\hat\theta_n-\theta_0\}\convl \mathbb G$ in $\mathbb D$ for some $r_n\uparrow\infty$; (ii) $\mathbb G$ is tight and its support is in $\mathbb D_0$;\footnote{The support of $\mathbb G$ is the set of points in $\mathbb D$ all of whose open neighborhoods have positive probability.} (iii) $\mathbb D_0$ is closed under vector addition, i.e., $h_1+h_2\in\mathbb D_0$ whenever $h_1,h_2\in\mathbb D_0$.
\end{ass}

Assumption \ref{Ass: HDD 2nd} formalizes the requirement that the map $\phi:\mathbb D_\phi \rightarrow \mathbb E$ be second order Hadamard directionally differentiable at $\theta_0$, and the defining feature of this paper, namely, degeneracy of the first order derivative. Assumption \ref{Ass: weak convergence for theta}(i) defines another key ingredient: there is an estimator $\hat\theta_n$ for $\theta_0$ that admits a weak limit $\mathbb G$ at a potentially non-$\sqrt n$ rate $r_n$; see Remark \ref{Rem: rate}. Assumption \ref{Ass: weak convergence for theta}(ii) ensures that the support of $\mathbb G$ is included in the domain of the derivative $\phi_{\theta_0}''$ so that $\phi_{\theta_0}''(\mathbb G)$ is well defined, while tightness of $\mathbb G$ is only a minimal requirement. Assumption \ref{Ass: weak convergence for theta}(iii) is a mild condition, which shall play a technical role in the proof of our bootstrap results.

Given Assumptions \ref{Ass: HDD 2nd} and \ref{Ass: weak convergence for theta}, we now present a second order Delta method building upon \citet{Shapiro2000inference} and \citet{Romish2004delta} but without requiring $\mathbb D_\phi$ to be convex.

\begin{thm}\label{Thm: delta method, 2nd order}
If Assumptions \ref{Ass: HDD 2nd}(i)(ii) and \ref{Ass: weak convergence for theta}(i)(ii) hold, then\footnote{The term $\phi_{\theta_0}''(r_n\{\hat\theta_n-\theta_0\})$ is interpreted as some continuous extension of $\phi_{\theta_0}''$ (which always exists in our setup) evaluated at $r_n\{\hat\theta_n-\theta_0\}$ whenever $r_n\{\hat\theta_n-\theta_0\}\notin \mathbb D_0$; see the comment preceding the proof of Theorem \ref{Thm: delta method, 2nd order}. Since \eqref{Eqn: delta method, 2nd, prob} is an asymptotic result, the choice of the continuous extension is irrelevant.}
\begin{align}\label{Eqn: delta method, 2nd, prob}
r_n^2\{\phi(\hat\theta_n)-\phi(\theta_0)-\phi_{\theta_0}'(\hat\theta_n-\theta_0)\}=\phi_{\theta_0}''(r_n\{\hat\theta_n-\theta_0\})+o_p(1)~.
\end{align}
and hence
\begin{align}\label{Eqn: delta method, 2nd}
r_n^2\{\phi(\hat\theta_n)-\phi(\theta_0)-\phi_{\theta_0}'(\hat\theta_n-\theta_0)\}\convl \phi_{\theta_0}''(\mathbb G)~.
\end{align}
\end{thm}

The essence of Theorem \ref{Thm: delta method, 2nd order} is in complete accord with that underlying the first order Delta method. In particular, the definition of second order Hadamard directional differentiability is engineered so that the second order Delta method is nothing more than a stochastic version of the Taylor expansion of order two, i.e.,
\begin{align*}
\phi(\theta_0+t_nh_n)=\phi(\theta_0)+t_n\phi_{\theta_0}'(h_n)+t_n^2\phi_{\theta_0}''(h)+o(t_n^2)~,
\end{align*}
where $t_n$ corresponds to $r_n^{-1}$, and $h_n$ to $r_n\{\hat\theta_n-\theta_0\}$. Note that Theorem \ref{Thm: delta method, 2nd order} is valid regardless of the nature of the differentiability (i.e., fully differentiable or directionally differentiable) and the presence of first order degeneracy. When $\phi_{\theta_0}'$ is degenerate, the convergence \eqref{Eqn: delta method, 2nd} simplifies to
\begin{align}\label{Eqn: delta method with FOD, 2nd}
r_n^2\{\phi(\hat\theta_n)-\phi(\theta_0)\}\convl \phi_{\theta_0}''(\mathbb G)~.
\end{align}
Finally, we note that higher order versions of the Delta method can be developed along the lines of Remark \ref{Rem: Higher Order}; see Remark \ref{Rem: Higher order, delta}.

\begin{rem}\label{Rem: Higher order, delta}
Suppose that Assumptions \ref{Ass: HDD 2nd}(i) and \ref{Ass: weak convergence for theta}(i)(ii) hold and $\phi$ is $p$-th order Hadamard directionally differentiable at $\theta_0\in\mathbb D_\phi$ tangentially to $\mathbb D_0$. It follows that
\[
r_n^p\big[\phi(\hat\theta_n)-\{\phi(\theta_0)+\sum_{j=1}^{p-1}\phi_{\theta_0}^{(j)}(\hat\theta_n-\theta_0)\}\big]=
\phi_{\theta_0}^{(p)}(r_n\{\hat\theta_n-\theta_0\})+o_p(1)~,
\]
and hence
\[
r_n^p\big[\phi(\hat\theta_n)-\{\phi(\theta_0)+\sum_{j=1}^{p-1}\phi_{\theta_0}^{(j)}(\hat\theta_n-\theta_0)\}\big]\convl\phi_{\theta_0}^{(p)}(\mathbb G)~. \tag*{$\qed$}
\]
\end{rem}

\section{The Bootstrap}\label{Sec: Bootsrtap}

Establishing asymptotic distributions as in Theorem \ref{Thm: delta method, 2nd order} is the first step towards conducting statistical inference on $\phi(\theta_0)$, the usefulness of which relies on our ability to accurately estimate the limiting law. In this section, we discuss how first order degeneracy of $\phi$ can complicate inference using the standard bootstrap based on first and especially second order asymptotics, and provide alternative consistent resampling schemes.

\subsection{Bootstrap Setup}

Throughout, we let $\hat \theta_n^*$ denote a ``bootstrapped version" of $\hat \theta_n$, which is defined as a function mapping the data $\{X_i\}_{i=1}^n$ and random weights $\{W_i\}_{i=1}^n$ that are independent of $\{X_i\}_{i=1}^n$ into the domain $\mathbb D_\phi$ of $\phi$. This general definition allows us to include diverse resampling schemes such as nonparametric, Bayesian, block, score, more generally multiplier and exchangeable bootstrap as special cases. Next, making sense of bootstrap consistency necessitates a metric that quantifies distances between probability measures. As is standard in the literature, we employ the bounded Lipschitz metric $d_{\bl}$ formalized by \citet{Dudley1966Baire,Dudley1968distance}: for two Borel probability measures $L_1$ and $L_2$ on $\mathbb D$, define
\[
d_{\bl}(L_1,L_2)\equiv\sup_{f\in\bl_1(\mathbb D)}|\int f\,dL_1-\int f\,dL_2| ~,
\]
where we recall that $\bl_1(\mathbb D)$ denotes the set of Lipschitz functionals whose absolute level and Lipschitz constant are bounded by one, i.e.,
\begin{equation*}
\text{BL}_1(\mathbb D) \equiv \{f : \mathbb D \rightarrow \mathbf R : \sup_{t\in \mathbb D} |f(t)| + \sup_{t_1,t_2\in \mathbb D, t_1\neq t_2}\frac{|f(t_1) - f(t_2)|}{\|t_1-t_2\|_{\mathbb D}}\le 1\} ~.
\end{equation*}

Since weak convergence in the Hoffmann-J{\o}rgensen sense to separable limits can be metrized by $d_{\bl}$ \citep{Dudley1990nonlinear,Vaart_Wellner1990prohorov}, we may now measure the distance between the ``conditional law'' of $\hat{\mathbb G}_n^*\equiv r_n\{\hat\theta_n^*-\hat\theta_n\}$ given $\{X_i\}$ and the limiting law of $r_n\{\hat\theta_n-\theta_0\}$ by
\begin{align}\label{Eqn: distance based on BL}
d_{\bl}(\hat{\mathbb G}_n^*,\mathbb G)=\sup_{f\in\bl_1(\mathbb D)}|E_W[f(r_n\{\hat\theta_n^*-\hat\theta_n\})]-E[f(\mathbb G)]| ~,
\end{align}
where $E_W$ denotes expectation with respect to the bootstrap weights $\{W_i\}_{i=1}^n$ holding the data $\{X_i\}_{i=1}^n$ fixed. Employing the distribution of $r_n\{\hat \theta_n^* - \hat \theta_n\}$ conditional on the data as an approximation to the distribution of $\mathbb G$ is then asymptotically justified if their distance, equivalently \eqref{Eqn: distance based on BL}, converges in probability to zero.

We formalize the above discussion by imposing the following assumptions on $\hat \theta_n^*$.

\begin{ass}\label{Ass: bootstrap for theta}
(i) $\hat \theta_n^* : \{X_i,W_i\}_{i=1}^n \rightarrow \mathbb D_\phi$ with $\{W_i\}_{i=1}^n$ independent of $\{X_i\}_{i=1}^n$; (ii) $\hat \theta_n^*$ satisfies $\sup_{f \in \text{BL}_1(\mathbb D)} |E_W[f(r_n\{\hat \theta_n^* - \hat \theta_n\})] - E[f(\mathbb G)]| = o_p(1)$.
\end{ass}

\begin{ass}\label{Ass: measurability}
(i) $E[f(r_n\{\hat \theta_n^* - \hat \theta_n\})^*]-E[f(r_n\{\hat \theta_n^* - \hat \theta_n\})_*]\to 0$ for all $f\in\text{BL}_1(\mathbb D)$ where $f(r_n\{\hat \theta_n^* - \hat \theta_n\})^*$ and $f(r_n\{\hat \theta_n^* - \hat \theta_n\})_*$ denote minimal measurable majorant and maximal measurable minorant (with respect to $\{X_i,W_i\}_{i=1}^n$ jointly) respectively; (ii) $f(r_n\{\hat \theta_n^* - \hat \theta_n\})$ is a measurable function of $\{W_i\}_{i=1}^n$ outer almost surely in $\{X_i\}_{i=1}^n$ for any continuous and bounded $f:\mathbb D \rightarrow \mathbf R$.
\end{ass}

Assumption \ref{Ass: bootstrap for theta}(i) formally defines the bootstrap analog $\hat \theta_n^*$ of $\hat\theta_n$, while Assumption \ref{Ass: bootstrap for theta}(ii) simply imposes the consistency of the ``law'' of $r_n\{\hat \theta_n^* - \hat \theta_n\}$ conditional on the data for the law of $\mathbb G$, i.e., the bootstrap ``works" for the estimator $\hat \theta_n$. Assumption \ref{Ass: measurability} is of technical concern. In particular, Assumption \ref{Ass: measurability}(i) can often be established as a result of bootstrap consistency \citep{Vaart1996}, while Assumption \ref{Ass: measurability}(ii) is easy to verify for particular resampling schemes. For example, if $\{W_i\}_{i=1}^n\mapsto f(r_n\{\hat \theta_n^* - \hat \theta_n\})$ is continuous, then Assumption \ref{Ass: measurability}(ii) is fulfilled. When $\theta_0$ is Euclidean-valued, i.e., $\mathbb D=\mathbf R^k$ with $k\in\mathbf N$, one can dispense with Assumption \ref{Ass: measurability}.

\subsection{Failures of the Standard Bootstrap}

We now turn to the challenges for inferences using the standard bootstrap caused by first order degeneracy. As is well known in the literature, the law of
\begin{align}\label{Eqn: standard bootstrap, first order}
r_n\{\phi(\hat\theta_n^*)-\phi(\hat\theta_n)\}
\end{align}
conditional on the data provides a consistent estimator of the law of $\phi_{\theta_0}'(\mathbb G)$ provided $\phi$ is Hadamard differentiable \citep{Vaart1996}, which in particular includes the case when $\phi_{\theta_0}'=0$. In other words, the standard bootstrap, meaning the law of \eqref{Eqn: standard bootstrap, first order} conditional on the data, is consistent for the law of $\phi_{\theta_0}'(\mathbb G)$ regardless of the presence of first order degeneracy.

Substantial difficulties, however, arise from using \eqref{Eqn: standard bootstrap, first order} for inferential purposes when first order degeneracy does occur. Ignoring the first order degeneracy or perhaps as a way to avoid ridiculous confidence intervals such as \eqref{Eqn: CI given 1st order degeneracy}, one might consider the following confidence interval for real-valued $\phi(\theta_0)$:
\begin{align}\label{Eqn: CI given 1st order degeneracy, II}
[\phi(\hat\theta_n)-\frac{\tilde c_{1-\alpha/2}}{r_n}, \phi(\hat\theta_n)-\frac{\tilde c_{\alpha/2}}{r_n}]~,
\end{align}
where $\tilde c_{1-\alpha}$ is the $(1-\alpha)$-th bootstrapped quantile for $\alpha\in(0,1)$ defined as
\[
\tilde c_{1-\alpha}\equiv \inf\{c\in\mathbf R: P_W(r_n\{\phi(\hat\theta_n^*)-\phi(\hat\theta_n)\}\le c)\ge 1-\alpha\}~.
\]
However, establishing the validity of \eqref{Eqn: CI given 1st order degeneracy, II} as a level $1-\alpha$ confidence interval for $\phi(\theta_0)$ is problematic because $\tilde c_{1-\alpha}\convp 0$ for all $\alpha\in(0,1)$ and $0$ is a discontinuity point of the cdf of the limit (see Lemma \ref{Lem: degenerate limit, cv}).

In fact, simple algebra reveals that \eqref{Eqn: CI given 1st order degeneracy, II} is numerically identical to
\begin{align}\label{Eqn: CI given 1st order degeneracy, III}
[\phi(\hat\theta_n)-\frac{\bar c_{1-\alpha/2}}{r_n^2}, \phi(\hat\theta_n)-\frac{\bar c_{\alpha/2}}{r_n^2}]~,
\end{align}
where $\bar c_\alpha$ is defined as
\[
\bar c_{1-\alpha}\equiv \inf\{c\in\mathbf R: P_W(r_n^2\{\phi(\hat\theta_n^*)-\phi(\hat\theta_n)\}\le c)\ge 1-\alpha\}~.
\]
In other words, $\bar c_\alpha$ is the $\alpha$-th bootstrapped quantile of the standard bootstrap based on second order asymptotics:
\begin{align}\label{Eqn: standard bootstrap, second order}
r_n^2\{\phi(\hat\theta_n^*)-\phi(\hat\theta_n)\}~.
\end{align}
As illustrated by \citet{Babu1984bootstrap} for the squared mean example, the conditional law of \eqref{Eqn: standard bootstrap, second order} is inconsistent for the law of $\phi_{\theta_0}''(\mathbb G)$ when $\theta_0=0$, the point at which first order degeneracy arises. We next demonstrate that the bootstrap failure in this simple example is a reflection of a deeper principle: the second order standard bootstrap is consistent if and only if $\phi_{\theta_0}''$ is degenerate, under regularity conditions.

\begin{thm}\label{Thm: bootstrap, iff}
Suppose that Assumptions \ref{Ass: HDD 2nd}, \ref{Ass: weak convergence for theta}, \ref{Ass: bootstrap for theta} and \ref{Ass: measurability} hold, and that $\mathbb G$ is centered Gaussian. Then $\phi_{\theta_0}''=0$ on the support of $\mathbb G$ if and only if
\begin{align}\label{Eqn: bootstrap, iff}
\sup_{f\in\bl_1(\mathbb E)}|E_W[f(r_n^2\{\phi(\hat\theta_n^*)-\phi(\hat\theta_n)\})]-E[f(\phi_{\theta_0}''(\mathbb G))]|=o_p(1)~.
\end{align}
If, in addition, $\phi$ is second order Hadamard differentiable, then the conclusion holds without requiring $\mathbb G$ to be centered Gaussian.
\end{thm}

The sufficiency part of the theorem is somewhat expected and not a deep result, while the necessity is perhaps surprising and has far-reaching implications for statistical inference as we shall detail shortly. The proof of the latter consists of two steps: in the first step, we show that bootstrap consistency as in \eqref{Eqn: bootstrap, iff} implies existence of a bilinear map $\Phi_{\theta_0}''$ corresponding to $\phi_{\theta_0}''$, in similar fashion as the proof of Theorem 3.1 in \citet{Fang_Santos2014HDD}; in the second step, we establish that $\Phi_{\theta_0}''$ and hence $\phi_{\theta_0}''$ is necessarily degenerate. Both steps involve the insights of equating distributions through their characteristic functionals as in \citet{Vaart1991differentibility} and \citet{Hirano_Porter2012}.

Theorem \ref{Thm: bootstrap, iff} implies that, in the presence of first order degeneracy, if the second order derivative $\phi_{\theta_0}''$ is nondegenerate, then the standard bootstrap based on second order asymptotics is necessarily inconsistent whenever $\mathbb G$ is centered Gaussian. If $\phi_{\theta_0}''$ is degenerate, we have a degenerate limiting distribution that can not be directly used for inference. We thus conclude that bootstrap failure is an inherent implication of models with first order degeneracy.

Heuristically, the reason why the standard bootstrap fails is that even though $r_n^2\phi_{\theta_0}'(\hat\theta_n-\theta_0)=0$ in the ``real world'', its bootstrap counterpart is non-negligible. To see this, consider the squared mean example. If $\theta_0=0$, then
\[
n\phi_{\hat\theta_n}'(\hat\theta_n^*-\hat\theta_n)=n 2\hat\theta_n\cdot\{\hat\theta_n^*-\hat\theta_n\}=2\sqrt{n}\{\hat\theta_n-\theta_0\}\cdot\sqrt n\{\hat\theta_n^*-\hat\theta_n\}=O_p(1)~.
\]
This is an emphatic reflection of \citet{Efron1979}'s caveat that the bootstrap, as well as other resampling schemes, provides frequency approximations rather than likelihood approximations. These heuristics suggest that the standard bootstrap might work if the first order term $r_n^2\phi_{\hat\theta_n}'(\hat\theta_n^*-\hat\theta_n)$ is included, which turns out to be true for sufficiently smooth maps; see Theorem \ref{Thm: Babu extension}.

It is worth noting that Theorem \ref{Thm: bootstrap, iff} holds even if $\phi$ is smooth. Consequently, first order degeneracy is a source of bootstrap inconsistency completely different from that discussed in \citet{Fang_Santos2014HDD}, i.e., nondifferentiability of $\phi$. In addition, we note that, without the qualifier that $\mathbb G$ is centered Gaussian, bootstrap consistency \eqref{Eqn: bootstrap, iff} holds if and only if $\phi_{\theta_0}''(\mathbb G+h)-\phi_{\theta_0}''(h)\overset{d}{=}\phi_{\theta_0}''(\mathbb G)$ for all $h\in\mathrm{Supp}(\mathbb G)$ under mild support conditions; see Theorem A.1 in \citet{Fang_Santos2014HDD}.

Finally, to further articulate the relations between the current work and that of \citet{Fang_Santos2014HDD}, we present a table that describes the scopes we work in.

\renewcommand\cellgape{\Gape[5pt]}
\begin{table}[!htbp]
\centering
\begin{threeparttable}
\caption{Comparison with \citet{Fang_Santos2014HDD}}
\begin{tabular}{cccc}
\hline\hline
\multirow{2}{*}{ } & \multirow{2}{*}{ } & \multicolumn{2}{c}{First Order Degeneracy (i.e.\ $\phi_{\theta_0}'=0$)} \\ 
\cmidrule{3-4}
 &  & Yes & No\\
\hline
\multirowcell{2}[-2ex]{ Nondifferentiability\\ (1st or 2nd order)} & Yes & \makecell{This paper\\ $r_n^2\{\phi(\hat\theta_n^*)-\phi(\hat\theta_n)\}\overset{L^*}{\nrightarrow}\phi_{\theta_0}''(\mathbb G)$} & \makecell{\citet{Fang_Santos2014HDD}\\ $r_n\{\phi(\hat\theta_n^*)-\phi(\hat\theta_n)\}\overset{L^*}{\nrightarrow}\phi_{\theta_0}'(\mathbb G)$} \\
\cmidrule{3-4}
 & No &  \makecell{This paper\\ $r_n^2\{\phi(\hat\theta_n^*)-\phi(\hat\theta_n)\}\overset{L^*}{\nrightarrow}\phi_{\theta_0}''(\mathbb G)$} & \makecell{Standard\\ $r_n\{\phi(\hat\theta_n^*)-\phi(\hat\theta_n)\}\overset{L^*}{\to}\phi_{\theta_0}'(\mathbb G)$}\\
\hline\hline
\end{tabular}\label{Tab: compare with FS}
\end{threeparttable}
\begin{tablenotes}
\item 1. $L^*$ signifies conditional weak convergence (made precise by, for example, $d_{\mathrm{BL}}$).
\item 2. It is assumed that $r_n\{\hat\theta_n-\theta_0\}\convl\mathbb G$ and $r_n\{\hat\theta_n^*-\hat\theta_n\}\overset{L^*}{\rightarrow}\mathbb G$.
\item 3. Since $\phi$ is first order differentiable when $\phi_{\theta_0}'=0$, the nondifferentiability is meant in the second order for the third column and the first order for the last column.
\end{tablenotes}
\end{table}

\subsection{The Babu Correction}

We now extend the Babu correction under our more general setup. We proceed by imposing the following assumption.

\begin{ass}\label{Ass: HDD 2nd, refined}
(i) The map $\phi: \mathbb D_\phi\subset\mathbb D\to\mathbb E$ is second order Hadamard differentiable at $\theta_0\in\mathbb D_\phi$ tangentially to $\mathbb D_0$; (ii) $\phi$ is first order Hadamard differentiable at every point in some neighborhood of $\theta_0$ tangentially to $\mathbb D_0$ such that \footnote{The appearance of the factor 2 is due to omission of the factor $1/2$ in Definition \ref{Defn: HDD 2nd}.}
\begin{align}\label{Eqn: HDD 2nd, refined}
\lim_{n\to\infty}\|\frac{\phi_{\theta_0+t_ng_n}'(h_n)-\phi_{\theta_0}'(h_n)}{t_n}-2\Phi_{\theta_0}''(g,h)\|_{\mathbb E}=0~,
\end{align}
for all sequences $\{g_n,h_n\}\subset\mathbb D$ and $\{t_n\}\subset\mathbf R^+$ such that $t_n\downarrow 0$, $(g_n,h_n)\to (g,h)\in\mathbb D_0\times\mathbb D_0$ as $n\to\infty$ and $\theta+t_ng_n, \theta+t_nh_n\in\mathbb D_\phi$ for all sufficiently large $n$, where $\Phi_{\theta_0}'': \mathbb D_0\times\mathbb D_0\to\mathbb E$ is the bilinear map underlying $\phi_{\theta_0}''$.
\end{ass}

Assumption \ref{Ass: HDD 2nd, refined}(i) defines the scope of the Babu correction: it shall be applied to smooth maps, which excludes, for example, the functional associated with the $J$-test in GMM settings when first order or global identification fails -- see Section \ref{Sec: Application}. Assumption \ref{Ass: HDD 2nd, refined}(ii) is stronger than $\phi$ being simply second order Hadamard differentiable, in that it requires the existence of first order derivative at all points in a neighborhood of $\theta_0$ such that \eqref{Ass: HDD 2nd, refined} holds. Assumption \ref{Ass: HDD 2nd, refined} is fulfilled for the setup considered in \citet{Babu1984bootstrap} and for Examples \ref{Ex: mean squared} and \ref{Ex: Cramer-von Mises}, but violated for the remaining examples.

Under Assumption \ref{Ass: HDD 2nd, refined}, the corrected bootstrap
\begin{align}\label{Eqn: Babu extension}
r_n^2\{\phi(\hat\theta_n^*)-\phi(\hat\theta_n)-\phi_{\hat\theta_n}'(\hat\theta_n^*-\hat\theta_n)\}
\end{align}
is consistent for the law of $\phi_{\theta_0}''(\mathbb G)$ regardless of the degeneracy of $\phi_{\theta_0}'$.

\begin{thm}\label{Thm: Babu extension}
If Assumptions \ref{Ass: HDD 2nd}(i)(ii), \ref{Ass: weak convergence for theta}, \ref{Ass: bootstrap for theta}, \ref{Ass: measurability} and \ref{Ass: HDD 2nd, refined} hold, then
\begin{align}\label{Eqn: Babu extension thm}
\sup_{f\in\bl_1(\mathbb E)}|E_W[f(r_n^2\{\phi(\hat\theta_n^*)-\phi(\hat\theta_n)-\phi_{\hat\theta_n}'(\hat\theta_n^*-\hat\theta_n)\})]-E[f(\phi_{\theta_0}''(\mathbb G))]|=o_p(1)~.
\end{align}
\end{thm}

Theorem \ref{Thm: Babu extension} generalizes \citet{Babu1984bootstrap} considerably in that it accommodates semiparametric and nonparametric models, and allows wider resampling schemes beyond the nonparametric bootstrap of \citet{Efron1979}. The Babu correction works nicely with smooth maps in the sense of Assumption \ref{Ass: HDD 2nd, refined}, but unfortunately is inadequate to handle nonsmooth ones. This is because when $\phi$ is only second order directionally differentiable, often times the derivative $\phi_{\theta_{0}}''$ is not ``continuous'' in $\theta_{0}$, implying that the Babu correction \eqref{Eqn: Babu extension} is unable to estimate $\phi_{\theta_{0}}''$ properly and in this way results in inconsistent estimates. For this reason, we next provide yet another resampling method which accommodates (second order) nondifferentiable maps.

\subsection{A Modified Bootstrap}\label{Sec: modified bootstrap}

In this subsection, we shall present a modified bootstrap following \citet{Fang_Santos2014HDD} that is consistent for the law of $\phi_{\theta_0}''(\mathbb G)$, and adaptive to both the presence of first order degeneracy and nondifferentiability of $\phi$.

The heuristics underlying our proposal, however, are connected to those in \citet{Fang_Santos2014HDD} in a subtle way. In the context of first order asymptotics where $\phi$ is only directionally differentiable, inconsistency of the standard bootstrap arises from its inability to properly estimate the directional derivative $\phi_{\theta_0}'$. In our setup, however, there are examples in which the derivative $\phi_{\theta_0}''$ is a known map; see Examples \ref{Ex: mean squared} and \ref{Ex: Cramer-von Mises} which are all differentiable maps. The standard bootstrap in these settings fails because there is a non-negligible term being neglected. However, in all other examples where $\phi$ is not smooth enough, \citet{Fang_Santos2014HDD}'s arguments will come into play as well.

In any case, the second order weak limit $\phi_{\theta_0}''(\mathbb G)$ is a composition of the derivative $\phi_{\theta_0}''$ and the limit $\mathbb G$ of $\hat\theta_n$, as is the first order limit $\phi_{\theta_0}'(\mathbb G)$. Thus, the law of $\phi_{\theta_0}''(\mathbb G)$ can be estimated by composing a suitable estimator $\hat\phi_n''$ for $\phi_{\theta_0}''$ with a consistent bootstrap approximation for the law of $\mathbb G$, in exactly the same fashion as the resampling scheme proposed by \citet{Fang_Santos2014HDD}. That is, we propose employing the law of
\begin{align}\label{Eqn: modified bootstrap0}
\hat\phi_n''(r_n\{\hat\theta_n^*-\hat\theta_n\})
\end{align}
conditional on the data as an approximation for the law of $\phi_{\theta_0}''(\mathbb G)$, where $\hat\phi_n'': \mathbb D\to\mathbb E$ is a suitable estimator of $\phi_{\theta_0}''$. Certainly, we would like $\hat\phi_n''$ to converge to $\phi_{\theta_0}''$ in some sense as $n\to\infty$. This can be made precise as follows.

\begin{ass}\label{Ass: derivative estimation}
$\hat \phi_n'' : \mathbb D \rightarrow \mathbb E$ is a function of $\{X_i\}_{i=1}^n$ satisfying that for every sequence $\{h_n\}\subset\mathbb D$ and every $h\in\mathbb D_0$ such that $h_n\to h$ as $n\to\infty$,
\begin{align}\label{Eqn: derivative estimation, II}
\hat\phi_n''(h_n)\convp \phi_{\theta_0}''(h)~.
\end{align}
\end{ass}

Assumption \ref{Ass: derivative estimation} says that $\hat\phi_n''$ converges in probability to $\phi_{\theta_0}''$ along any convergent sequence $h_n\to h$ as $n\to\infty$. In cases when $\phi_{\theta_0}''$ is a known map, we may simply set $\hat\phi_n''=\phi_{\theta_0}''$ for all $n\in\mathbf N$. It is worth noting that Assumption \ref{Ass: derivative estimation} is equivalent to requiring: for every compact set $K\subset\mathbb D_0$ and every $\epsilon>0$,
\begin{align}\label{Eqn: derivative estimation}
\lim_{\delta\darrow 0}\limsup_{n\rightarrow \infty} P\Big( \sup_{h \in K^{\delta}} \| \hat \phi_n''(h) - \phi_{\theta_0}''(h)\|_{\mathbb E} > \epsilon\Big) = 0~,
\end{align}
where $K^{\delta} \equiv \{a \in \mathbb D : \inf_{b \in K} \|a - b\|_{\mathbb D} < \delta\}$; see Lemma \ref{Lem: derivative estimation, equiv}. Condition \eqref{Eqn: derivative estimation} was employed in \citet{Fang_Santos2014HDD} who also provided several sufficient conditions for it to hold. For example, if $\hat \phi_n'' : \mathbb D \rightarrow \mathbb E$ is Lipschitz continuous, then pointwise consistency of $\hat \phi_n''$ suffices for \eqref{Eqn: derivative estimation}. Unfortunately, second order derivatives often lack uniform continuity and hence those sufficient conditions are inapplicable. Nonetheless, condition \eqref{Eqn: derivative estimation, II} is straightforward to verify in all our examples.

Given the equivalence of conditions \eqref{Eqn: derivative estimation, II} and \eqref{Eqn: derivative estimation}, consistency of our modified bootstrap \eqref{Eqn: modified bootstrap0} follows from Theorem 3.2 in \citet{Fang_Santos2014HDD}.

\begin{thm}\label{Thm: modified bootstrap}
Under Assumptions \ref{Ass: HDD 2nd}(i)(ii), \ref{Ass: weak convergence for theta}, \ref{Ass: bootstrap for theta}, \ref{Ass: measurability} and \ref{Ass: derivative estimation}, it follows that
\begin{equation}\label{Eqn: modified bootstrap}
\sup_{f \in \bl_1(\mathbb E)} |E_W[f(\hat \phi_n''(r_n\{\hat \theta_n^* - \hat \theta_n\}))] - E[f(\phi_{\theta_0}''(\mathbb G))]| = o_p(1) ~.
\end{equation}
\end{thm}

Theorem \ref{Thm: modified bootstrap} shows that the law of $\hat \phi_n''(r_n\{\hat \theta_n^* - \hat \theta_n\})$ conditional on the data is indeed consistent for the law of $\phi_{\theta_0}''(\mathbb G)$, regardless of the degree of smoothness of $\phi$ and degeneracy of $\phi_{\theta_0}'$. Interestingly, the resampling scheme in Theorem \ref{Thm: modified bootstrap} is a mixture of the classical bootstrap and analytical asymptotic approximations. Finally, we note that Assumption \ref{Ass: derivative estimation} allows us to think of Theorem \ref{Thm: modified bootstrap} as a variant of the extended continuous mapping theorem.

Theorems \ref{Thm: Babu extension} and \ref{Thm: modified bootstrap} are useful for hypothesis testing. Specifically, consider
\begin{align}\label{Eqn: hypothesis}
\mathrm H_0: \phi(\theta_0)=0 \hspace{0.5in} \mathrm H_1: \phi(\theta_0)>0~.
\end{align}
Under first order degeneracy, as is the case in all our examples, we employ the test of rejecting $\mathrm H_0$ if $r_n^2\phi(\hat\theta_n)>\hat c_{1-\alpha}$ where $\hat c_{1-\alpha}$ is the critical value constructed from the Babu correction or our proposed bootstrap, i.e.,
\begin{align}\label{Eqn: critical values, Babu}
\hat c_{1-\alpha} = \inf\{ c\in\mathbf R : P_W(r_n^2\{\phi(\hat\theta_n^*)-\phi(\hat\theta_n)-\phi_{\hat\theta_n}'(\hat\theta_n^*-\hat\theta_n)\}\leq c) \geq 1-\alpha\} ~,
\end{align}
or
\begin{align}\label{Eqn: critical values, modified}
\hat c_{1-\alpha} = \inf\{ c\in\mathbf R : P_W(\hat \phi_n''(r_n\{\hat \theta_n^* - \hat \theta_n\}) \leq c) \geq 1-\alpha\} ~.
\end{align}
Note that $\hat c_{1-\alpha}$ is generally infeasible but can be estimated by Monte Carlo simulations \citep{Efron1979,Hall1992bootstrap,HorowitzBoot}. The pointwise size control of our test then follows according to Theorems \ref{Thm: Babu extension} and \ref{Thm: modified bootstrap}. In fact, under additional restrictions, it can provide local size control. This property is particularly attractive because of the irregularity arising from nondifferentiability of $\phi$. In this case, pointwise asymptotic approximations can be misleading \citep{Imbens_Manski,AndrewsandGuggen2009ETA}. Interestingly, it turns out that there is another source of irregularity due to the nature of first order degeneracy (see Lemma \ref{Lem: weak limit under perturbations}). We relegate the detailed discussions to Appendix \ref{Sec: Testing} in order to make our presentation concise.

We now briefly compare the Babu correction, the above composition procedure and the recentered bootstrap \citep{Hall_Horowitz1996bootstrap,HorowitzBoot}. In some cases (for instance, Example \ref{Ex: mean squared} and the regular $J$-test), they coincide with each other. However, the Babu correction applies to general smooth functionals, rather than just quadratic forms, and hence can be thought of as a generalization of the recentered bootstrap. The composition procedure, which works for an even larger class of functionals, is a direct approach by exploiting the structure of the limits, and hence is more tractable.

\begin{rem}\label{Rem: rate}
Examples where the convergence rate is not $\sqrt n$ include inference based on kernel estimators with undersmoothing \citep{Hall1992bootstrap}, smoothed maximum score estimators \citep{Horowitz2002maxscore}, and cointegration regressions \citep{ChangParkSong2006BootCoint}. For nonstandard convergence rates, however, the bootstrap process $r_n\{\hat\theta_n^*-\hat\theta_n\}$ can fail to consistently estimate the law of $\mathbb G$, violating Assumption \ref{Ass: bootstrap for theta}(ii). Fortunately, as far as Theorem \ref{Thm: modified bootstrap} is concerned, any consistent estimator, which need not satisfy Assumption \ref{Ass: bootstrap for theta}(ii), will do. For example, in cube-root estimation problems, one could instead employ some smoothed bootstrap $r_n\{\tilde\theta_n^*-\tilde\theta_n\}$ where $\tilde\theta_n^*$ and $\tilde\theta_n$ are some smoothed estimators, or $m$ out of $n$ resampling (or subsampling) $m_n\{\hat\theta_{m_n}^*-\hat\theta_{n}\}$ where $\hat\theta_{m_n}^*$ is a bootstrap estimator based on subsamples of size $m_n$. In the context of estimating nonincreasing density functions, see \citet{Kosorok2008Grenander} and \citet{Sen_Banerjee_Woodroofe2010}; for bootstrapping the maximum score estimators, see \citet{Delgado_Poo_Wolf2001} and \citet{Patra_Seijo_Sen2015}.\qed
\end{rem}

\subsection{Estimation of the Derivative}\label{Sec: estimation of derivative}

Given the posited bootstrap consistency for the law of $\mathbb G$, the remaining crucial piece towards consistent bootstrap for the law of $\phi_{\theta_0}''(\mathbb G)$ based on Theorem \ref{Thm: modified bootstrap} is then an estimator $\hat\phi_n''$ of the derivative $\phi_{\theta_0}''$ that satisfies Assumption \ref{Ass: derivative estimation}. There are two general approaches for estimation of $\phi_{\theta_0}''$: one by exploiting the structure of $\phi_{\theta_0}''$, and the other one based on numerical differentiation as we describe now.

When first order degeneracy occurs, we have
\begin{align}
\phi_{\theta_0}''(h)=\lim_{n\to\infty}\frac{\phi(\theta_0+t_nh)-\phi(\theta_0)}{t_n^2}~.
\end{align}
We may thus estimate $\phi_{\theta_0}''$ via numerical differentiation as follows: for any $h\in\mathbb D$,
\begin{align}\label{Eqn: derivative estimation, 2nd HDD}
\hat\phi_n''(h)=\frac{\phi(\hat\theta_n+t_nh)-\phi(\hat\theta_n)}{t_n^2}~.
\end{align}
If $t_n$ tends to zero at a suitable rate, the sense of which is made precise by the following assumption, then $\hat\phi_n''$ is a good estimator for $\phi_{\theta_0}''$ in the sense of Assumption \ref{Ass: derivative estimation}.
\begin{ass}\label{Ass: rate tn}
$\{t_n\}_{n=1}^\infty$ is a sequences of scalars such that $t_n\darrow 0$ and $r_{n}t_{n}\to\infty$.
\end{ass}

Assumption \ref{Ass: rate tn} allows a wide range of tuning parameters that can deliver first order validity of our method. The optimal choice of $t_n$ is challenging and beyond the scope of the present paper, which we hope to address in future. The next proposition confirms the validity of the numerical estimator \eqref{Eqn: derivative estimation, 2nd HDD}.

\begin{pro}[\citet{Hong_Li2015numericaldelta}]\label{Pro: numerical estimation}
If Assumptions \ref{Ass: HDD 2nd}, \ref{Ass: weak convergence for theta}(i)(ii), and \ref{Ass: rate tn} hold, then the numerical estimator $\hat\phi_n''$ in \eqref{Eqn: derivative estimation, 2nd HDD} satisfies Assumption \ref{Ass: derivative estimation}.
\end{pro}

The numerical differentiation approach of estimating the derivatives, in the context of the Delta method, dates back to at least \citet{Dumbgen1993} in his proposal of the rescaled bootstrap. However, the way it was presented is quite implicit in revealing this, and so the bootstrap procedure is sometimes misunderstood as the $m$ out of $n$ resampling. Effectively, the rescaled bootstrap amounts to estimating the derivative numerically and the law of $\mathbb G$ using $n$ bootstrap samples; see \citet{Beare_Fang2016Grenander} for more details. The recent work of \citet{Hong_Li2015numericaldelta} provided a range of extensions of the numerical Delta method that have wide applications in econometrics.

Proposition \ref{Pro: numerical estimation} provides a way of estimating the derivative $\phi_{\theta_0}''$ that is tractable in the sense that there is no need to explore the particular structures of $\phi$ or $\phi_{\theta_0}''$ as long as the tuning parameter $t_n$ is properly chosen. On the other hand, the expression of $\phi_{\theta_0}''$ itself often suggests an intuitive estimator as we elaborate in the next subsection.

\subsubsection{Examples Revisited}\label{Sec: estimation of derivative, II}

Examples \ref{Ex: mean squared} is trivial since $\phi_{\theta_0}''$ is a known map and hence one can simply set $\hat\phi_{n}''=\phi_{\theta_0}''$ for all $n\in\mathbf N$. Example \ref{Ex: Overidentification Test} is more complicated.

\begin{exctd}[\ref{Ex: Overidentification Test}]
In the classical case when $\Gamma_0(\theta)$ is singleton, we may estimate $\phi_{\theta_0}''$ based on the GMM estimator $\hat\gamma_n$ and the estimated Jacobian matrix $\hat J_n$. Generally, there are two unknown objects involved in the second order derivative: the identified set $\Gamma_0(\theta)$ and $J(\cdot)$. Let $\mathbf M^{m\times k}$ be the space of $m\times k$ matrices. Suppose that $\hat\Gamma_n\subset\Gamma$ is a $d_H$-consistent estimator for $\Gamma_0(\theta)$, and $\hat J_n: \Gamma\to\mathbf M^{m\times k}$ an estimator for $J: \Gamma\to\mathbf M^{m\times k}$ such that $\sup_{\gamma\in\Gamma}\|\hat J_n(\gamma)-J(\gamma)\|\convp 0$. Then we may estimate $\phi_{\theta_0}''$ by
\begin{align}
\hat\phi_n''(h)=\min_{\gamma\in\hat\Gamma_n}\min_{v\in B_n}\{h(\gamma)-\hat J_n(\gamma)v\}^\intercal W\{h(\gamma)-\hat J_n(\gamma)v\}~,
\end{align}
where $B_n\equiv\{v\in\mathbf R^k: \|v\|\le t_n^{-1}\}$ for $t_n\darrow 0$ satisfying $t_n\sqrt n\to\infty$. Consistency of $\hat\Gamma_n$ can be established by appealing to \citet{CHT2007}, while uniform consistency of $\hat J_n$ can be derived using Glivenko-Cantelli type arguments. Following the proof of Lemma \ref{Lemma: Common CH, derivative convergence}, it is straightforward to show that $\hat\phi_n''$ satisfies Assumption \ref{Ass: derivative estimation}. \qed
\end{exctd}

\section{Application: Testing for Common CH Features}\label{Sec: Application}

In this section, we apply our framework to develop a robust test of common conditionally heteroskedastic (CH) factor structure by allowing multiple common CH features. Let $\{Y_t\}_{t=1}^T$ be a $k$-dimensional time series. According to \citet{Engle_Kozicki1993CF}, a feature that is present in each component of $Y_t$ is said to be common to $Y_t$ if there exists a linear combination of $Y_t$ that fails to have the feature. A canonical example is the notion of cointegration developed by \citet{Engle_Granger1987Co-In} in order to characterize the common feature of stochastic trend.

\subsection{The Setup}

Following \citet{Engle_Ng_Rothschild1990asset} and \citet{Dovonon_Renault2013testing}, suppose that the $k$-dimensional process $\{Y_t\}$ satisfies
\begin{align}\label{Eqn: Common CH, defn}
\var(Y_{t+1}|\mathcal F_t)=\Lambda D_t\Lambda^\intercal+\Omega~,
\end{align}
where $\Lambda$ is a $k\times p$ matrix of full column rank with $p\le k$, $D_t$ a $p\times p$ diagonal matrix with diagonal (random) elements $\sigma _{jt}^{2}$ for $j=1,\ldots ,p$, $\Omega $ a $k\times k$ positive semidefinite matrix, and $\{\mathcal F_t\}_{t=1}^\infty$ a filtration to which $\{Y_t\}_{t=1}^\infty$ and $\{\sigma_{jt}^2: j=1,\ldots,p\}_{t=1}^\infty$ are adapted. By \citet{Engle_Kozicki1993CF}, we say that $\{Y_t\}$ has a common CH feature if there exists some nonzero $\gamma_0\in\mathbf R^k$ such that $\var(\gamma_0^\intercal Y_t|\mathcal F_t)$ is constant. The conditional covariance structure \eqref{Eqn: Common CH, defn} has some attractive properties that help to understand, for example, asset excess returns in a parsimonious way \citep{Engle_Ng_Rothschild1990asset}. Thus, tests of common CH features can be used to detect the underlying common factor structures that simplify capturing interrelations of economic and financial variables under consideration.

With the help of instrumental variables, a common CH feature can be reformulated by unconditional moments that fit into the classical GMM framework. The following assumption is taken directly from \citet{Dovonon_Renault2013testing}.

\begin{ass}\label{Ass: Common CH, setup} 
(i) $\Lambda$ is of full column rank; (ii) $\var(\sigma_t^2)$ is nonsingular for $\sigma_t^2\equiv(\sigma_{1t}^2,\ldots,\sigma_{pt}^2)^\intercal$; (iii) $E[Y_{t+1}|\mathcal F_t]=0$; (iv) $Z_t$ is an $m\times 1$ $\mathcal F_t$-measurable random vector such that $\var(Z_t)$ is nonsingular; (v) $\cov(Z_t,\sigma_t^2)$ has full column rank $p$; (vi) $\{Y_t,Z_t\}$ is stationary and ergodic such that $E[\|Z_t\|^2]<\infty$ and $E[\|Y_t\|^4]<\infty$.
\end{ass}

Assumption \ref{Ass: Common CH, setup}(i)-(ii) ensure that there are exactly $k-p$ linearly independent vectors $\gamma_0$, spanning the null space of $\Lambda^\intercal$, such that $\var(\gamma_0^\intercal Y_t|\mathcal F_t)$ is constant. In other words, the common CH features $\gamma_0$ are nonzero solutions of the equation $\Lambda^\intercal \gamma_0=0$.\footnote{If $\gamma_0$ is a common CH feature, so is $a\gamma_0$ for any nonzero $a\in\mathbf R$. For mathematical purpose, however, the number of common CH features is defined to the dimension of the null space of $\Lambda^\transpose$.} Assumption \ref{Ass: Common CH, setup}(iii) is a normalization condition that helps to simplify the exposition. Assumption \ref{Ass: Common CH, setup}(iv) defines the instrument $Z_t$ formed from the information set $\mathcal F_t$, while Assumption \ref{Ass: Common CH, setup}(v) implicitly requires that the number of instruments is no less than that of factors. Assumption \ref{Ass: Common CH, setup}(vi) further specifies the data generating process. We refer the readers to \citet{Dovonon_Renault2013testing} for further details on Assumption \ref{Ass: Common CH, setup}.

Assumption \ref{Ass: Common CH, setup} allows us to characterize common CH features as nonzero $\gamma_0$ satisfying the vector of unconditional moment equalities \citep{Dovonon_Renault2013testing}:
\begin{align}\label{Eqn: Common CH, moment condition}
E[Z_{t}\{( \gamma_0^\intercal Y_{t+1}) ^{2}-c( \gamma_0 )\}]=0~,
\end{align}
where $c(\gamma_0)=E[( \gamma_0^\intercal Y_{t+1})^2]$. It is then tempting to employ Hansen's $J$ statistic to test the existence of common CH features \citep{Engle_Kozicki1993CF}. Unfortunately, as noted by \citet{Dovonon_Renault2013testing}, the Jacobian matrix evaluated at the truth is degenerate at zero, rendering standard theory inapplicable. Though, as shall be illustrated, such degeneracy is of a nature different from first order degeneracy. By expanding the moment function to the second order, \citet{Dovonon_Renault2013testing} showed that the asymptotic distribution of the $J$ statistic is highly nonstandard. Nonetheless, \citet{Dovonon_Goncalves2017bootstrapping} developed a corrected bootstrap that can consistently estimate the limiting law when the bootstrap of \citet{Hall_Horowitz1996bootstrap} fails to do so.

However, a key assumption in previous studies is that there exists a unique nonzero $\gamma_0$ such that \eqref{Eqn: Common CH, moment condition} is satisfied, ensured by exclusion restrictions and linear normalization $\sum_{j=1}^{k}\gamma_0^{(j)}=1$ \citep{Dovonon_Renault2013testing,Dovonon_Goncalves2017bootstrapping,Lee_Liao2017LocalIDfailure}. This is undesirable for the following reasons. First, it is unknown {\it a priori} how many (linearly independent) CH features are common to the series under consideration. Second, as pointed out by \citet{Engle_Ng_Rothschild1990asset} in the context of asset pricing, empirical work often considers large numbers of assets and the numbers of common CH features are expected to be large as well. Third, the linear normalization may in fact lead to no $\gamma_0$ satisfying \eqref{Eqn: Common CH, moment condition} (i.e.\ non-existence). For example, suppose $\Lambda=[1,1]^\intercal$. Then any common CH feature $\gamma_0$ must satisfy $\gamma_0^{(1)}+\gamma_0^{(2)}=0$, contradicting the linear normalization $\gamma_0^{(1)}+\gamma_0^{(2)}=1$ proposed in \citet{Dovonon_Renault2013testing}. Fourth, in addition to the possibility that exclusion restrictions may be hard to form, the linear normalization is not susceptible of a unique common CH feature (i.e.\ non-uniqueness). To see this, suppose $\Lambda=[1,-1,-1]^\intercal$. Then for any common CH feature satisfying the normalization, we must have $\gamma_0^{(1)}-\gamma_0^{(2)}-\gamma_0^{(3)}=0$ and $\gamma_0^{(1)}+\gamma_0^{(2)}+\gamma_0^{(3)}=1$, which admit infinitely many solutions, i.e., the uniqueness is undermined in this case. These arguments motivate us to modify the $J$-test in a way that accommodates partial identification as well as degenerate Jacobian matrices. Such an extension is nontrivial because the second order (and hence global) identification,\footnote{Given first order identification failure, second order identification is equivalent to global identification in the current context because the moment function is quadratic in $\gamma_0$.} a condition that \citet{Dovonon_Renault2013testing} and \citet{Dovonon_Goncalves2017bootstrapping} heavily rely on, fails.


\subsection{A Modified $J$ Test}

To exclude the zero solution and avoid falsely excluding the existence of CH features, we employ the following normalization
\begin{align}\label{Eqn: Common CH, normalization}
\gamma\in\mathbb S^k\equiv\{\gamma'\in\mathbf R^k: \|\gamma'\|=1\}~.
\end{align}
Next, to map the current setup into our developed framework, we define a function $\phi: \prod_{j=1}^m \ell^\infty(\mathbb S^k)\to\mathbf R$ by: for any $\theta\in \prod_{j=1}^m \ell^\infty(\mathbb S^k)$,
\begin{align}\label{Eqn: Common CH, functional}
\phi(\theta)\equiv\inf_{\gamma\in\mathbb S^k}\|\theta(\gamma)\|^2~.
\end{align}
Then in view of the moment conditions \eqref{Eqn: Common CH, moment condition}, the hypothesis that there exists at least one common CH feature can be reformulated as
\begin{align}
\mathrm H_0: \phi(\theta_0)=0\qquad \mathrm H_1: \phi(\theta_0)>0~,
\end{align}
where $\theta_0: \mathbb S^k\to \mathbf R^m$ is defined as $\theta_0(\gamma)\equiv E[Z_{t}\{( \gamma ^\intercal Y_{t+1}) ^{2}-c( \gamma )\}]$. In this formulation, we have taken the identity matrix $I_m$ as the weighting matrix for simplicity.

Given our treatment of Example \ref{Ex: Overidentification Test}, one might next try appealing to the results developed there. Unfortunately, they are not directly applicable. First, the parameter space $\Gamma$ of $\gamma_0$ is required to have nonempty interior (see Lemma \ref{Lem: over ID, HDD}), whereas in the current context $\Gamma=\mathbb S^k$ which has empty interior. Second, there is a technical condition there that prevents the Jacobian matrix from being degenerate even when there does exist a unique common CH feature; see Remark \ref{Rem: over ID, HDD} for details. Consequently, we have to re-verify the differentiability conditions for the map \eqref{Eqn: Common CH, functional}. By Lemma \ref{Lem: Common CH, HDD2}, under the null, $\phi$ is Hadamard differentiable with degenerate derivative, and second order Hadamard directionally differentiable at $\theta_0$ tangentially to $\prod_{j=1}^m C(\mathbb S^k)$ with the derivative
\begin{align}\label{Eqn: Common CH, Derivative}
\phi_{\theta_0}''(h)=\min_{\gamma_0\in\Gamma_0}\min_{v\in\mathbf R^k}\|h(\gamma_0)+G\vecop(vv^\intercal)\|^2~,
\end{align}
for any $h\in\prod_{j=1}^m C(\mathbb S^k)$, where $\Gamma_0=\{\gamma_0\in\mathbb S^k: \theta_0(\gamma_0)=0\}$ is the identified set of $\gamma_0$, and $G\in\mathbf M^{m\times k^2}$ with the $j$th row given by $\vecop(\Delta_j)^\intercal$ and
\[
\Delta_j=E[Z_t^{(j)}(Y_{t+1}Y_{t+1}^\intercal-E[Y_{t+1}Y_{t+1}^\intercal])]~.
\]

We now make some remarks before proceeding further. First, we stress that first order degeneracy refers to the first order derivative $\phi_{\theta_0}'$ of the functional $\phi$, mapping from the function space $\prod_{j=1}^m \ell^\infty(\mathbb S^k)$ to $\mathbf R$, being degenerate, while the degeneracy \citet{Dovonon_Renault2013testing} focused on refers to degeneracy of the Jacobian matrix $J(\gamma_0)\equiv\frac{d}{d\gamma}\theta_0(\gamma)|_{\gamma=\gamma_0}$ of the moment function $\theta_0$ that maps from the parameter space $\Gamma\subset\mathbf R^k$ of $\gamma_0$ to $\mathbf R^m$. Thus, the two types of degeneracy are conceptually different. Second, perhaps more importantly, they are also different in terms of the consequences. By Theorem \ref{Thm: bootstrap, iff} and in view of \eqref{Eqn: Common CH, Derivative}, $\phi$ being first order degenerate means that the second order standard bootstrap is inconsistent regardless of whether the Jacobian matrix is degenerate or not, while degeneracy of the Jacobian matrix generates the {\it additional} complication that $\phi$ is second order nondifferentiable as reflected by the inside minimization in \eqref{Eqn: Common CH, Derivative}. Third, further allowing multiple (linearly independent) common CH features reinforces the nondifferentiability of $\phi$ as can be seen from the outside minimization in \eqref{Eqn: Common CH, Derivative}.

Next, let the estimator $\hat\theta_T: \mathbb S^k\to\mathbf R^m$ be defined by $\hat\theta_T(\gamma)=\frac{1}{T}\sum_{t=1}^T Z_t\{(\gamma^\intercal Y_{t+1})^2-\hat c(\gamma)\}$ with $\hat c(\gamma)=\frac{1}{T}\sum_{t=1}^T(\gamma^\intercal Y_{t+1})^2$. Given the established differentiability of $\phi$, the asymptotic distribution of $\phi(\hat\theta_T)$ is then an immediate consequence of Theorem \ref{Thm: delta method, 2nd order} provided $\hat\theta_T$ converges weakly. Towards this end, we impose the following assumption as in \citet{Dovonon_Renault2013testing}.

\begin{ass}\label{Ass: Common CH, CLT}
$Z_{t}$, $\vecop(Y_{t}Y_{t}^\intercal)$ and $\vecop(Y_{t}Y_{t}^\intercal) \otimes Z_{t}$ fulfill CLT.\footnote{The symbol $\otimes$ denotes Kronecker product.}
\end{ass}

Assumptions \ref{Ass: Common CH, setup} and \ref{Ass: Common CH, CLT} together imply that
\begin{align}
\sqrt T \{\hat\theta_T-\theta_0\} \convl \mathbb G \text{ in } \prod_{j=1}^m \ell^\infty(\mathbb S^k)~,
\end{align}
where $\mathbb{G}$ is a zero mean Gaussian process with the covariance functional satisfying: for any $\gamma_{1}$, $\gamma_{2}\in\Gamma_{0}$ and $\mu_z\equiv E[Z_t]$,
\begin{align*}
E[\mathbb{G}(\gamma_{1})\mathbb{G}(\gamma_{2})]=E[(Z_{t}-\mu_{z})(Z_{t}-\mu_{z})^\intercal\{(\gamma_{1}^\intercal Y_{t+1})^{2}-c(\gamma_{1})\}\{(\gamma_{2}^\intercal Y_{t+1})^{2}-c(\gamma_{2})\}] ~.
\end{align*}

The proposition below delivers the limiting distribution of test statistic $T\phi(\hat\theta_T)$.

\begin{pro}\label{Pro: Common CH, weak limit}
Let Assumptions \ref{Ass: Common CH, setup} and \ref{Ass: Common CH, CLT} hold. Then we have under $\mathrm H_0$
\begin{equation}\label{Eqn: Common CH, weak limit}
 T\min_{\gamma \in \mathbb S^k}\Vert\hat{\theta}_T(\gamma)\Vert^{2}\overset{L}{\rightarrow}\min_{\gamma_0 \in \Gamma_0}\min_{v\in \mathbf{R}^k}\Vert \mathbb{G}(\gamma_0) + G \vecop(vv^\intercal)\Vert ^{2}~.
\end{equation}
\end{pro}

The asymptotic distribution in \eqref{Eqn: Common CH, weak limit} is a highly nonlinear functional of the Gaussian process $\mathbb G$ in general, which turns out to be consistent with the limits obtained in \citet{Dovonon_Renault2013testing} and \citet{Dovonon_Goncalves2017bootstrapping} whenever their second order identification (and global) condition holds; see Remark \ref{Rem: Common CH, compare}. In the latter setting, \citet{Dovonon_Goncalves2017bootstrapping} showed that the recentered bootstrap of \citet{Hall_Horowitz1996bootstrap} is inconsistent and thus proposed corrected versions of the standard GMM bootstrap. Unfortunately, their methods are not directly applicable to our setup that allows multiple common CH features (i.e.\ partial identification), because they crucially rely on the second order and global identification.

We next demonstrate how our bootstrap works. First, let $\{Y_{t+1}^*,Z_t^*\}_{t=1}^T$ be a bootstrap sample, which can be obtained by block bootstrap, nonoverlapping or overlapping \citep{Carlstein1986subseries,Kunsch1989Jackknife}. Because the limiting process $\{\mathbb G(\gamma): \gamma\in\Gamma_0\}$ is determined by a martingale difference sequence indexed by $\gamma\in\Gamma_0$, the dependence structure of the data does not enter into the limit and we may thus employ  \citet{Efron1979}'s nonparametric bootstrap or more general bootstrap schemes. In any case, we set
\begin{align}\label{Eqn: Common CH, bootstrap for theta}
\hat\theta_T^*(\gamma)=\frac{1}{T}\sum_{t=1}^T Z_t^*\{(\gamma^\intercal Y_{t+1}^*)^2-\hat c^*(\gamma)\}~,\,\hat c^*(\gamma)=\frac{1}{T}\sum_{t=1}^T(\gamma^\intercal Y_{t+1}^*)^2~.
\end{align}
To accommodate diverse resampling schemes, we simply impose the high level condition that $\hat\theta_T^*$ satisfies Assumptions \ref{Ass: bootstrap for theta} and \ref{Ass: measurability} \citep{DehlingMikoschSorensen2002EPDep}.

It remains to estimate the derivative \eqref{Eqn: Common CH, Derivative}. The numerical differentiation approach can be implemented as in the beginning of Section \ref{Sec: estimation of derivative}. That is, we estimate $\phi_{\theta_0}''$ by
\begin{align}\label{Eqn: Common CH, derivative estimate0}
\hat\phi_T''(h)=\frac{\inf_{\gamma\in\mathbb S^k}\|\hat\theta_T(\gamma)+\kappa_Th(\gamma)\|^2- \min_{\gamma\in\mathbb S^k}\|\hat\theta_T(\gamma)\|^2}{\kappa_T^2}~,
\end{align}
where $\kappa_T$ satisfies Assumption \ref{Ass: rate tn}. We now describe how to estimate $\phi_{\theta_0}''$ by exploiting its structure. Let $B_T\equiv\{v\in\mathbf R^k: \|v\|\le \kappa_T^{-1/2}\}$ and $\hat\Gamma_T\equiv \{\gamma\in\mathbb S^k:\Vert\hat{\theta}_T(\gamma)\Vert^{2}-\phi(\hat{\theta}_T)\le \kappa_{T}^2\}$,\footnote{One can theoretically ignore $\phi(\hat\theta_T)$ in the expression of $\hat\Gamma_T$. As pointed out by \citet{CHT2007}, however, such a modification helps avoid an empty set of solutions and improve power.} where $\kappa_T$ is to be specified. Then we may estimate $\phi_{\theta_0}''(h)$ by:
\begin{equation}\label{Eqn: Common CH, derivative estimate}
\hat{\phi} _{T }^{\prime \prime }(h) =\inf_{\gamma \in \hat\Gamma_T}\min_{v\in B_T}\Vert h(\gamma) + \hat{G} \vecop(vv^\intercal)\Vert ^{2}~,
\end{equation}
where $\hat G\in\mathbf M^{m\times k^2}$ with its $j$th row given by $\vecop(\hat\Delta_j)^\intercal$ for
\[
\hat\Delta_j=\frac{1}{T}\sum_{t=1}^T Z_t^{(j)}Y_{t+1}Y_{t+1}^\intercal-\frac{1}{T}\sum_{t=1}^T Z_t^{(j)}\frac{1}{T}\sum_{t=1}^T Y_{t+1}Y_{t+1}^\intercal~.
\]
In fact, we may further restrict the bounded set $B_T$ to reduce the computation burden for $\hat{\phi}_{T}^{\prime \prime }$; see Remark \ref{Rem: Common CH, derivative est}. Clearly, the sequence $\{\kappa_T\}$ should tend to zero at a suitable rate as $T\to\infty$. This is made precise as follows.

\begin{ass}\label{Ass: Common CH, tuning parameter}
$\{\kappa_T\}$ satisfies (i) $\kappa_T\darrow 0$, and (ii) $\sqrt{T}\kappa_T\to\infty$.
\end{ass}

Assumption \ref{Ass: Common CH, tuning parameter} regulates the rates at which the tuning parameters $\kappa_T$ should approach zero, in order to deliver first order validity of our bootstrap inference procedure. The optimal choice of $\kappa_T$ is concerned with higher order accuracy of our method, which we do not touch in this paper. Combining the bootstrap $\hat\theta_T^*$ in \eqref{Eqn: Common CH, bootstrap for theta} and the derivative estimator, we are then able to consistently estimate the law of the weak limit in \eqref{Eqn: Common CH, weak limit} following Theorem \ref{Thm: modified bootstrap}, which in turn allows us to construct critical values. Specifically, let $\hat c_{1-\alpha}$ be the $1-\alpha$ quantile of  $\hat \phi_T''(\sqrt{T}\{\hat \theta_T^* - \hat \theta_T\})$ conditional on the data:\footnote{As usual, $P_W$ denotes the probability taken with respect to the bootstrap weights $\{W_T\}$, though in the current setup they are implicitly defined. Alternatively, one can think of $P_W$ as the probability with respect to the bootstrap sample $\{Z_t^*,Y_{t+1}^*\}$ holding data fixed.}
\begin{align}\label{critical values: application}
\hat c_{1-\alpha} \equiv \inf\{ c\in\mathbf R : P_W(\hat \phi_T''(\sqrt{T}\{\hat \theta_T^* - \hat \theta_T\}) \leq c) \geq 1-\alpha\} ~.
\end{align}

The following proposition confirms that the test of rejecting the existence of common CH features when $T\phi(\hat\theta_T)>\hat c_{1-\alpha}$ is valid.

\begin{pro}\label{Pro: Common CH, test validity}
Suppose Assumptions \ref{Ass: bootstrap for theta}, \ref{Ass: measurability}, \ref{Ass: Common CH, setup}, \ref{Ass: Common CH, CLT}, and \ref{Ass: Common CH, tuning parameter} hold. If the cdf of the limit in \eqref{Eqn: Common CH, weak limit} is continuous and strictly increasing at its $1-\alpha$ quantile for $\alpha\in(0,1)$, then we have under $\mathrm H_0$,
\[\lim_{T\rightarrow\infty}P(T\min_{\gamma \in \mathbb S^k}\Vert\hat{\theta}_T(\gamma)\Vert^{2}> \hat c_{1-\alpha})=\alpha~.
\]
\end{pro}

Proposition \ref{Pro: Common CH, test validity} implies our test has pointwise asymptotic exact size $\alpha$ and thus is not conservative (in the pointwise sense). Establishing local size control, unfortunately, is challenging in this case, because asymptotic distributions of the statistic under local perturbations do not have definitive relations (to us) to the corresponding pointwise limits in terms of first order dominance. It appears that the problem of developing (at least) locally valid {\it and} non-conservative overidentification tests is prevalent in the literature of partial identification \citep{CHT2007,AndrewsandSoares2010}.

Finally, we stress that the quadratic structure of the moment function plays no essential roles in our framework. Building upon Example \ref{Ex: Overidentification Test}, one may work with a general moment function that admits a zero Jacobian matrix, but without the requirement that the parameter space have nonempty interior. It is also possible to deal with GMM problems with a rank deficient but possibly nonzero Jacobian matrix. For example, consider testing whether a matrix $\Pi_0\in\mathbf M^{m\times k}$ with $m\ge k$ has rank $k$. This amounts to testing
\begin{align}
\mathrm H_0: \Pi_0 \gamma=0 \text{ for some }\gamma\in\mathbb S^k\quad  \text{ v.s. } \quad \mathrm H_0: \Pi_0 \gamma\neq 0 \text{ for any }\gamma\in\mathbb S^k~.
\end{align}
Here, the moment function is $\gamma\mapsto \theta_0(\gamma)\equiv\Pi_0\gamma$ which is non-quadratic and whose Jacobian matrix, namely, $\Pi_0$, may have rank less than or equal to $k-1$. Note also that the parameter space $\mathbb S^k$ of $\gamma$ has empty interior. We refer the reader to \citet{ChenFang2016Rank} for more detailed discussions.

\begin{rem}\label{Rem: Common CH, compare}
The weak limit in Proposition \ref{Pro: Common CH, weak limit} is consistent with the one in \citet{Dovonon_Renault2013testing}, when there does exist a unique common CH feature which satisfies their linear normalization and when the weighting matrix is the identity matrix (for reasons we have mentioned at the beginning of this section) -- otherwise the two are not comparable. At the first sight, our testing statistic is different from \citet{Dovonon_Renault2013testing}'s because we adopted a different normalization, resulting in a different parameter space.\footnote{\citet{Dovonon_Renault2013testing} also recentered $Z_t$ in their construction, though this does not change the statistic numerically.} Close inspection, however, shows that the asymptotic distributions are in fact identical, up to a multiplicative constant. Specifically, let $\gamma_0$ be the (nonzero) unique CH feature such that $\sum_{j=1}^{k}\gamma_0^{(j)}=1$. Then $\Gamma_0=\{\pm\gamma_0/\|\gamma_0\|\}$ and so by Proposition 5.1, the asymptotic distribution of our $J$-statistic is simply the law of
\begin{multline}\label{Eqn: Common CH, DR comparison}
\min_{v\in\mathbf R^k}\|\mathbb G(\pm\gamma_{0}/\|\gamma_0\|)+G\mathrm{vec}(vv^\intercal)\|^2\overset{d}{=}\|\gamma_0\|^{-4}\min_{v\in\mathbf R^k} \big\{\mathbb G(\gamma_{0})^\intercal \mathbb G(\gamma_{0})\\+\mathbb G(\gamma_0)^\intercal G\mathrm{vec}(vv^\intercal)+\frac{1}{4}(\mathrm{vec}(vv^\intercal))^\intercal G^\intercal G\mathrm{vec}(vv^\intercal)\big\}~,
\end{multline}
where we simply replaced $v$ with $v/(\sqrt{2}\|\gamma_0\|^2)$. By Theorem 3.1 and Corollary 3.1 in \citet{Dovonon_Renault2013testing}-- see also \citet{Dovonon_Goncalves2017bootstrapping}, their $J$-statistic (with $W$ being the identity matrix) converges in law to
\begin{align}\label{Eqn: Common CH, DR comparison1}
&\min_{u\in\mathbf R^{k-1}} \{\mathbb G(\gamma_0)^\intercal \mathbb G(\gamma_0)+\mathbb G(\gamma_0)^\intercal \bar G\mathrm{vec}(uu^\intercal)+\frac{1}{4}(\mathrm{vec}(uu^\intercal))^\intercal \bar G^\intercal \bar G\mathrm{vec}(uu^\intercal)\}~,
\end{align}
where $\bar G\in\mathbf M^{m\times (k-1)^2}$ with the $j$th row $\mathrm{vec}(A\Delta_jA^{\intercal})^\intercal$ for $A = [I_{k-1},-\jmath_{k-1}]$ and $\jmath_{k-1}$ the $(k-1)\times 1$ vector of ones. By Lemma \ref{Lem: Common CH, compare with DR}, however, the two limits in \eqref{Eqn: Common CH, DR comparison} and \eqref{Eqn: Common CH, DR comparison1} differ only by the multiplicative constant $\|\gamma_0\|^{-4}$, establishing the claimed consistency. If the common CH feature also satisfies our normalization, i.e., $\|\gamma_0\|=1$, then the two limits are identical. We reiterate that the our main motivation is to build upon \citet{Dovonon_Renault2013testing} by allowing multiple common CH features and adopting a normalization that would not falsely exclude the existence of any common features.\footnote{Any other linear normalization $c^\intercal \gamma_0=r$ for known $c\in\mathbf R^k$ and $r\in\mathbf R$ would share the same deficiency as the linear normalization, which includes, for example, $\gamma_0^{(1)}=1$ -- see our next section.}  \qed
\end{rem}

\subsection{Simulation Studies} \renewcommand{\arraystretch}{1.5}

In this section, we examine the finite sample performance of our framework based on Monte Carlo simulations, and show how the identification assumption in \citet{Dovonon_Renault2013testing} and \citet{Dovonon_Goncalves2017bootstrapping} may suffer from their linear normalization. One may then try the multiple testing versions of these tests by testing a few linearly independent linear restrictions, but we show they may be too conservative.

As in \citet{Dovonon_Renault2013testing} and \citet{Dovonon_Goncalves2017bootstrapping}, we consider the following CH factor model:
\begin{equation}\label{Eqn: Common CH, simulation model}
Y_{t}=\Lambda F_{t}+U_{t}~,
\end{equation}
where $Y_t$ is a $k\times 1$ vector that can be thought of asset returns, $F_t$ is a $p\times 1$ vector of CH factors, $\Lambda $ is a $k\times p$ matrix of factor loadings,  and $U_t$ is a vector of idiosyncratic shocks independent of $F_t$. Following \citet{Dovonon_Renault2013testing} and \citet{Dovonon_Goncalves2017bootstrapping}, we let $\{U_t\}$ be an i.i.d.\ sequence from $N(0,I_k/2)$, and the $j$th component $f_{j,t+1}$ of $F_{t+1}$ follow a Gaussian-GARCH(1,1) model such that
\begin{equation*}
f_{j,t+1}=\sigma _{j,t}\epsilon _{j,t+1}~,\,\sigma _{j,t}^{2}=\omega_{j} +\alpha_{j} f_{j,t}^{2}+\beta_{j} \sigma _{j,t-1}^{2}~,
\end{equation*}
where $\omega_{j}, \alpha_{j}, \beta_{j} >0$, $\{\epsilon_{j,t}\}\sim N(0,1)$ i.i.d.\ across both $j$ and $t$, and $\{\sigma_{j0}\}$ are independent across $j$ and of $\{\epsilon_{j,t}\}$. It follows that $\{f_{j,t}\}$ are independent across $j$ for each $t$. The remaining specifications are detailed in Table \ref{Table: simulation designs}. Our designs are the same as those in \citet{Dovonon_Renault2013testing} and \citet{Dovonon_Goncalves2017bootstrapping} except that different values for $\Lambda$ are used to illustrate the restrictiveness of the linear normalization. Designs D1 and D2 generate two assets while Designs D3, D4 and D5 generate three assets. In Designs D1, D3 and D4, the factor loading matrices $\Lambda$ ensure the existence of common CH features and thus serves for investigation of size performance, while no common CH features exist in Designs D2 and D5, which help us inspect power performance.

\begin{table}[!htbp]
\caption{Simulation Designs}\label{Table: simulation designs}
\begin{small}
\begin{center}
\begin{tabular}{ccccc}
\hline\hline
Design  & \# of Assets & \# of Factors & GARCH Parameters &  Factor Loadings\\
\hline
 D1 & $k=2$ & $p=1$ & $(\omega_1,\alpha_1,\beta_1)=(0.2,0.2,0.6)$ & $\Lambda=(1,1)^\intercal$\\
\multirow{2}{*}{D2} & \multirow{2}{*}{$k=2$} & \multirow{2}{*}{$p=2$} & $(\omega_1,\alpha_1,\beta_1)=(0.2,0.2,0.6)$ & \multirow{2}{*}{$\Lambda=I_2$}\\
 & & & $(\omega_2,\alpha_2,\beta_2)=(0.2,0.4,0.4)$ & \\
D3 & $k=3$ & $p=1$ & $(\omega_1,\alpha_1,\beta_1)=(0.2,0.2,0.6)$ & $\Lambda=(1,1,1)^\intercal$\\
\multirow{2}{*}{D4} & \multirow{2}{*}{$k=3$} & \multirow{2}{*}{$p=2$} & $(\omega_1,\alpha_1,\beta_1)=(0.2,0.2,0.6)$ & \multirow{2}{*}{$\Lambda=\begin{bmatrix} 1 & 1 & 1\\ -1 & 0 & 1 \end{bmatrix}^\intercal$}\\
 & & & $(\omega_2,\alpha_2,\beta_2)=(0.2,0.4,0.4)$ & \\
\multirow{3}{*}{D5} & \multirow{3}{*}{$k=3$} & \multirow{3}{*}{$p=3$} & $(\omega_1,\alpha_1,\beta_1)=(0.2,0.2,0.6)$ & \multirow{3}{*}{$\Lambda=I_3$}\\
 & & & $(\omega_2,\alpha_2,\beta_2)=(0.2,0.4,0.4)$ & \\
 & & & $(\omega_3,\alpha_3,\beta_3)=(0.1,0.1,0.8)$ & \\
\hline\hline
\end{tabular}
\end{center}
\end{small}
\end{table}

The tests are implemented with $m=2$ and instruments $Z_{t}=(Y_{1,t}^{2}, Y_{2,t}^{2})^\intercal$ for Designs D1 and D2, and with $m=3$ and $Z_{t}=(Y_{1,t}^{2}, Y_{2,t}^{2}, Y_{3,t}^{2})^\intercal$ for Designs D3, D4 and D5. For derivative estimation, we set the tuning parameters $\kappa_{T}=T^{-1/4}, T^{-1/3}, T^{-2/5}$ for both the derivative estimator in \eqref{Eqn: Common CH, derivative estimate} and the numerical derivative estimator as in \eqref{Eqn: Common CH, derivative estimate0} respectively. These choices are meant to satisfy Assumption \ref{Ass: Common CH, tuning parameter}. Again, we do not touch the issue of optimality in this paper, but instead hope to make the point that, even with these crude choices, our methods show substantial improvement over existing ones. The results corresponding to the two sets of choices are denoted as CF1 and CF2. To show the restrictiveness of the linear normalization $\gamma\in\{\gamma' \in \mathbf{R}^{k}: \sum_{i=1}^{k}\gamma _{i}'=1\}$ as in \citet{Dovonon_Renault2013testing}, \citet{Dovonon_Goncalves2017bootstrapping} and \citet{Lee_Liao2017LocalIDfailure}, we report the results based on \citet{Dovonon_Goncalves2017bootstrapping}'s corrected and continuously-corrected bootstrap as well as those based on the asymptotic test of \citet{Dovonon_Renault2013testing}, denoted as DG1, DG2 and DR respectively. The sample sizes are $T=1,000,$ $2,000$, $5,000,$ $10,000$, $20,000$, $40,000$ and $50,000$. To minimize the initial value effect, the data are obtained by generating $T+100$ samples and dropping the first $100$ samples. We conduct $2,000$ Monte Carlo replications with $200$ empirical bootstrap repetitions for each replication. The nominal level is $5\%$ throughout.

The results are summarized in Tables \ref{Tab: size, D1}-\ref{Tab: power, D5}. As expected, \citet{Dovonon_Goncalves2017bootstrapping}'s resampling methods exhibit substantial size distortion, often close to or over $50\%$; so does the asymptotic test DR. This does not appear to be a finite sample issue as the distortion is especially severe in large samples. Rather, it is because the linear normalization excludes common CH features that actually exist in the data and in this way leads to wrong conclusions. Our tests considerably reduce the null rejection rates for all the chosen tuning parameters, though both CF1 and CF2 exhibit some degrees of over- and under-rejection, due to the issue of tuning parameters. Another interesting finding is that our bootstrap based on numerical differentiation (CF2) appears to be more sensitive to the choice of tuning parameters, which is somewhat expected because the structural method (CF1) exploits more information of the derivative. We leave a thorough comparison between these two methods for future study.

Alternatively, one may test a few linearly independent linear restrictions by adopting multiple testing versions of the DG and the DR tests, so as to avoid falsely excluding the existence of common CH features. One then rejects the existence of common CH features if {\it all} the restrictions are rejected at level $\alpha=5\%$.\footnote{Since the null is a union of ``sub-nulls'', no Bonferroni-type correction is needed.} However, the resulting tests, though valid, may be too conservative. To illustrate, we test the null that $\gamma_0$ satisfies (i) $\gamma_0^{(1)}+\gamma_0^{(2)} = 1$ or (ii) $\gamma_0^{(1)} = 1$ for D1 and D2, and satisfies (i) $\gamma_0^{(1)}+\gamma_0^{(2)} + \gamma_0^{(3)} = 1$, (ii) $\gamma_0^{(1)} = 1$, or (iii) $\gamma_0^{(2)} = 1$ for D3, D4 and D5. We implement the multiple testing procedures based on \citet{Dovonon_Renault2013testing} with optimal weighting matrix and \citet{Dovonon_Goncalves2017bootstrapping} with the identity weighting matrix, and respectively label them as M-DG1, M-DG2 and M-DR. As expected, the M-DR test suffers from substantial under-rejection for D1, D3 and D4 even in large samples. M-DG1 and M-DG2 improve the situation somewhat, but the under-rejection is still significant for D3. Tables \ref{Tab: power, D2} and \ref{Tab: power, D5} indicate that our tests are more powerful than M-DG1, M-DG2 and M-DR in all cases. In particular, for D5 the rejection rates of our tests are close to one when $T$ is large while those of M-DG1 and M-DG2 are not. Results for multiple testing procedures based on \citet{Dovonon_Goncalves2017bootstrapping} with optimal weighting matrix share similar patterns and are available upon request. We reiterate that the multiple testing procedure would not help with partial identification, and both \citet{Dovonon_Renault2013testing} and \citet{Dovonon_Goncalves2017bootstrapping} crucially rely on point identification.

\setlength{\tabcolsep}{2pt}

\begin{table}[htbp]
\caption{Rejection rates under the null: Design D1}\label{Tab: size, D1}
\begin{footnotesize}
\begin{center}
\begin{tabular}{ccccccccccccc}
\hline\hline
\multirow{2}{*}{$T\backslash\text{Tests}$} & \multicolumn{3}{c}{CF1} & \multicolumn{3}{c}{CF2} & \multicolumn{4}{c}{DG}& \multicolumn{2}{c}{DR}\\
\cmidrule{2-13}
 & $T^{-1/4}$ & $T^{-1/3}$ & $T^{-2/5}$ & $T^{-1/4}$ & $T^{-1/3}$ & $T^{-2/5}$ & DG1 & DG2& M-DG1 & M-DG2 & DR & M-DR\\
\hline
$1000$ & 0.0850 & 0.0640 &  $0.0420$& 0.0395 & $0.0185$ & 0.0100 &  $0.3975$ & $0.4015$ &0.0140&0.0160&0.1740&0.0075\\
$2000$ &  0.0940 & 0.0715 & $0.0530$& 0.0550 & $0.0320$ & 0.0120 &  $0.5060$ & $0.5045$ &0.0290&0.0315&0.2855&0.0125\\
$5000$ & 0.1010 & 0.0740 & $0.0515$& 0.0505 & $0.0290$ & 0.0075 &  $0.6215$ & $0.6185$  &0.0485&0.0510&0.3805&0.0185\\
$10000$ & 0.1010  &0.0820 & $0.0585$& 0.0550 & $0.0285$ & 0.0090 & $0.6375$ & $0.6270$  &0.0480&0.0545&0.4005&0.0240\\
$20000$ & 0.1005 & 0.0725 & $0.0525$& 0.0495 & $0.0285$ & 0.0115 &  $0.6750$ & $0.6705$ &0.0425&0.0550&0.4405&0.0225\\
$40000$ &  0.1180 & 0.0900 &$0.0670$& 0.0700 & $0.0410$ & 0.0165 & $0.6865$ & $0.6845$  &0.0635&0.0625&0.4710&0.0400\\
$50000$ & 0.1070 & 0.0830 & $0.0660$ & 0.0665 & $0.0410$ & 0.0145 & $0.6895$ & $0.6870$ &0.0425&0.0515&0.4430&0.0335\\
\hline\hline
\end{tabular}
\end{center}
\end{footnotesize}
\end{table}

{\setlength\intextsep{0.1in}
\begin{table}[htbp]
\caption{Rejection rates under the null: Design D3}\label{Tab: size, D3}
\begin{footnotesize}
\begin{center}
\begin{tabular}{ccccccccccccc}
\hline\hline
\multirow{2}{*}{$T\backslash\text{Tests}$} & \multicolumn{3}{c}{CF1} & \multicolumn{3}{c}{CF2} & \multicolumn{4}{c}{DG}& \multicolumn{2}{c}{DR}\\
\cmidrule{2-13}
 & $T^{-1/4}$ & $T^{-1/3}$ & $T^{-2/5}$ & $T^{-1/4}$ & $T^{-1/3}$ & $T^{-2/5}$ & DG1 & DG2& M-DG1 & M-DG2 & DR & M-DR\\
\hline
$1000$ & 0.0605 &	0.0390  &  $0.0285$& 0.0660 & $0.0605$ & 0.0430 & $0.2300$&$0.2400$  &0.0025&0.0030 &0.0305 &0.0000 \\
$2000$ & 0.0645	& 0.0385 & $0.0280$& 0.0655 & $0.0570$ & 0.0380 & $0.3425$ &$0.3470$   &0.0040&0.0040  &0.0565 &0.0005   \\
$5000$ & 0.0520 & 0.0385 & $0.0315$& 0.0505 & $0.0455$ & 0.0275 &  $0.3970$ &$0.3965$  &0.0025&0.0015  &0.0715 &0.0000   \\
$10000$ & 0.0690 & 0.0565 & $0.0450$& 0.0830 & $0.0665$ & 0.0320 & $0.4385$ & $0.4415$ &0.0030&0.0040  &0.0960 &0.0000   \\
$20000$ & 0.0660  & 0.0600 & $0.0490$& 0.0850 & $0.0660$ & 0.0335 &  $0.4765$ &$0.4790$&0.0070&0.0065  &0.1145 &0.0005   \\
$40000$ & 0.0520 & 0.0460 &$0.0390$& 0.0645 & $0.0475$ & 0.0225 & $0.5030$ & $0.5065$  &0.0025&0.0040  &0.1175 &0.0000    \\
$50000$ & 0.0745 & 0.0670 & $0.0585$ & 0.0920 & $0.0635$ & 0.0395 &  $ 0.5255$ &$0.5290$&0.0065&0.0040  &0.1540&0.0005  \\
\hline\hline
\end{tabular}
\end{center}
\end{footnotesize}
\end{table}
}

{\setlength\intextsep{0pt}
\begin{table}[htbp]
\caption{Rejection rates under the null: Design D4}\label{Tab: size, D4}
\begin{footnotesize}
\begin{center}
\begin{tabular}{ccccccccccccc}
\hline\hline
\multirow{2}{*}{$T\backslash\text{Tests}$} & \multicolumn{3}{c}{CF1} & \multicolumn{3}{c}{CF2} & \multicolumn{4}{c}{DG}& \multicolumn{2}{c}{DR}\\
\cmidrule{2-13}
 & $T^{-1/4}$ & $T^{-1/3}$ & $T^{-2/5}$ & $T^{-1/4}$ & $T^{-1/3}$ & $T^{-2/5}$ & DG1 & DG2& M-DG1 & M-DG2 & DR & M-DR\\
\hline
$1000$ & 0.0715	& 0.0445 &  $0.0265$& 0.1305 & $0.0915$ & 0.0415 &  $0.4795$ & $0.4870$ &0.0240&0.0240&0.1795&0.0010  \\
$2000$ & 0.0895	& 0.0515 & $0.0380$& 0.1485 & $0.0935$ & 0.0330 &  $0.6380$ & $0.6515$  &0.0345&0.0335&0.3210&0.0055  \\
$5000$ & 0.1055 & 0.0720 & $0.0545$& 0.1590 & $0.0960$ & 0.0300 &  $0.7810$ & $0.7820$  &0.0400&0.0400&0.4625&0.0075  \\
$10000$ & 0.1135 & 0.0615 & $0.0485$& 0.1440 & $0.0750$ & 0.0290 & $0.8055$ & $0.8030$  &0.0445&0.0370&0.4840&0.0080  \\
$20000$ & 0.1155 & 0.0715 & $0.0555$& 0.1530 & $0.0960$ & 0.0290 &  $0.8495$ & $0.8485$ &0.0565&0.0460&0.5555&0.0170  \\
$40000$ & 0.1280 & 0.0810 &$0.0640$& 0.1655 & $0.0900$ & 0.0300 &  $0.8650$ & $0.8670$  &0.0635&0.0700&0.5650&0.0145   \\
$50000$ & 0.1150 & 0.0775 & $0.0660$ & 0.1650 & $0.0855$ & 0.0260 & $0.8610$ & $0.8590$ &0.0535&0.0685&0.5980&0.0125   \\
\hline\hline
\end{tabular}
\end{center}
\end{footnotesize}
\end{table}
}

{\setlength\intextsep{0pt}
\begin{table}[!htbp]
\caption{Rejection rates under the alternative: Design D2}\label{Tab: power, D2}
\begin{footnotesize}
\begin{center}
\begin{tabular}{cccccccccc}
\hline\hline
\multirow{2}{*}{$T\backslash\text{Tests}$} & \multicolumn{3}{c}{CF1} & \multicolumn{3}{c}{CF2} & \multicolumn{2}{c}{DG}& \multicolumn{1}{c}{DR}\\
\cmidrule{2-10}
 & $T^{-1/4}$ & $T^{-1/3}$ & $T^{-2/5}$ & $T^{-1/4}$ & $T^{-1/3}$ & $T^{-2/5}$& M-DG1 & M-DG2 & M-DR\\
\hline
$1000$  &0.6450 & 0.5915 & $0.5050$& 0.7255 & $0.6890$ & 0.5570 &0.2420&0.2170&0.3740  \\
$2000$  &0.9410 & 0.9185 & $0.8805$& 0.9530 & $0.9365$ & 0.8785 &0.4935&0.3945&0.8325  \\
$5000$  &0.9975 & 0.9975 & $0.9960$& 0.9995 & $0.9990$ & 0.9950 &0.9070&0.9180&0.9940  \\
$10000$ &0.9980 & 0.9980 & $0.9975$& 0.9985 & $0.9985$ & 0.9985 &0.9995&0.9995&0.9985  \\
$20000$ &0.9985 & 0.9990 & $0.9985$& 0.9995 & $0.9995$ & 0.9985 &1.0000&1.0000&0.9985  \\
$40000$ &0.9995 & 0.9995 & $0.9995$& 1.0000 & $1.0000$ & 1.0000 &1.0000&1.0000&0.9950   \\
$50000$ &0.9995 & 0.9995 & $0.9995$& 0.9995 & $0.9995$ & 0.9995 &1.0000&1.0000&0.9995   \\
\hline\hline
\end{tabular}
\end{center}
\end{footnotesize}
\end{table}
}

{\setlength\intextsep{0pt}
\begin{table}[!htbp]
\caption{Rejection rates under the alternative: Design D5}\label{Tab: power, D5}
\begin{footnotesize}
\begin{center}
\begin{tabular}{cccccccccc}
\hline\hline
\multirow{2}{*}{$T\backslash\text{Tests}$} & \multicolumn{3}{c}{CF1} & \multicolumn{3}{c}{CF2} & \multicolumn{2}{c}{DG}& \multicolumn{1}{c}{DR}\\
\cmidrule{2-10}
 & $T^{-1/4}$ & $T^{-1/3}$ & $T^{-2/5}$ & $T^{-1/4}$ & $T^{-1/3}$ & $T^{-2/5}$& M-DG1 & M-DG2 & M-DR\\
\hline
$1000$  &  0.1240 & 0.0740 & $0.0630$& 0.3990 & $0.3645$ & 0.3000 &0.0385&0.0395&0.0140  \\
$2000$  &  0.3520 & 0.2710 & $0.2300$& 0.6975 & $0.6675$ & 0.5570 &0.1065&0.0870&0.1295  \\
$5000$  &  0.8250 & 0.7710 & $0.7255$& 0.9610 & $0.9460$ & 0.8885 &0.3470&0.3365&0.6675  \\
$10000$ &  0.9865 & 0.9850 & $0.9755$& 0.9995 & $0.9985$ & 0.9955 &0.5945&0.6765&0.9420  \\
$20000$ &  0.9980 & 0.9970 & $0.9955$& 1.0000 & $1.0000$ & 1.0000 &0.6385&0.6005&0.9665  \\
$40000$ &  1.0000 & 1.0000 & $0.9985$& 1.0000 & $1.0000$ & 1.0000 &0.7225&0.7135&0.9710   \\
$50000$ &  0.9995 & 0.9995 & $0.9990$& 1.0000 & $1.0000$ & 1.0000 &0.7755&0.7445&0.9765   \\
\hline\hline
\end{tabular}
\end{center}
\end{footnotesize}
\end{table}
}

\section{Conclusion}\label{Sec: conclusion}

In this paper, we developed a general statistical framework for conducting inference on functionals exhibiting first order degeneracy, i.e., the first order derivative of the parameter is zero. Our first contribution implies that the standard bootstrap necessarily fails to work in these settings. In light of this failure, we provided two general solutions: one generalizes the Babu correction, and the other one is a modified bootstrap following \citet{Fang_Santos2014HDD}. Our framework includes many existing results as special cases. To further demonstrate the applicability of our theory, we developed a test of common CH features studied by \citet{Dovonon_Renault2013testing} but under weaker assumptions that allow the existence of more than one common CH features.

\addcontentsline{toc}{section}{References}
\putbib
\end{bibunit}

\clearpage

\newpage

\begin{bibunit}

\setcounter{page}{1}
\begin{appendices}
\titleformat{\section}{\Large\center}{{\sc Appendix} \thesection}{1em}{}
\setcounter{equation}{0}
\numberwithin{equation}{section}
\numberwithin{table}{section}
\setcounter{table}{0}
\renewcommand{\thetable}{\thesection.\arabic{table}}
\setcounter{figure}{0}
\renewcommand\thefigure{\thesection.\arabic{figure}}

\emptythanks

\phantomsection
\pdfbookmark[1]{Appendix Title}{title1}
\title{Online Supplemental Appendix to ``Inference on Functionals under First Order Degeneracy''}
\author{
Qihui Chen\\ School of Management and Economics\\The \hspace{-0.03cm}Chinese \hspace{-0.03cm}University \hspace{-0.03cm}of \hspace{-0.03cm}Hong \hspace{-0.03cm}Kong, \hspace{-0.03cm}Shenzhen\\ qihuichen@cuhk.edu.cn
\and
Zheng Fang \\ Department of Economics \\ Texas A\&M University\\ zfang@tamu.edu}
\date{\today}
\maketitle

The following list includes notation that will be used throughout the supplement.

\renewcommand{\arraystretch}{1}
\begin{table}[h]
\begin{center}
\begin{tabular}{cl}
$ a \lesssim b$                     & $a\leq Mb$ for some constant $M$ that is universal in the proof.\\
$A^\epsilon$                     & For $A$ in a metric space $(T,d)$, $A^\epsilon\equiv\{t\in T: \inf_{a\in A}d(t,a)\le\epsilon\}$.\\
$\mathbf M^{m\times k}$ & The space of $m\times k$ real matrices.\\
$\ell^\infty(T)$            & For a set $T$, $\ell^\infty(T) \equiv \{f: T \rightarrow \mathbf R : \sup_{t\in T} |f(t)| < \infty\}$.\\
$C(T)$             & For a set $T$, $C(T) \equiv \{f:T \rightarrow \mathbf R : \sup_{t\in T} |f(t)| < \infty \text{ and } f \text{ is continuous}\}$. \\
$C^1(T)$ & For a set $T\subset\mathbf R^k$, $C^1(T)$ is the set of continuously differentiable functions on $T$.\\
$d_H(\cdot,\cdot)$ & For sets $A,B$, $d_H(A,B)$ is the Hausdorff distance between $A$ and $B$.
\end{tabular}
\end{center}
\end{table}

\section{Local Analysis}\label{Sec: Testing}

In this appendix, we show how our bootstrap procedures can provide local size control. We start by characterizing local perturbations of the data generating process and their implications for the testing statistic $r_n^2\phi(\hat\theta_n)$.

\subsection{Local Perturbations}

We first introduce relevant concepts following \citet{BKRW993Efficient}. In what follows we specialize our setup to the the i.i.d.\ setting for simplicity.\footnote{Generally, we may consider models that are locally asymptotically quadratic \citep{Vaart1998,Ploberger_Phillips2012optimal}.} In particular, the data $\{X_i\}_{i=1}^n$ is presumed to have a common probability measure $P\in\mathcal P$, where $\mathcal P$ is a collection of Borel probability measures that possibly generate the data. Further, we think of the parameter $\theta_0$ as a map $\theta:\mathcal P\to\mathbb D_\phi$, i.e., $\theta_0=\theta(P)$. Formally, we impose the following:

\begin{ass}\label{Ass: iid}
(i) $\{X_i\}_{i=1}^n$ is an i.i.d.\ sequence with each $X_i\in \mathbf R^{d_x}$ distributed according to $P\in \mathcal P$; (ii) $\theta_0 \equiv \theta(P)$ for some known map $\theta : \mathcal P \rightarrow \mathbb D_\phi$ and $\phi(\theta_{0})=0$.
\end{ass}

Given the model $\mathcal P$ defined in Assumption \ref{Ass: iid}, we now formalize the notion of local perturbations to the true probability measure $P$. Intuitively, a local perturbation can be thought as a sequence of probability measures contained in $\mathcal P$ that approaches $P$. Since the set of probability measures is not a vector space, an appropriate embedding is needed to make precise sense of this idea. This is simplified by considering one dimensional parametric models containing $P$ and contained in $\mathcal P$ \citep{Stein1956efficient}.

\begin{defn}
A function $t\mapsto P_t$ mapping a neighborhood $(-\epsilon,\epsilon)$ of zero into $\mathcal P$ is called a differentiable path passing through $P$ if $P_0=P$ and for some $h: \mathbf R^{d_x}\to\mathbf R$,
\begin{align}\label{Eqn: differentiable path}
\lim_{t\to 0}\int\left[\frac{dP_t^{1/2}-dP^{1/2}}{t}-\frac{1}{2}h\,dP^{1/2}\right]^2= 0~.
\end{align}
\end{defn}

Intuitively, a differentiable path is just a parametric model in $\mathcal P$ and indexed by $t\in(-\epsilon,\epsilon)$ such that it is getting close to $P$ sufficiently fast as $t\to 0$. The function $h$ is referred to as the score function of $P$ and satisfies $\int h\,dP=0$ and $h\in L^2(P)$.

The perturbations on $P$ are fundamental in that they affect everything that is built on the model, which in particular includes the parameter $\theta: \mathcal P\to\mathbb D_\phi$ and the estimator $\hat\theta_n:\{X_i\}_{i=1}^n\to\mathbb D_\phi$. In this paper, we shall only consider $\theta$ and $\hat\theta_n$ that are well behaved with respect to these local perturbations. This is formalized by the following assumption.

\begin{ass}\label{Ass: regular}
(i) For every differentiable path $\{P_t\}$ in $\mathcal P$ with score function $h$, $\theta:\mathcal P\to\mathbb D_\phi$ is regular in the sense that there exists $\theta_0'(h)\in \mathbb D_0$ such that $\|\theta(P_t) - \theta(P) - t\theta_0'(h)\|_{\mathbb D} = o(t)$ (as $t \rightarrow 0$); (ii) $\hat \theta_n$ is a regular estimator for $\theta(P)$.\footnote{Formally, $\hat \theta_n$ is a regular estimator if for every differentiable path $\{P_t\}$ in $\mathcal P$ with score function $h$, we have $r_n\{\hat \theta_n - \theta(P_n)\} \stackrel{L_n}{\rightarrow } \mathbb G$, where $P_n \equiv P_{1/r_n}$ and $L_n$ denotes the law under $\prod_{i=1}^n P_n$.}
\end{ass}

Assumption \ref{Ass: regular}(i) is a smoothness condition on the parameter $\theta:\mathcal P\to\mathbb D_\phi$ and the model $\mathcal P$, which rules out parameters defined by, for example, densities or conditional densities with jumps \citep{Ibragimov_Hasminskii1981,Chernozhukov_Hong2004nonregular}. In our examples, $\theta_0$ takes the form of expectations, so Assumption \ref{Ass: regular}(i) is met under standard conditions as long as the model $\mathcal P$ is sufficiently rich to include differentiable paths \citep{BKRW993Efficient,Brown_Newey1998expectation}. Assumption \ref{Ass: regular}(ii) means that $\hat\theta_n$ is asymptotically invariant to local perturbations, excluding superefficient estimators such as Hodges's estimator or Stein's estimator \citep{Vaart1997Superefficiency}. Since $\theta_0$ are population means in all our examples, Assumption \ref{Ass: regular}(ii) is satisfied if we take $\hat\theta_n$ to be the corresponding sample averages; see, for example, Theorem 3.10.12 in \citet{Vaart1996} and \citet{Jeganathan1995LANtimeseries}. Assumption \ref{Ass: regular}(i) and (ii) in fact are closely related, though themselves alone do not imply one another. In particular, regularity of $\hat\theta_n$ plus a mild condition implies regularity of $\theta:\mathcal P\to\mathbb D_\phi$, and vice versa \citep{Vaart1991differentibility,Hirano_Porter2012}.

The local behaviors of our test statistic can now be characterized as follows.

\begin{lem}\label{Lem: weak limit under perturbations}
Let $\{P_t\}$ be a differentiable path with score function $h$. Suppose that Assumptions \ref{Ass: HDD 2nd}, \ref{Ass: weak convergence for theta}, \ref{Ass: iid} and \ref{Ass: regular} hold. Then,
\begin{align}
r_n^2\phi(\hat\theta_n)\convln \phi_{\theta_0}''(\mathbb G+\theta_0'(h))~,
\end{align}
where $L_n$ denotes the law under $\prod_{i=1}^n P_n$ with $P_n\equiv P_{1/r_n}$ by abuse of notation.
\end{lem}

Lemma \ref{Lem: weak limit under perturbations} indicates that the asymptotic distribution of $r_n^2\phi(\hat\theta_n)$ varies as a function of the score $h$, and in this sense exhibits second order irregularity, even if the map $\phi$ is both first and second order differentiable and $\hat\theta_n$ is regular. This is perhaps surprising {\it ex ante} and yet somewhat expected {\it ex post}. One important implication of Lemma \ref{Lem: weak limit under perturbations} is that one should carefully evaluate how sensitive the statistical procedures under consideration is, in the presence of first order degeneracy.

\subsection{Local Size and Power}\label{Sec: Local Size and Power}

Having derived the asymptotic distributions of $r_n^2\phi(\hat\theta_n)$ under local perturbations, we are now in a position to establish local power performance and local size control of our test. We consider differentiable paths $\{P_t\}$ in $\mathcal P$ that also belong to the set
\[
\mathcal H \equiv \{ \{P_t\} : \text{ (i) } \phi(\theta(P_{t}))= 0 \text{ if } t \leq 0,  \text{ and (ii) } \phi(\theta(P_{t})) > 0 \text{ if } t > 0 \} ~.
\]
Thus, a path $\{P_t\} \in \mathcal H$ is such that $\{P_t\}$ satisfies the null hypothesis whenever $t \leq 0$, but switches to satisfying the alternative hypothesis at all $t >0$. One can think of $\mathcal H$ as a simple device to study local size and power in a compact way. Further, we denote the power function at sample size $n$ for the test that rejects whenever $r_n^2\phi(\hat\theta_n) > \hat c_{1-\alpha}$ by
\[
\pi_n(P_{\eta/r_n}) \equiv P_n^n( r_n^2\phi(\hat\theta_n)> \hat c_{1-\alpha}) ~,
\]
where we write $P_n\equiv P_{\eta/r_n}$ and $P_n^n\equiv\prod_{i=1}^n P_n$. The following additional assumption ensures local size control of our test.

\begin{ass}\label{Ass: local}
(i) $\mathbb E = \mathbf R$; (ii) The cdf of $\phi_{\theta_0}''(\mathbb G)$ is strictly increasing and continuous at its $(1-\alpha)$-th quantile $c_{1-\alpha}$; (iii) There exists a strictly increasing function $\tau: \phi_{\theta_0}''(\mathbb D_0)\to\mathbf R$ such that $\tau(0)=0$ and $\tau\circ\phi_{\theta_0}'': \mathbb D_0\to\mathbf R$ is subadditive.
\end{ass}

Assumption \ref{Ass: local}(i) formalizes the requirement that $\phi$ be scalar valued. Assumption \ref{Ass: local}(ii) requires strict monotonicity of the cdf of $\phi_{\theta_0}''(\mathbb G)$ at $c_{1-\alpha}$ which ensures consistency of the critical value $\hat c_{1-\alpha}$, and continuity which ensures the test controls size at least pointwise in $P$. Subadditivity of $\tau\circ\phi_{\theta_0}''$ as required in Assumption \ref{Ass: local}(iii) is crucial for establishing local size control of our test. This condition was imposed directly on the first order derivative in \citet{Fang_Santos2014HDD}. In our setup, $\phi_{\theta_0}''$ itself often violates subadditivity because it is closely related to quadratic forms. Nonetheless, in all but Example \ref{Ex: Overidentification Test}, $\tau\circ\phi_{\theta_0}''$ is subadditive for $\tau: \mathbf R^+\to\mathbf R^+$ given by $\tau(\nu)=\sqrt{\nu}$.\footnote{For Example \ref{Ex: Overidentification Test}, it turns out that $\sqrt{\phi_{\theta_0}''(\cdot)}$ is subadditive when $\gamma_0$ is point identified, though the main motivation for us being general there is to accommodate partial identification as well as the Jacobian matrix being degenerate.}

The following theorem derives the asymptotic limits of the power function $\pi_n(P_{\eta/r_n})$.

\begin{thm}\label{Thm: local analysis}
Let Assumptions \ref{Ass: HDD 2nd}, \ref{Ass: weak convergence for theta}, \ref{Ass: bootstrap for theta}, \ref{Ass: measurability}, \ref{Ass: derivative estimation}, \ref{Ass: iid}, \ref{Ass: regular} and \ref{Ass: local}(i)(ii) hold. It then follows that for any differentiable path $\{P_t\}$ in $\mathcal H$ with score function $h$, and every $\eta\in \mathbf R$ we have
\begin{equation}\label{Eqn: local analysis, power}
\liminf_{n\rightarrow \infty} \pi_n(P_{\eta/r_n}) \geq P(\phi_{\theta_0}''(\mathbb G + \theta_0^\prime(\eta h)) > c_{1-\alpha}) ~.
\end{equation}
If in addition Assumption \ref{Ass: local}(iii) also holds, then we can conclude that for any $\eta \leq 0$
\begin{equation}\label{Eqn: local analysis, size}
\limsup_{n\rightarrow \infty} \pi_n(P_{\eta/r_n}) \leq \alpha ~.
\end{equation}
\end{thm}

The first claim of the theorem establishes a lower bound for the power function under local perturbations to the null which includes in particular local alternatives. In fact, the lower bound is sharp whenever $c_{1-\alpha}$ is a continuity point of the cdf of $\phi_{\theta_0}''(\mathbb G + \eta \theta_0'(h))$, in which case \eqref{Eqn: local analysis, power} holds with equality. The role of Assumption \ref{Ass: local}(iii) can be seen from \eqref{Eqn: local analysis, power} and the inequalities
\begin{align*}
P(\phi_{\theta_0}''(\mathbb G + \eta \theta_0'(h)) > c_{1-\alpha}) &=P(\tau\circ\phi_{\theta_0}''(\mathbb G + \theta_0'(\eta h)) > \tau(c_{1-\alpha}))\\
&\leq P(\tau\circ\phi_{\theta_0}''(\mathbb G) + \tau\circ\phi_{\theta_0}''(\theta_0^\prime(\eta h)) > \tau (c_{1-\alpha}))\\
&= P(\tau\circ\phi_{\theta_0}''(\mathbb G)  > \tau(c_{1-\alpha}))\\
&=P(\phi_{\theta_0}''(\mathbb G)  > c_{1-\alpha}) \le \alpha ~,
\end{align*}
where the second equality is due to $\phi_{\theta_0}''(\theta_0^\prime(\eta h))=0$ and $\tau(0)=0$.\footnote{This is because $\phi_{\theta_0}''(\eta \theta_0^\prime(h)) = \lim_{n\rightarrow \infty} n \{\phi(\theta(P_n)) - \phi(\theta(P))\} = 0$ by Assumption \ref{Ass: HDD 2nd} and $\{P_n\}$ being a local perturbation under the null.}

To conclude this section, we note that it is possible to develop a testing procedure adaptive to potential first order degeneracy, that is, in settings where $\phi$ is not always first order degenerate under the null. We emphasize that $r_n^2\phi(\hat\theta_n)$ fails to be a valid statistic since it diverges to infinity at those nondegenerate points, and so does
\[
r_n^2\{\phi(\hat\theta_n)-\phi_{\theta_0}'(\hat\theta_n-\theta_0)\}~,
\]
because $\theta_0$ might not be identified given $\phi(\theta_0)=0$. By introducing an appropriate selection rule, we can combine first and second order asymptotics to provide a more general testing procedure; see Remark \ref{Rem: Test adaptive}. Development of adaptiveness not only serves to maintain generality of our theory, but also is necessary when constructing confidence sets for $\phi(\theta_0)$; see Remark \ref{Rem: CI adaptive}.

\begin{rem}\label{Rem: Test adaptive}
If $\phi_{\theta_0}'$ is only degenerate at some but not all points under the null, then one may employ the statistic
\begin{align*}
T_n\equiv r_n\phi(\hat\theta_n) \cdot 1\{\frac{r_n\phi(\hat\theta_n)}{\kappa_n}>1\}
+r_n^2\phi(\hat\theta_n)\cdot 1\{\frac{r_n\phi(\hat\theta_n)}{\kappa_n}\le 1\}~,
\end{align*}
where $\kappa_n\darrow 0$ satisfying $\kappa_nr_n\to\infty$ as $n\to\infty$. Heuristically, if $\phi_{\theta_0}'$ is nondegenerate, then $r_n\phi(\hat\theta_n)/\kappa_n=O_p(1)/o_p(1)\convp\infty$ and thus with probability approaching one $T_n= r_n\phi(\hat\theta_n)$ which has nondegenerate weak limit $\phi_{\theta_0}'(\mathbb G)$. If $\phi_{\theta_0}'$ is degenerate, then $r_n\phi(\hat\theta_n)/\kappa_n=r_n^2\phi(\hat\theta_n)/\kappa_nr_n=O_p(1)/\kappa_nr_n\convp 0$ and therefore with probability approaching one $T_n= r_n^2\phi(\hat\theta_n)$ which has nondegenerate weak limit $\phi_{\theta_0}''(\mathbb G)$. Accordingly we may construct the corresponding critical value as
\begin{align}\label{Eqn: Test, cv}
\hat c_{1-\alpha}^*\equiv \tilde c_{1-\alpha}\cdot 1\{\frac{r_n\phi(\hat\theta_n)}{\kappa_n}>1\}+\hat c_{1-\alpha}\cdot 1\{\frac{r_n\phi(\hat\theta_n)}{\kappa_n}\le 1\}~,
\end{align}
where for $\alpha\in(0,1)$ and some estimator $\hat\phi_n'$ of $\phi_{\theta_0}'$,
\begin{align*}
\tilde c_{1-\alpha}\equiv\inf\{c\in\mathbf R: P_W(\hat\phi_n'(r_n\{\hat\theta_n^*-\hat\theta_n\})\le c)\ge 1-\alpha\}~.
\end{align*}
The indicator functions above serve as a rule for selecting proper statistics based on degeneracy of (a finite sample analogue of) $\phi_{\theta_0}'$. \qed
\end{rem}

\begin{rem}\label{Rem: CI adaptive}
Confidence regions for $\nu_0\equiv\phi(\theta_0)\in\mathbb{E}$ can be constructed by test inversion based on the statistic
\begin{align}\label{Eqn: CI, feasible}
T_n(\nu_0)\equiv r_n\psi(\hat\theta_n) \cdot 1\{\frac{r_n\psi(\hat\theta_n)}{\kappa_n}>1\}+r_n^2\psi(\hat\theta_n)\cdot 1\{\frac{r_n\psi(\hat\theta_n)}{\kappa_n}\le 1\}~,
\end{align}
where $\psi:\mathbb{D}_{\phi}\to\mathbf{R}$ is given by $\psi(\theta)\equiv\|\phi(\theta)-\nu_{0}\|_{\mathbb{E}}$. Critical values can be constructed in a similar fashion as in Remark \ref{Rem: Test adaptive}. By the chain rule \citep[Proposition 3.6]{Shapiro1990}, it is straightforward to see that $\psi'_{\theta_{0}}=\|\phi'_{\theta_{0}}\|_{\mathbb{E}}$ and so $\phi'_{\theta_{0}}=0$ if and only if $\psi'_{\theta_{0}}=0$. Moreover, $\psi''_{\theta_{0}}=\|\phi''_{\theta_{0}}\|_{\mathbb{E}}$ when $\psi'_{\theta_{0}}=0$. In general, confidence regions thus constructed are less conservative than the plug-in type confidence regions $\phi(\mathcal C_{n,\theta})$ with $\mathcal C_{n,\theta}$ some level $1-\alpha$ confidence region for $\theta_0$. Pointwise validity of $\mathcal C_{n,\theta}$ is straightforward to establish, but the local properties appear to be challenging to develop. \qed
\end{rem}

Finally, we present the proofs of Lemma \ref{Lem: weak limit under perturbations} and Theorem \ref{Thm: local analysis}.

\noindent{\sc Proof of Lemma \ref{Lem: weak limit under perturbations}:} By Assumptions \ref{Ass: weak convergence for theta}(i)(ii), \ref{Ass: iid} and \ref{Ass: regular}, we have for $P_n\equiv P_{1/r_n}$,
\begin{align}\label{Eqn: weak limit under perturbation, aux1}
r_n\{\hat\theta_n-\theta(P)\}=r_n\{\hat\theta_n-\theta(P_n)\}+r_n\{\theta(P_n)-\theta(P)\}\convln \mathbb G+\theta_0'(h)~.
\end{align}
Combination of Assumptions \ref{Ass: HDD 2nd}(i)(ii), $\phi(\theta(P))=\phi_{\theta_0}'=0$, and result \eqref{Eqn: weak limit under perturbation, aux1} allows us to invoke the second order Delta method to conclude that
\begin{align}
r_n^2\phi(\hat\theta_n)=r_n^2\{\phi(\hat\theta_n)-\phi(\theta(P))-\phi_{\theta_0}'(\hat\theta_n-\theta(P))\}\convln \phi_{\theta_0}''(\mathbb G+\theta_0'(h))~.
\end{align}
This completes the proof of the lemma.\qed

\noindent{\sc Proof of Theorem \ref{Thm: local analysis}:} Under the assumptions in Theorem \ref{Thm: modified bootstrap} and Assumptions \ref{Ass: local}(i)(ii), we can show following the proof of Corollary 3.2 in \citet{Fang_Santos2014HDD} that $\hat c_{1-\alpha}\convp c_{1-\alpha}$ under $P^n$. By Theorem 12.2.3 and Corollary 12.3.1 in \citet{TSH2005}, $P_n^n$ and $P^n$ are mutually contiguous. It follows that
\begin{align}\label{Eqn: local, aux1}
\hat c_{1-\alpha}\convp c_{1-\alpha} \text{ under }P_n^n~.
\end{align}
Lemma \ref{Lem: weak limit under perturbations}, Assumption \ref{Ass: local}(i)(ii) and result \eqref{Eqn: local, aux1} allow us to conclude by the portmanteau theorem that
\begin{align}
\liminf_{n\to\infty}\pi_n(P_{\eta/r_n})\ge P(\phi_{\theta_0}''(\mathbb G+\theta_0'(\eta h))>c_{1-\alpha})~.
\end{align}
This establishes the first claim of the theorem.

For the second claim, note that if $\eta\le 0$, then
\begin{align}\label{Eqn: local, aux2}
0=\lim_{n\to\infty}r_n^2\{\phi(\theta(P_n))-\phi(\theta(P))\}=\phi_{\theta_0}''(\theta_0'(\eta h))~,
\end{align}
where we exploited $\phi(\theta(P))=\phi(\theta(P_n)=0$ for all $n$ and Assumption \ref{Ass: HDD 2nd}(iii). Hence,
\begin{align}
\limsup_{n\to\infty}\pi_n(P_{\eta/r_n})&\equiv \limsup_{n\to\infty}P_n^n (r_n^2\phi(\hat\theta_n)>\hat c_{1-\alpha})\le \limsup_{n\to\infty}P_n^n (r_n^2\phi(\hat\theta_n)\ge \hat c_{1-\alpha})\notag\\
&\le P(\phi_{\theta_0}''(\mathbb G+\theta_0'(\eta h))\ge c_{1-\alpha})= P(\tau\circ \phi_{\theta_0}''(\mathbb G+\theta_0'(\eta h))\ge \tau(c_{1-\alpha}))\notag\\
&\le P(\tau\circ \phi_{\theta_0}''(\mathbb G)+\tau\circ \phi_{\theta_0}''(\theta_0'(\eta h))\ge \tau(c_{1-\alpha}))\notag\\
&=P(\phi_{\theta_0}''(\mathbb G)\ge c_{1-\alpha})=\alpha~,
\end{align}
where the second inequality is due to the Lemma \ref{Lem: weak limit under perturbations}, result \eqref{Eqn: local, aux1} and the portmanteau theorem, the second equality is by $\tau$ being strictly increasing, the third inequality is by $\tau\circ\phi_{\theta_0}''$ being subadditive, and the third equality is due to result \eqref{Eqn: local, aux2}, $\tau(0)=0$ and $\tau$ being strictly increasing. This proves the second claim of the theorem.\qed

\section{Proofs of Main Results}

By Assumption \ref{Ass: weak convergence for theta}(ii), the support $\mathbb D_L$ of $\mathbb G$ satisfies $\mathbb D_L\subset \mathbb D_0$. Since only the differentiability of $\phi$ on $\mathbb D_L$ is relevant, we may assume without loss of generality that $\mathbb D_0=\mathbb D_L$ in what follows. Moreover, By Proposition I.3.3 in \citet{Vakhania_Tarieladze_Chobanyan1987probability}, the support $\mathbb D_0$ of $\mathbb G$ is closed. It then follows from Theorem 4.1 in \citet{Dugundji1951extension} and Assumption \ref{Ass: HDD 2nd}(i),  $\phi_{\theta_0}''$ can be continuously extended from $\mathbb D_0$ to $\mathbb D$. Throughout the appendix, we thus interpret $\phi_{\theta_0}''$ as its continuous extension whenever it takes arguments $h\in\mathbb D\backslash\mathbb D_0$ with $\mathbb D_0$ being the support of $\mathbb G$.

\noindent{\sc Proof of Theorem \ref{Thm: delta method, 2nd order}:}  The second claim follows from the first by the Slutsky theorem and the continuous mapping theorem, in view of Assumption \ref{Ass: weak convergence for theta}(i)(ii) and continuity of $\phi_{\theta_0}''$ on $\mathbb D$ (interpreted as some continuous extension). Nonetheless, for pedagogical purposes, we go backwards and start by proving the second claim first. For each $n\in\mathbf N$, let $\mathbb D_n\equiv\{h\in\mathbb D: \theta_0+h/r_n\in\mathbb D_\phi\}$ and define $g_n: \mathbb D_n\to\mathbb E$ by
\[
g_n(h_n)\equiv r_n^2\{\phi(\theta_0 +r_n^{-1} h_n)-\phi(\theta_0)-r_n^{-1}\phi_{\theta_0}'(h_n)\}\text{ for any }h_n\in\mathbb D_n~.
\]
By Assumption \ref{Ass: HDD 2nd}(ii), $\|g_n(h_n)-\phi_{\theta_0}''(h)\|_{\mathbb E}\to 0$ whenever $h_n\to h\in\mathbb D_0$. Moreover, $\mathbb G\in\mathbb D_0$ (almost surely) is separable since it is tight by Assumption \ref{Ass: weak convergence for theta}(ii). The second claim then follows by Theorem 1.11.1(i) in \citet{Vaart1996}.

As for the first claim, define $f_n: \mathbb D_n\times\mathbb D\to\mathbb E\times\mathbb E$ by
\[
f_n(h_n,h)\equiv(g_n(h_n),\phi_{\theta_0}''(h))\text{ for any }(h_n,h)\in\mathbb D_n\times\mathbb D~.
\]
Assumption \ref{Ass: HDD 2nd}(ii) then allows us to conclude again by Theorem 1.11.1(i) in \citet{Vaart1996} that
\begin{align}\label{Eqn: HDD 2nd, aux1}
\begin{bmatrix}
r_n^2\{\phi(\hat\theta_n)-\phi(\theta_0)-\phi_{\theta_0}'(\hat\theta_n-\theta_0)\}\\
\phi_{\theta_0}''(r_n\{\hat\theta_n-\theta_0\})
\end{bmatrix}\convl \begin{bmatrix}
\phi_{\theta_0}''(\mathbb G)\\
\phi_{\theta_0}''(\mathbb G)
\end{bmatrix} \text{ in }\mathbb E\times\mathbb E~.
\end{align}
By the continuous mapping theorem applied to result \eqref{Eqn: HDD 2nd, aux1}, we have
\begin{align}\label{Eqn: HDD 2nd, aux2}
r_n^2\{\phi(\hat\theta_n)-\phi(\theta_0)-\phi_{\theta_0}'(\hat\theta_n-\theta_0)\}-\phi_{\theta_0}''(r_n\{\hat\theta_n-\theta_0\})\convl 0~.
\end{align}
The first claim then follows from result \eqref{Eqn: HDD 2nd, aux2} and Lemma 1.10.2(iii) in \citet{Vaart1996}. \qed

\noindent{\sc Proof of Theorem \ref{Thm: bootstrap, iff}:} Inspecting the structure of the problem, we see that the bootstrap consistency \eqref{Eqn: bootstrap, iff} is equivalent to $\phi_{\theta_0}''(\mathbb G+h)-\phi_{\theta_0}''(h)\overset{d}{=}\phi_{\theta_0}''(\mathbb G)$ for all $h\in\mathbb D_0$ by exactly the same arguments as the proof of Theorem A.1 in \citet{Fang_Santos2014HDD}. Thus, it boils down to showing that $\phi_{\theta_0}''(\mathbb G+h)-\phi_{\theta_0}''(h)\overset{d}{=}\phi_{\theta_0}''(\mathbb G)$ for all $h\in\mathbb D_0$ if and only if $\phi_{\theta_0}''(h)=0$ for $h\in\mathbb D_0$. One direction is immediate since if latter holds, then both $\phi_{\theta_0}''(\mathbb G+h)-\phi_{\theta_0}''(h)$ and $\phi_{\theta_0}''(\mathbb G)$ are degenerate at $0$ for all $h\in\mathbb D_0$, and hence are equal in distribution. The converse consists of two steps.

To begin with, note that by Assumption \ref{Ass: weak convergence for theta}(ii), $\mathbb G$ being centered Gaussian and Lemma A.7 in \citet{Fang_Santos2014HDD}, we may assume without loss of generality that the support of $\mathbb G$ is $\mathbb D$ and that $\mathbb D$ is separable. Since $\mathbb D$ is separable, it follows that the Borel $\sigma$-algebra, the $\sigma$-algebra generated by the weak topology, and the cylindrical $\sigma$-algebra coincide by Theorem 2.1 in \citet{Vakhania_Tarieladze_Chobanyan1987probability}. Furthermore, by Theorem 7.1.7 in \cite{Bogachev2007}, $P$ is Radon with respect to the Borel $\sigma$-algebra, and hence also with respect to the cylindrical $\sigma$-algebra. Finally, let $P$ be the law of $\mathbb G$ on $\mathbb D$.

\noindent\underline{\sc Step 1:} Show that $\phi_{\theta_0}''$ corresponds to a bilinear map if $\phi_{\theta_0}''(\mathbb G+h)-\phi_{\theta_0}''(h)\overset{d}{=}\phi_{\theta_0}''(\mathbb G)$ for all $h\in\mathbb D$.

For completeness, we introduce additional notation following Section 3.7 in \citet{Davydov1998local}. First, let $\mathbb D^*$ denote the dual space of $\mathbb D$, and $\langle x,x^*\rangle_{\mathbb D} = x^*(x)$ for any $x\in \mathbb D$ and $x^*\in \mathbb D^*$. Similarly denote the dual space of $\mathbb E$ by $\mathbb E^*$ and the corresponding bilinear form by $\langle \cdot,\cdot\rangle_{\mathbb E}$. Since $\mathbb G$ is Gaussian, $\mathbb D^*\subset L^2(P)$ \citep[p.42]{Bogavcev1998gaussian}. We may thus embed $\mathbb D^*$ into $L^2(P)$. Denote by $\mathbb D_P'$ the closure of $\mathbb D^*$, viewed as a subset of $L^2(P)$. By some abuse of notation write $x^\prime(x) = \langle x^\prime, x\rangle_{\mathbb D}$ for any $x^\prime \in \mathbb D^\prime_P$ and $x\in \mathbb D$.  Finally, for each $h\in \mathbb D$ we let $P^h$ denote the law of $\mathbb G + h$, write $P^h \ll P$ whenever $P^h$ is absolutely continuous with respect to $P$, and define the set:
\begin{equation*}
\mathbb H_P \equiv \{h \in \mathbb D : P^{rh} \ll P \text{ for all } r\in \mathbf R\} ~.
\end{equation*}

Since $P$ is Radon with respect to the cylindrical $\sigma$-algebra of $\mathbb D$, it follows by Theorem 7.1 in \cite{Davydov1998local} that there exists a continuous linear map $I:\mathbb H_P \rightarrow \mathbb D^\prime_P$ satisfying for every $h\in \mathbb H_P$:
\begin{equation}\label{Eqn: bootstrap iff, aux1}
\frac{dP^h}{dP}(x) = \exp\Big\{ \langle x,Ih\rangle_{\mathbb D} - \frac{1}{2} \sigma^2(h)\Big\} \hspace{0.5 in} \sigma^2(h) \equiv \int_{\mathbb D} \langle x,Ih \rangle_{\mathbb D}^2\,P(dx) ~.
\end{equation}

Fix an arbitrary $e^* \in \mathbb E^*$ and $h \in \mathbb H_P$. Since $\phi_{\theta_0}''(\mathbb G+h)-\phi_{\theta_0}''(h)\overset{d}{=}\phi_{\theta_0}''(\mathbb G)$ for all $h\in\mathrm{Supp}(\mathbb G)$, it follows that $\langle e^*, \phi_{\theta_0}''(\mathbb G + rh) - \phi_{\theta_0}''(rh)\rangle_{\mathbb E}$ and $\langle e^*, \phi_{\theta_0}''(\mathbb G)\rangle_{\mathbb E}$ must be equal in distribution for all $r\in \mathbf R$.\footnote{The proof of Lemma A.3 in \citet{Fang_Santos2014HDD} never exploits that $\phi_{\theta_0}'$ is a first order derivative beyond continuity of $\phi_{\theta_0}'$ and $\phi_{\theta_0}'(0)=0$ which are satisfied by $\phi_{\theta_0}''$.} In particular, their characteristic functions must equal each other, and hence for all $r\geq 0$ and $t\in \mathbf R$:
\begin{multline}\label{Eqn: bootstrap iff, aux2}
E[\exp\{it\langle e^*,\phi_{\theta_0}''(\mathbb G)\rangle_{\mathbb E}\}] = E[\exp\{it\{\langle e^*, \phi_{\theta_0}''(\mathbb G + rh) - \phi_{\theta_0}''(rh)\rangle_{\mathbb E}\}\}] \\
 = \exp\{-itr^2\langle e^*,\phi_{\theta_0}''(h)\rangle_{\mathbb E}\} E[\exp\{it\langle e^*, \phi_{\theta_0}''(\mathbb G + rh)\rangle_{\mathbb E}\}]~,
\end{multline}
where in the second equality we have exploited $\phi_{\theta_0}''$ being positively homogenous of degree two. Setting $C(t) \equiv  E[\exp\{it\langle e^*,\phi_{\theta_0}''(\mathbb G)\rangle_{\mathbb E}\}]$, we have by \eqref{Eqn: bootstrap iff, aux2} that
\begin{align}\label{Eqn: bootstrap iff, aux3}
\exp\{itr^2\langle e^*,\phi_{\theta_0}''(h)\rangle_{\mathbb E}\} C(t)=E[\exp\{it\langle e^*, \phi_{\theta_0}''(\mathbb G + rh)\rangle_{\mathbb E}\}]~,
\end{align}
for all $r\geq 0$ and $t\in \mathbf R$.

We next aim to equate second order right derivatives of both sides in the identity \eqref{Eqn: bootstrap iff, aux3}. The second order right derivative of the left hand side at $r=0$ is given by
\begin{align}\label{Eqn: bootstrap iff, aux4}
2it C(t)\langle e^*,\phi_{\theta_0}''(h)\rangle_{\mathbb E}~.
\end{align}
On the other hand, exploiting result \eqref{Eqn: bootstrap iff, aux1}, linearity of $I:\mathbb H_P\rightarrow \mathbb D^\prime_P$ and that $h\in \mathbb H_P$ implies $rh \in \mathbb H_P$ for all $r\in \mathbf R$ and in particular for all $r\in[0,1]$, we may rewrite the right hand side of \eqref{Eqn: bootstrap iff, aux3} as
\begin{align}\label{Eqn: bootstrap iff, aux5}
E[\exp\{it\langle e^*, \phi_{\theta_0}''(\mathbb G &+ rh)\rangle_{\mathbb E}\}]=\int_{\mathbb D} \exp\{it\langle e^*, \phi_{\theta_0}''(x)\rangle_{\mathbb E}\}\frac{dP^{rh}}{dP}(x)P(dx)\notag\\
&=\int_{\mathbb D} \exp\Big\{it\langle e^*,\phi_{\theta_0}''(x)\rangle_{\mathbb E} + r\langle x,Ih\rangle_{\mathbb D} - \frac{r^2}{2}\sigma^2(h)\Big\}P(dx)~.
\end{align}
The integrand on the right hand side of \eqref{Eqn: bootstrap iff, aux5} is differentiable with respect to $r$ for all $r\in[0,1]$ and the resulting derivative is dominated by $\exp\{|\langle x,Ih\rangle_\mathbb D|\}\times\{|\langle x,Ih\rangle_\mathbb D|+\sigma^2(h)\}$ which is integrable against $P$ since $\langle \mathbb G,Ih\rangle_{\mathbb D}\sim N(0,\sigma^2(h))$ by Proposition 2.10.3 in \cite{Bogavcev1998gaussian} and $Ih \in \mathbb D^\prime_P$. Thus by Theorem 2.27(ii) in \citet{Folland1999}, the first order derivative of the right hand side in \eqref{Eqn: bootstrap iff, aux5} at $r\in[0,1]$ exists and is given by
\begin{align}\label{Eqn: bootstrap iff, aux6}
\int_{\mathbb D} \exp\Big\{it\langle e^*,\phi_{\theta_0}''(x)\rangle_{\mathbb E} + r\langle x,Ih\rangle_{\mathbb D} - \frac{r^2}{2}\sigma^2(h)\Big\}\{\langle x,Ih\rangle_{\mathbb D} - r\sigma^2(h)\}P(dx)~.
\end{align}
In turn, result \eqref{Eqn: bootstrap iff, aux6} allows us to conclude that the second order right derivative of the right hand side in \eqref{Eqn: bootstrap iff, aux5} at $r=0$ exists and is given by
\begin{align}\label{Eqn: bootstrap iff, aux7}
\int_{\mathbb D}\exp\{it\langle e^*,\phi_{\theta_0}''(x)\rangle_{\mathbb E}\}[\langle x,Ih\rangle_{\mathbb D}^2-\sigma^2(h)]P(dx)~.
\end{align}
Since equation \eqref{Eqn: bootstrap iff, aux3} holds for all $r\ge 0$ and $t\in\mathbf R$, it follows from results \eqref{Eqn: bootstrap iff, aux4} and \eqref{Eqn: bootstrap iff, aux7} that for all $t\in\mathbf R$:
\begin{align}\label{Eqn: bootstrap iff, aux8}
2it C(t)\langle e^*,\phi_{\theta_0}''(h)\rangle_{\mathbb E}=\int_{\mathbb D}\exp\{it\langle e^*,\phi_{\theta_0}''(x)\rangle_{\mathbb E}\}[\langle x,Ih\rangle_{\mathbb D}^2-\sigma^2(h)]P(dx)~.
\end{align}

Note that $t\mapsto C(t)$ is the characteristic function of $\langle e^*,\phi_{\theta_0}''(\mathbb G)\rangle_{\mathbb E}$ and hence it is continuous. Thus, since $C(0) = 1$ there exists a $t_0 > 0$ such that $C(t_0)t_0 \neq 0$. For such $t_0$ it follows from \eqref{Eqn: bootstrap iff, aux8} that
\begin{equation}\label{Eqn: bootstrap iff, aux9}
\langle e^*,\phi_{\theta_0}''(h)\rangle_{\mathbb E} =-\frac{i E[\exp\{it_0\langle e^*,\phi_{\theta_0}''(\mathbb G)\rangle_{\mathbb E}\}\{\langle \mathbb G,Ih\rangle_{\mathbb D}^2-\sigma^2(h)\}]}{2t_0C(t_0)} ~.
\end{equation}
Define a map $\Phi_{\theta_0}'': \mathbb D\times\mathbb D\to\mathbb E$ by
\begin{equation}\label{Eqn: bootstrap iff, aux10}
\Phi_{\theta_0}''(h,g)\equiv\frac{1}{4}[\phi_{\theta_0}''(h+g)-\phi_{\theta_0}''(h-g)]~.
\end{equation}
It then follows from \eqref{Eqn: bootstrap iff, aux9} that, for any $e^*\in\mathbb E^*$ and any $g,h\in\mathbb D$,
\begin{equation}\label{Eqn: bootstrap iff, aux11}
\langle e^*,\Phi_{\theta_0}''(g,h)\rangle_{\mathbb E} =-\frac{i E[\exp\{it_0\langle e^*,\phi_{\theta_0}''(\mathbb G)\rangle_{\mathbb E}\}\{\langle \mathbb G,Ig\rangle_{\mathbb D}\langle \mathbb G,Ih\rangle_{\mathbb D}-\sigma(g,h)\}]}{2t_0C(t_0)} ~,
\end{equation}
where $\sigma(g,h)\equiv E[\langle \mathbb G,Ig\rangle \langle \mathbb G,Ih\rangle]$. Since $I: \mathbb H_P\to \mathbb D_P'$ is linear, $(h,g)\mapsto\langle e^*,\Phi_{\theta_0}''(g,h)\rangle_{\mathbb E}$ is bilinear on $\mathbb H_P\times\mathbb H_P$. Moreover, $(h,g)\mapsto\langle e^*,\Phi_{\theta_0}''(g,h)\rangle_{\mathbb E}$ is continuous on $\mathbb H_P\times\mathbb H_P$ due to continuity of $\phi_{\theta_0}''$ (and hence $\Phi_{\theta_0}''$) and $e^*\in\mathbb E^*$. We thus conclude from $\mathbb H_P$ being a dense subspace of $\mathbb D$ by Proposition 7.4(ii) in \citet{Davydov1998local} that $(h,g)\mapsto\langle e^*,\Phi_{\theta_0}''(g,h)\rangle_{\mathbb E}$ is continuous and bilinear on $\mathbb D\times\mathbb D$. Since $e^*\in\mathbb E^*$ is arbitrary, it follows from Lemma A.2 in \citet{Vaart1991differentibility} that $\Phi_{\theta_0}'': \mathbb D\times\mathbb D\to\mathbb E$ is bilinear and continuous. By identity \eqref{Eqn: bootstrap iff, aux10}, we have $\phi_{\theta_0}''(h)=\Phi_{\theta_0}''(h,h)$ for all $h\in\mathbb D$. Hence, $\phi_{\theta_0}''$ is a quadratic form corresponding to the bilinear map $\Phi_{\theta_0}''$.

\noindent\underline{\sc Step 2:} Conclude that $\phi_{\theta_0}''=0$ on the support of $\mathbb G$. Note that if $\phi$ is second order Hadamard differentiable, then one can directly start with Step 2.

By Lemma A.3 in \citet{Fang_Santos2014HDD}, for all $h\in\mathbb D$,
\begin{align}\label{Eqn: bootstrap iff, aux13}
\phi_{\theta_0}''(\mathbb G)&\overset{d}{=}\phi_{\theta_0}''(\mathbb G+h)-\phi_{\theta_0}''(h)\notag\\
&=\Phi_{\theta_0}''(\mathbb G+h,\mathbb G+h)-\Phi_{\theta_0}''(h,h)\notag\\
&=\Phi_{\theta_0}''(\mathbb G,\mathbb G)+2\Phi_{\theta_0}''(\mathbb G,h)=\phi_{\theta_0}''(\mathbb G)+2\Phi_{\theta_0}''(\mathbb G,h)~,
\end{align}
where the third equality exploited bilinearity of $\Phi_{\theta_0}''$. Fix an arbitrary $e^*\in\mathbb E^*$. By result \eqref{Eqn: bootstrap iff, aux13}, we have for all $r\in\mathbf R$ and $h\in\mathbb D$,
\begin{align}
E[\exp\{it\langle e^*, \phi_{\theta_0}''(\mathbb G)\rangle_{\mathbb E}\}]&=E[\exp\{it\langle e^*, \phi_{\theta_0}''(\mathbb G)+2\Phi_{\theta_0}''(\mathbb G,rh)\rangle_{\mathbb E}\}]\notag\\
&=E[\exp\{it\langle e^*, \phi_{\theta_0}''(\mathbb G)\rangle_{\mathbb E}\}\exp\{2irt\langle e^*,\Phi_{\theta_0}''(\mathbb G,h)\rangle_{\mathbb E}\}]~,
\end{align}
where the last step used linearity of $\Phi_{\theta_0}''$ in its second argument. We now equate second derivatives of both sides at $r=0$. The second derivative of the left hand side is trivially zero, while that of the right hand side, by the recursive use of dominated convergence arguments, is given by $E[\exp\{it\langle e^*, \phi_{\theta_0}''(\mathbb G)\rangle_{\mathbb E}\}\{2it\langle e^*,\Phi_{\theta_0}''(\mathbb G,h)\rangle_{\mathbb E}\}^2]$. Thus we have
\begin{align*}
E[\exp\{it\langle e^*, \phi_{\theta_0}''(\mathbb G)\rangle_{\mathbb E}\}\{2it\langle e^*,\Phi_{\theta_0}''(\mathbb G,h)\rangle_{\mathbb E}\}^2]=0~,
\end{align*}
for all $t\in\mathbf R$, which in turn implies that for all $t\in\mathbf R\setminus\{0\}$,
\begin{align}\label{Eqn: bootstrap iff, aux14}
E[\exp\{it\langle e^*, \phi_{\theta_0}''(\mathbb G)\rangle_{\mathbb E}\}\langle e^*,\Phi_{\theta_0}''(\mathbb G,h)\rangle_{\mathbb E}^2]=0~.
\end{align}
Picking a sequence $t_n\darrow 0$, replacing $t$ with $t_n$ in \eqref{Eqn: bootstrap iff, aux14} and letting $n\to\infty$ leads to, by the dominated convergence theorem: for all $e^*\in\mathbb E^*$ and all $h\in\mathbb D$,
\begin{align}
E[\langle e^*,\Phi_{\theta_0}''(\mathbb G,h)\rangle_{\mathbb E}^2]=0~.
\end{align}
Consequently, $\langle e^*,\Phi_{\theta_0}''(g,h)\rangle_{\mathbb E}=0$ for all $h\in\mathbb D$ and $P$-almost surely $g\in\mathbb D$. Since $e^*$ is arbitrary, we conclude by Lemma 6.10 in \citet{AliprantisandBorder2006} that $\Phi_{\theta_0}''(g,h)=0$ for all $h\in\mathbb D$ and $P$-almost $g\in\mathbb D$. Hence, $\phi_{\theta_0}''(h)=0$ for $P$-almost $h\in\mathbb D$.

Finally, denote by $\Omega$ the collection of all $h\in\mathbb D$ such that $\phi_{\theta_0}''(h)=0$. Then we have $P(\Omega)=1$ by Assumption \ref{Ass: weak convergence for theta}(ii) and the above discussion. We claim that $\Omega$ is dense in $\mathbb D$. To see this, suppose otherwise and then there must exist some $h_0\in \mathbb D$ and some $\delta>0$ such that $B(h_0,\delta)\cap\Omega =\emptyset$. Note that i) $P(B(h_0,\delta))>0$ since $h_0\in\mathrm{Supp}(P)=\mathbb D$, and ii) $\phi_{\theta_0}''(h)\neq 0$ for all $h\in B(h_0,\delta)$ by the definition of $\Omega$. These contradict the fact $P(\Omega)=1$. Since $\phi_{\theta_0}''$ is continuous $\mathbb D$, we may conclude from $\Omega$ being dense in $\mathrm{Supp}(P)$ and $\phi_{\theta_0}''=0$ on $\Omega$ that $\phi_{\theta_0}''=0$ on $\mathbb D$. \qed

\comment{
\noindent{\sc Addendum to Theorem \ref{Thm: bootstrap, iff}:} In this addendum we sketch the proof of Remark \ref{Rem: bootstrap, iff, higher order}. In the setup of previous proof, it boils down to showing that $\phi_{\theta_0}^{(p)}$ is $\mathbb G$-translation invariant if and only if $\phi_{\theta_0}^{(p)}(h)=0$ for all $h\in\mathbb D$. First, for $C(t)=E[\exp\{it\langle e^*, \phi_{\theta_0}^{(p)}(\mathbb G)\rangle_{\mathbb E}\}]$, Lemma A.3 in \citet{Fang_Santos2014HDD} implies that
\begin{align}\label{Eqn: bootstrap iff, addendum, aux1}
\exp\{itr^p\langle e^*,\phi_{\theta_0}^{(p)}(h)\rangle_{\mathbb E}\} C(t)=E[\exp\{it\langle e^*, \phi_{\theta_0}^{(p)}(\mathbb G + rh)\rangle_{\mathbb E}\}]~,
\end{align}
for all $r\geq 0$ and $t\in \mathbf R$. By elementary calculus, the $p$-th order left derivative of the left hand side in \eqref{Eqn: bootstrap iff, addendum, aux1} at $r=0$ is $p!it\langle e^*,\phi_{\theta_0}^{(p)}(h)\rangle_{\mathbb E} C(t)$. By \eqref{Eqn: bootstrap iff, aux1} and the dominated convergence arguments, the $p$-th order right derivative of the right hand side at $r=0$ is given by
}

\noindent{\sc Proof of Theorem \ref{Thm: Babu extension}:} Let $\mathbb D_n\equiv\{h\in\mathbb D: \theta_0+h/r_n\in\mathbb D_\phi\}$ and define for each $n\in\mathbf N$ the map $\Psi_n: \mathbb D_{n}\times\mathbb D_n\to\mathbb E$ by
\[
\Psi_n(g_n,h_n)\equiv r_n^{-2}[\phi(\theta_0+r_n^{-1}h_n)-\phi(\theta_0+r_n^{-1}g_n)-\phi_{\theta_0+r_n^{-1}g_n}'(r_n^{-1}\{h_n-g_n\})]~.
\]
If $\{g_n,h_n\}_{n=1}^\infty \subset\mathbb D_n$ satisfies $(g_n,h_n)\to (g,h)\in\mathbb D_0\times\mathbb D_0$ as $n\to\infty$, then Assumption \ref{Ass: HDD 2nd, refined} allows us to conclude that
\begin{align}\label{Eqn: Babu extension, aux1}
&\Psi_n(g_n,h_n)\equiv r_n^{-2}[\phi(\theta_0+r_n^{-1}h_n)-\phi(\theta_0+r_n^{-1}g_n)-\phi_{\theta_0+r_n^{-1}g_n}'(r_n^{-1}\{h_n-g_n\})]\notag\\
&=r_n^{-2}[\{\phi(\theta_0+r_n^{-1}h_n)-\phi(\theta_0)-r_n^{-1}\phi_{\theta_0}'(h_n)\}-\{\phi(\theta_0+r_n^{-1}g_n)-\phi(\theta_0)-r_n^{-1}\phi_{\theta_0}'(g_n)\}]\notag\\
&\quad-r_n^{-1}[\{\phi_{\theta_0+r_n^{-1}g_n}'(h_n)-\phi_{\theta_0}'(h_n)\}-\{\phi_{\theta_0+r_n^{-1}g_n}'(g_n)-\phi_{\theta_0}'(g_n)\}]\notag\\
&\to \Phi_{\theta_0}''(h,h)-\Phi_{\theta_0}''(g,g)-2\Phi_{\theta_0}''(g,h)+2\Phi_{\theta_0}''(g,g)\notag\\
&=\Psi(g,h)\equiv\Phi_{\theta_0}''(h,h)+\Phi_{\theta_0}''(g,g)-2\Phi_{\theta_0}''(g,h)~.
\end{align}
Since $\phi_{\theta_0}''$ admits a continuous extension on $\mathbb D$, by the corresponding extension of $\Phi_{\theta_0}''$ according to equation \ref{Eqn: bootstrap iff, aux10}, it follows from \eqref{Eqn: Babu extension, aux1} that
\begin{align}\label{Eqn: Babu extension, aux2}
\Psi_n(g_n,h_n)-\Psi(g_n,h_n)=\Psi_n(g_n,h_n)-\Psi(g,h)-\{\Psi(g_n,h_n)-\Psi(g,h)\}\to 0~.
\end{align}

Next, let $\mathbb G_n\equiv r_n\{\hat\theta_n-\theta_0\}$, $\mathbb G_n^*\equiv r_n\{\hat\theta_n^*-\hat\theta_n\}$ and $\mathbb G_n^\dag\equiv r_n\{\hat\theta_n^*-\theta_0\}=\mathbb G_n^*+\mathbb G_n$. By Assumption \ref{Ass: HDD 2nd}, \ref{Ass: weak convergence for theta}, \ref{Ass: bootstrap for theta} and \ref{Ass: measurability}(i), it follows from Lemma A.2 in \citet{Fang_Santos2014HDD} that for $\mathbb G_1,\mathbb G_2$ independently distributed according to $\mathbb G$,
\begin{align}\label{Eqn: Babu extension, aux3}
(\mathbb G_n,\mathbb G_n^*)\convl (\mathbb G_1,\mathbb G_2)~.
\end{align}
By the continuous mapping theorem and result \eqref{Eqn: Babu extension, aux3} we have
\begin{align}\label{Eqn: Babu extension, aux4}
(\mathbb G_n,\mathbb G_n^\dag)=(\mathbb G_n,\mathbb G_n^*+\mathbb G_n)\convl (\mathbb G_1, \mathbb G_1+\mathbb G_2)~.
\end{align}
Combining the separability of $\mathbb G_1$ and $\mathbb G_2$ by Assumption \ref{Ass: weak convergence for theta}(ii), results \eqref{Eqn: Babu extension, aux2} and \eqref{Eqn: Babu extension, aux4}, we conclude by Theorem 1.11.1(i) in \citet{Vaart1996} that
\begin{align}\label{Eqn: Babu extension, aux5}
\Psi_n(\mathbb G_n, \mathbb G_n^\dag)-\Psi(\mathbb G_n, \mathbb G_n^\dag)\convl 0~.
\end{align}
By Lemma 1.10.2 in \citet{Vaart1996} we have from \eqref{Eqn: Babu extension, aux5} that
\begin{align}\label{Eqn: Babu extension, aux6}
\Psi_n(\mathbb G_n, \mathbb G_n^\dag)-\Psi(\mathbb G_n, \mathbb G_n^\dag)=o_p(1)~.
\end{align}

Now fix $\epsilon>0$. Note that
\begin{align}\label{Eqn: Babu extension, aux7}
\sup_{f\in\bl_1(\mathbb E)}|E_W^*[f(\Psi_n(\mathbb G_n, \mathbb G_n^\dag))]&-E_W^*[f(\Psi(\mathbb G_n, \mathbb G_n^\dag))]|\notag\\
& \le\epsilon+2P_W^*(\|\Psi_n(\mathbb G_n, \mathbb G_n^\dag)-\Psi(\mathbb G_n, \mathbb G_n^\dag)\|_{\mathbb E}>\epsilon)~.
\end{align}
By Lemma 1.2.6 in \citet{Vaart1996},
\begin{align}\label{Eqn: Babu extension, aux8}
E_X^*[P_W^*(\|\Psi_n(\mathbb G_n, \mathbb G_n^\dag)-\Psi(\mathbb G_n, \mathbb G_n^\dag)\|_{\mathbb E}>\epsilon)]\le P^*(\|\Psi_n(\mathbb G_n, \mathbb G_n^\dag)-\Psi(\mathbb G_n, \mathbb G_n^\dag)\|_{\mathbb E}>\epsilon)~.
\end{align}
Results \eqref{Eqn: Babu extension, aux6}, \eqref{Eqn: Babu extension, aux7} and \eqref{Eqn: Babu extension, aux8}, together with $\epsilon$ being arbitrary, then yield
\begin{align}\label{Eqn: Babu extension, aux9}
\sup_{f\in\bl_1(\mathbb E)}|E_W^*[f(\Psi_n(\mathbb G_n, \mathbb G_n^\dag))]-E_W^*[f(\Psi(\mathbb G_n, \mathbb G_n^\dag))]|=o_p(1)~.
\end{align}

Result \eqref{Eqn: Babu extension, aux4} and Assumption \ref{Ass: weak convergence for theta}(ii) implies that $(\mathbb G_n,\mathbb G_n^\dag)$ is asymptotically measurable and asymptotically tight. In turn, Lemmas 1.4.3 and 1.4.4 in \citet{Vaart1996} implies that  $(\mathbb G_n,\mathbb G_n^\dag, \mathbb G_1, \mathbb G_1+\mathbb G_2)$ is asymptotically tight and asymptotically measurable. Fix an arbitrary subsequence $\{n_k\}$. Then Theorem 1.3.9 in \citet{Vaart1996} implies that $(\mathbb G_n,\mathbb G_n^\dag, \mathbb G_1, \mathbb G_1+\mathbb G_2)$ converges weakly along a further subsequence of $\{n_k\}$ to a tight Borel law in $\prod_{j=1}^4\mathbb D$, which is equal to $(\mathbb G_1, \mathbb G_1+\mathbb G_2,\mathbb G_1, \mathbb G_1+\mathbb G_2)$ by marginal convergence. This is a weak limit where the dependence structure between the first two components and last two components is known and in fact unique. Since $n_k$ is arbitrary, it follows that
\begin{align}\label{Eqn: Babu extension, aux10}
(\mathbb G_n,\mathbb G_n^\dag, \mathbb G_1, \mathbb G_1+\mathbb G_2)\convl (\mathbb G_1, \mathbb G_1+\mathbb G_2,\mathbb G_1, \mathbb G_1+\mathbb G_2)~.
\end{align}
Since $\Psi: \mathbb D\times\mathbb D\to\mathbb E$ and hence $(\Psi,\Psi): \prod_{j=1}^4\mathbb D\to\prod_{j=1}^2\mathbb E$ is continuous, it follows from result \eqref{Eqn: Babu extension, aux10} and the continuous mapping theorem that
\begin{align}\label{Eqn: Babu extension, aux11}
(\Psi(\mathbb G_n, \mathbb G_n^\dag), \Psi(\mathbb G_1, \mathbb G_1+\mathbb G_2))\convl (\Psi(\mathbb G_1, \mathbb G_1+\mathbb G_2),\Psi(\mathbb G_1, \mathbb G_1+\mathbb G_2))~.
\end{align}
Combination of the continuous mapping theorem and Lemma 1.10.2(iii) in \citet{Vaart1996} yields that
\begin{align}\label{Eqn: Babu extension, aux12}
\Psi(\mathbb G_n, \mathbb G_n^\dag)-\Psi(\mathbb G_1, \mathbb G_1+\mathbb G_2)=o_p(1)~.
\end{align}
By the triangle inequality, we have
\begin{align}\label{Eqn: Babu extension, aux13}
\sup_{f\in\bl_1(\mathbb E)}|E_W^*[f(\Psi&(\mathbb G_n, \mathbb G_n^\dag))]-E[f(\Psi(\mathbb G_1, \mathbb G_1+\mathbb G_2))]|\notag\\
&\le \epsilon+ 2P_W^*(\|\Psi(\mathbb G_n, \mathbb G_n^\dag)-\Psi(\mathbb G_1, \mathbb G_1+\mathbb G_2)\|_{\mathbb E}>\epsilon)~.
\end{align}
By Lemma 1.2.6 in \citet{Vaart1996} and result \eqref{Eqn: Babu extension, aux12}
\begin{align}\label{Eqn: Babu extension, aux14}
E_X^*P_W^*(\|\Psi(\mathbb G_n, \mathbb G_n^\dag)-&\Psi(\mathbb G_1,\mathbb G_1+\mathbb G_2)\|_{\mathbb E}>\epsilon)\notag\\
&\le P^*(\|\Psi(\mathbb G_n, \mathbb G_n^\dag)-\Psi(\mathbb G_1, \mathbb G_1+\mathbb G_2)\|_{\mathbb E}>\epsilon)=o(1)~.
\end{align}
Combination of \eqref{Eqn: Babu extension, aux9}, \eqref{Eqn: Babu extension, aux13}, \eqref{Eqn: Babu extension, aux14} and the triangle inequality leads to
\begin{align}\label{Eqn: Babu extension, aux15}
\sup_{f\in\bl_1(\mathbb E)}|E_W^*[f(\Psi_n(\mathbb G_n, \mathbb G_n^\dag))]-E[f(\Psi(\mathbb G_1, \mathbb G_1+\mathbb G_2))]|=o_p(1)~.
\end{align}
The theorem follows by combining \eqref{Eqn: Babu extension, aux9} and \eqref{Eqn: Babu extension, aux15} and noticing that
\[
\Psi_n(\mathbb G_n, \mathbb G_n^\dag)=r_n^2\{\phi(\hat\theta_n^*)-\phi(\hat\theta_n)-\phi_{\hat\theta_n}'(\hat\theta_n^*-\hat\theta_n)\}\text{ and } \Psi(\mathbb G_1, \mathbb G_1+\mathbb G_2)=\phi_{\theta_0}''(\mathbb G_2)~,
\]
where the second equality is due to bilinearity of $\Phi_{\theta_0}''$.\qed

\noindent{\sc Proof of Theorem \ref{Thm: modified bootstrap}:} Inspecting the proof of Theorem 3.2 in \citet{Fang_Santos2014HDD}, we see that $\phi_{\theta_0}'$ being a first order derivative is actually never exploited there. The conclusion of the theorem then follows in view of Lemma \ref{Lem: derivative estimation, equiv} when combined with exactly the same arguments in \citet{Fang_Santos2014HDD}. \qed

\noindent{\sc Proof of Proposition \ref{Pro: numerical estimation}:} Let $\{h_n\}\subset\mathbb D$ and $h\in\mathbb D_0$ such that $h_n\to h$. Since $\phi_{\theta_0}'=0$ by Assumption \ref{Ass: HDD 2nd}(iii), we may rewrite $\hat\phi_n''(h_n)$:
\begin{align}\label{Eqn: numerical estimation, aux1}
&\hat\phi_n''(h_n)=\frac{\phi(\hat\theta_n+t_nh_n)-\phi(\hat\theta_n)-t_n\phi_{\theta_0}'(h_n)}{t_n^2}\notag\\
&=\frac{\phi(\theta_0+t_ng_n)-\phi(\theta_0)-t_n\phi_{\theta_0}'(g_n)}{t_n^2}
-\frac{r_n^2\{\phi(\hat\theta_n)-\phi(\theta_0)-\phi_{\theta_0}'(\hat\theta_n-\theta_0)\}}{(r_nt_n)^2}
~,
\end{align}
where $g_n\equiv (t_nr_n)^{-1}r_n\{\hat\theta_n-\theta_0\}+h_n$. By Assumptions \ref{Ass: weak convergence for theta}(i), \ref{Ass: rate tn}, Lemma 1.10.2 in \citet{Vaart1996} and $h_n\to h$, we have $g_n\convp h$. By Assumptions \ref{Ass: HDD 2nd}(ii), \ref{Ass: weak convergence for theta}(ii) and Theorem 1.11.1(ii) in \citet{Vaart1996}, we thus have
\begin{align}\label{Eqn: numerical estimation, aux2}
\frac{\phi(\theta_0+t_ng_n)-\phi(\theta_0)-t_n\phi_{\theta_0}'(g_n)}{t_n^2}\convp \phi_{\theta_0}''(h)~.
\end{align}
By Assumption \ref{Ass: HDD 2nd} and \ref{Ass: weak convergence for theta}, it follows from Theorem \ref{Thm: delta method, 2nd order} and $r_nt_n\to\infty$ that
\begin{align}\label{Eqn: numerical estimation, aux3}
\frac{r_n^2\{\phi(\hat\theta_n)-\phi(\theta_0)-\phi_{\theta_0}'(\hat\theta_n-\theta_0)\}}{(r_nt_n)^2}\convp 0~.
\end{align}
Combining results \eqref{Eqn: numerical estimation, aux1}, \eqref{Eqn: numerical estimation, aux2} and \eqref{Eqn: numerical estimation, aux3} we thus arrive at the desired conclusion. \qed

\begin{lem}\label{Lem: degenerate limit, cv}
Suppose that Assumptions \ref{Ass: weak convergence for theta}(i)(ii) and \ref{Ass: bootstrap for theta}(ii) hold, and that $\phi:\mathbb D_\phi\subset\mathbb D\to\mathbb E\equiv\mathbf R$ is Hadamard differentiable at $\theta_0\in\mathbb D_\phi$ tangentially to $\mathbb D_0$ with $\phi_{\theta_0}'$ satisfying Assumption \ref{Ass: HDD 2nd}(iii). Then $\hat c_{1-\alpha}\convp 0$, where for $\alpha\in(0,1)$,
\[
\hat c_{1-\alpha}\equiv \inf\{c\in\mathbf R: P_W(r_n\{\phi(\hat\theta_n^*)-\phi(\hat\theta_n)\}\le c)\ge 1-\alpha\}~.
\]
\end{lem}
\noindent{\sc Proof:} This lemma is somewhat similar to Lemma 5 in \citet{AndrewsandGuggen2010ET} and we include the proof here only for completeness. Fix $\alpha\in(0,1)$ and let $c_{1-\alpha}\equiv \inf\{c\in\mathbf R: P(\phi_{\theta_0}'(\mathbb G)\le c)\ge 1-\alpha\}$. Note that $c_{1-\alpha}=0$ for all $\alpha\in(0,1)$. Since $\phi$ is Hadamard differentiable at $\theta_0\in\mathbb D_\phi$ tangentially to $\mathbb D_0$, it follows by Theorem 3.9.15 in \citet{Vaart1996} that
\begin{align}\label{Eqn: degenerate limit, cv, aux1}
\sup_{f\in\bl_1(\mathbb D)}|E_W[f(r_n\{\phi(\hat\theta_n^*)-\phi(\hat\theta_n)\})]-E[f(\phi_{\theta_0}'(\mathbb G))]|\convp 0~.
\end{align}
This, together with Lemma 10.11 in \citet{Kosorok2008}, give us: for all $t\in\mathbf R\setminus\{0\}$,
\begin{align}\label{Eqn: degenerate limit, cv, aux2}
P_W(r_n\{\phi(\hat\theta_n^*)-\phi(\hat\theta_n)\}\le t)\convp P(\phi_{\theta_0}'(\mathbb G)\le t)~.
\end{align}
Fix $\epsilon>0$. Clearly, $c_{1-\alpha}\pm\epsilon\in\mathbf R\setminus\{0\}$ for all $\epsilon>0$ and all $\alpha\in(0,1)$. Hence, by \eqref{Eqn: degenerate limit, cv, aux2},
\begin{equation}\label{Eqn: degenerate limit, cv, aux3}
\begin{aligned}
P_W(r_n\{\phi(\hat\theta_n^*)-\phi(\hat\theta_n)\}\le c_{1-\alpha}-\epsilon)\convp P(\phi_{\theta_0}'(\mathbb G)\le c_{1-\alpha}-\epsilon)=0<1-\alpha~,\\
P_W(r_n\{\phi(\hat\theta_n^*)-\phi(\hat\theta_n)\}\le c_{1-\alpha}+\epsilon)\convp P(\phi_{\theta_0}'(\mathbb G)\le c_{1-\alpha}+\epsilon)=1>1-\alpha~.
\end{aligned}
\end{equation}
By definition of $\hat c_{1-\alpha}$, it follows from \eqref{Eqn: degenerate limit, cv, aux3} that
\begin{align}\label{Eqn: degenerate limit, cv, aux4}
P(-\epsilon\le \hat c_{1-\alpha}\le \epsilon)=P(c_{1-\alpha}-\epsilon\le \hat c_{1-\alpha}\le c_{1-\alpha}+\epsilon)\to 1~.
\end{align}
Since $\epsilon$ is arbitrary, the conclusion of the lemma then follows from result \eqref{Eqn: degenerate limit, cv, aux4}.\qed

\begin{lem}\label{Lem: derivative estimation, equiv}
Let Assumptions \ref{Ass: HDD 2nd} hold, and $\hat \phi_n'' : \mathbb D \rightarrow \mathbb E$ be an estimator depending on $\{X_i\}_{i=1}^n$. Then the following are equivalent:
\begin{itemize}
\item[(i)] For every compact set $K\subset\mathbb D_0$ and every $\epsilon>0$,
\begin{align}\label{Eqn: derivative estimation, lem aux1}
\lim_{\delta\darrow 0}\limsup_{n\rightarrow \infty} P\Big( \sup_{h \in K^{\delta}} \| \hat \phi_n''(h) - \phi_{\theta_0}''(h)\|_{\mathbb E} > \epsilon\Big) = 0~.
\end{align}
\item[(ii)] For every compact set $K\subset\mathbb D_0$, every $\delta_n\darrow 0$ and every $\epsilon>0$,
\begin{align}\label{Eqn: derivative estimation, lem aux2}
\limsup_{n\rightarrow \infty} P\Big( \sup_{h \in K^{\delta_n}} \| \hat \phi_n''(h) - \phi_{\theta_0}''(h)\|_{\mathbb E} > \epsilon\Big) = 0~.
\end{align}
\item[(iii)] For every sequence $\{h_n\}\subset\mathbb D$ and every $h\in\mathbb D_0$ such that $h_n\to h$ as $n\to\infty$,
\begin{align}\label{Eqn: derivative estimation, lem aux3}
\hat\phi_n''(h_n)\convp \phi_{\theta_0}''(h)~.
\end{align}
\end{itemize}
\end{lem}
\noindent{\sc Proof:} The equivalence between (i) and (ii) is intuitive and straightforward to establish. Suppose that (i) holds. Fix a compact set $K\subset\mathbb D_0$, a sequence $\{\delta_n\}$ with $\delta_n\darrow 0$, and $\epsilon,\eta>0$. We want to show that there exists some $N_0>0$ such that for all $n\ge N_0$,
\begin{align}\label{Eqn: derivative estimation, equiv, aux1}
P\Big( \sup_{h \in K^{\delta_n}} \| \hat \phi_n''(h) - \phi_{\theta_0}''(h)\|_{\mathbb E} > \epsilon\Big)\le\eta~.
\end{align}
But from (i) we know that there is some $\delta_0>0$ such that
\begin{align}\label{Eqn: derivative estimation, equiv, aux2}
\limsup_{n\rightarrow \infty} P\Big( \sup_{h \in K^{\delta_0}} \| \hat \phi_n''(h) - \phi_{\theta_0}''(h)\|_{\mathbb E} > \epsilon\Big)<\eta~,
\end{align}
which in turn implies that there is some $N_1$ satisfying for all $n\ge N_1$
\begin{align}\label{Eqn: derivative estimation, equiv, aux3}
P\Big( \sup_{h \in K^{\delta_0}} \| \hat \phi_n''(h) - \phi_{\theta_0}''(h)\|_{\mathbb E} > \epsilon\Big)<\eta~.
\end{align}
Since $\delta_n\darrow 0$, there exists some $N_2$ such that $\delta_n\le \delta_0$ for all $n\ge N_2$ and hence
\begin{align}\label{Eqn: derivative estimation, equiv, aux4}
P\Big( \sup_{h \in K^{\delta_n}} \| \hat \phi_n''(h) - \phi_{\theta_0}''(h)\|_{\mathbb E} > \epsilon\Big)\le P\Big( \sup_{h \in K^{\delta_0}} \| \hat \phi_n''(h) - \phi_{\theta_0}''(h)\|_{\mathbb E} > \epsilon\Big)~.
\end{align}
Setting $N_0\equiv\max\{N_1,N_2\}$, we see that \eqref{Eqn: derivative estimation, equiv, aux1} follows from \eqref{Eqn: derivative estimation, equiv, aux3} and \eqref{Eqn: derivative estimation, equiv, aux4}.

Conversely, suppose that (ii) holds, fix a compact set $K\subset\mathbb D_0$ and $\epsilon>0$, and we aim to establish (i) or equivalently, there exists some $\delta_0>0$ such that \eqref{Eqn: derivative estimation, equiv, aux3} holds. Pick a sequence $\delta_n\darrow 0$. Then there exists some $N_0$ such that \eqref{Eqn: derivative estimation, equiv, aux1} holds with ``$\le$'' replaced by ``$<$''. Setting $\delta_0\equiv\delta_{N_0}$, we may then conclude \eqref{Eqn: derivative estimation, equiv, aux3} from \eqref{Eqn: derivative estimation, equiv, aux1}.

Now suppose (ii) (and hence (i)) holds again and let $\{h_n\}\subset\mathbb D$ such that $h_n\to h\in\mathbb D_0$. Fix $\delta>0$. There must be some $N_1$ such that $\|h_n-h\|_{\mathbb D}<\delta$ for all $n\ge N_1$. By the triangle inequality we have: for all $n\ge N_1$,
\begin{align}\label{Eqn: derivative estimation, equiv, aux5}
\|\hat\phi_n''(h_n)-\phi_{\theta_0}''(h)\|_{\mathbb E}&\le \|\hat\phi_n''(h_n)-\phi_{\theta_0}''(h_n)\|_{\mathbb E}+\|\phi_{\theta_0}''(h_n)-\phi_{\theta_0}''(h)\|_{\mathbb E}\notag\\
&\le \sup_{h\in K^\delta}\|\hat\phi_n''(h)-\phi_{\theta_0}''(h)\|_{\mathbb E}+\|\phi_{\theta_0}''(h_n)-\phi_{\theta_0}''(h)\|_{\mathbb E}~.
\end{align}
Part (iii) then follows from \eqref{Eqn: derivative estimation, equiv, aux5} and part (i).

Finally, suppose that (iii) holds. Fix a compact set $K\subset\mathbb D_0$ and $\epsilon>0$. Let $\delta_n\darrow 0$. Note that if $\sup_{h\in K^{\delta_n}}\|\hat\phi_n''(h)-\phi_{\theta_0}''(h)\|_{\mathbb E}>\epsilon$, then there must exist some $h_n\in K^{\delta_n}$ such that $\|\hat\phi_n''(h_n)-\phi_{\theta_0}''(h_n)\|_{\mathbb E}>\epsilon$ and this is true for all $n\in\mathbf N$. It follows that
\begin{align}\label{Eqn: derivative estimation, equiv, aux6}
P\Big( \sup_{h \in K^{\delta_n}} \| \hat \phi_n''(h) - \phi_{\theta_0}''(h)\|_{\mathbb E} > \epsilon\Big)\le P(\|\hat\phi_n''(h_n)-\phi_{\theta_0}''(h_n)\|_{\mathbb E}>\epsilon)~.
\end{align}
Note that $h_n\in K^{\delta_n}$ is possibly random and satisfies $d(h_n,K)\equiv\inf_{a\in K}\|h_n-a\|_{\mathbb D}\le \delta_n \to 0$ as $n\to\infty$. Fix an arbitrary subsequence $\{n_k\}$. Since $K$ is compact, it follows by Lemma A.6 in \citet{Fang2014Plugin} that there exists a further subsequence $\{n_{k_j}\}$ and some deterministic $h\in K$ such that $h_{n_{k_j}}\convp h$ as $j\to\infty$. By the triangle inequality,
\begin{align}\label{Eqn: derivative estimation, equiv, aux7}
P(\|\hat\phi_n''(h_n)-\phi_{\theta_0}''(h_n)\|_{\mathbb E}>\epsilon)\le & P(\|\hat\phi_n''(h_n)-\phi_{\theta_0}''(h)\|_{\mathbb E}>\frac{\epsilon}{2})\notag\\
&+P(\|\phi_{\theta_0}''(h_n)-\phi_{\theta_0}''(h)\|_{\mathbb E}>\frac{\epsilon}{2})~.
\end{align}
Since $h_{n_{k_j}}\convp h$ as $j\to\infty$, the first term on the right hand side above tends to zero along $\{n_{k_j}\}$ by (iii) and Lemma \ref{Lem: extended CMT}, while the second term tends to zero along $\{n_{k_j}\}$ by Theorem 1.9.5 in \citet{Vaart1996}. Since $\{n_k\}$ is arbitrary, combination of results \eqref{Eqn: derivative estimation, equiv, aux6} and \eqref{Eqn: derivative estimation, equiv, aux7} then leads to (ii). \qed

\begin{lem}[Extended Continuous Mapping Theorem]\label{Lem: extended CMT}
Let $\mathbb D$ and $\mathbb E$ be metric spaces equipped with metrics $d$ and $\rho$ respectively, $g_n: \mathbb D_n\subset\mathbb D\to\mathbb E$ a possibly random map for each $n\in\mathbf N$, and $g: \mathbb D_0\subset\mathbb D\to\mathbb E$ a nonrandom map. Suppose that $g_n(x_n)\convp g(x)$ whenever $x_n\to x$ for $x_n\in\mathbb D_n$ and $x\in\mathbb D_0$. If $X_n\convp X$ such that $X$ is Borel measurable, separable and satisfies $P(X\in\mathbb D_0)=1$, then $g_n(X_n)\convp g(X)$.
\end{lem}
\noindent{\sc Proof:} We closely follow the proof of Proposition A.8.6 in \citet{BKRW993Efficient} (see also \citet{Vaart_Wellner1990prohorov}). Fix $\epsilon>0$ throughout. First, we show that $g: \mathbb D_0\to\mathbb E$ is continuous.
By assumption, for each $x\in\mathbb D_0$ we have
\begin{align}\label{Eqn: Extended CMT, aux1}
\lim_{\delta\darrow 0}\limsup_{n\to\infty}P(\text{Osc}_{g_n}(B(x,\delta))>\epsilon)=0~,
\end{align}
where $\text{Osc}_{g_n}(B(x,\delta))\equiv\sup_{y,z\in B(x,\delta)}\rho(g_n(y),g_n(z))$ for $B(x,\delta)\equiv \{y\in\mathbb D_n: d(y,x)<\delta\}$. This can be easily seen by the triangle inequality:
\begin{align*}
\lim_{\delta\darrow 0}\limsup_{n\to\infty}P(\text{Osc}_{g_n}(B(x,\delta))>\epsilon)\le &\lim_{\delta\darrow 0}\limsup_{n\to\infty}P(\sup_{y\in B(x,\delta)}\rho(g_n(y),g(x))>\frac{\epsilon}{2})\\
&+\lim_{\delta\darrow 0}\limsup_{n\to\infty}P(\sup_{z\in B(x,\delta)}\rho(g_n(z),g(x))>\frac{\epsilon}{2})\\
&=0~.
\end{align*}
Notice that again by assumption, the triangle inequality and result \eqref{Eqn: Extended CMT, aux1} we have
\begin{align}
\rho(g(y),g(x))&\le \rho(g(y),g_n(y))+\rho(g(x),g_n(x))+\rho(g_n(y),g_n(x))\notag\\
&\le \rho(g(y),g_n(y))+\rho(g(x),g_n(x))+\text{Osc}_{g_n}(B(x,d(x,y)))\notag\\
&\convp 0~,
\end{align}
as $n\to\infty$ followed by $d(x,y)\to\infty$. Since $g$ is a nonrandom function, we must have $\rho(g(y),g(x))\to 0$ as $d(y,x)\to 0$ and hence $g$ is continuous on $\mathbb D_0$.

Next, for $x\in\mathbb D_0$ define
\[
k(x,\epsilon)\equiv\min\{k: \text{for }\forall\,y\text{ with }d(y,x)<\frac{1}{k}\text{ and all }n\ge k, P(\rho(g_n(y),g(x))\le\epsilon)\ge 1-\epsilon\}~.
\]
This is well defined by a simple {\it reductio ad absurdum} argument as in \citet{BKRW993Efficient}. We now show that $k(\cdot, \epsilon): \mathbb D_0\to\mathbf N$ is measurable. This is done by proving that $k(\cdot, \epsilon)$ is lower semicontinuous, i.e., $x_m\to x$ for $\{x,x_m\}\subset\mathbb D_0$ implies
\begin{align}\label{Eqn: Extended CMT, aux2}
\liminf_{m\to\infty} k(x_m,\epsilon)\ge k(x,\epsilon)~.
\end{align}
Fix $x\in\mathbb D_0$ and $\{x_m\}\subset\mathbb D_0$ such that $x_m\to x$ as $m\to\infty$. Then there must exist some subsequence $\{m'\}$ of $\{m\}$ such that $\liminf_{m\to\infty} k(x_m,\epsilon)=\lim_{m'\to\infty} k(x_{m'},\epsilon)$. Since $k(\cdot, \epsilon)$ is integer valued, we further have $\liminf_{m\to\infty} k(x_m,\epsilon)= k(x_{m'},\epsilon)\equiv k'$ for all $m'$ sufficiently large. If $k'=\infty$, then the inequality \eqref{Eqn: Extended CMT, aux2} follows trivially. Otherwise, suppose that $k'<\infty$. For any $y$ with $d(x,y)<1/k'$, there exists an $m_0$ such that $d(x_{m'},y)<1/k'$ for all $m'\ge m_0$. By definition of $k(x,\epsilon)$, it follows that for all $n\ge k'$,
\begin{align}\label{Eqn: Extended CMT, aux3}
P(\rho(g_n(y),g(x_{m'}))\le\epsilon)\ge 1-\epsilon~.
\end{align}
Letting $m'\uarrow\infty$, we have by $x_{m'}\to x$ and continuity of $g$ and $P$ that for all $n\ge k'$,
\begin{align}\label{Eqn: Extended CMT, aux3}
P(\rho(g_n(y),g(x))\le\epsilon)\ge 1-\epsilon~.
\end{align}
Hence, $k(x,\epsilon)\le k'=\liminf_{m\to\infty} k(x_m,\epsilon)$ and hence $k(\cdot,\epsilon)$ is Borel measurable.

Since $P(X\in\mathbb D_0)=1$, we may assume without loss of generality that $X$ takes values in $\mathbb D_0$. In turn, it follows that $k(X,\epsilon)$ is a Borel $\mathbf N$-valued random variable. Thus there exists some $k_0\equiv k_0(\epsilon)$ such that
\begin{align}\label{Eqn: Extended CMT, aux4}
P(k(X,\epsilon)>k_0)<\epsilon~.
\end{align}
Since $X_n\convp X$, there exists some $n_0\equiv n_0(\epsilon)$ such that for all $n\ge n_0(\epsilon)$,
\begin{align}\label{Eqn: Extended CMT, aux5}
P(d(X_n,X)>\frac{1}{k_0})<\epsilon~.
\end{align}
Now define
\[
B_n\equiv \{\rho(g_n(X_n),g(X))>\epsilon\}~,\,C_n\equiv \{d(X_n,X)>\frac{1}{k_0}\}~,\,D\equiv\{k(X,\epsilon)>k_0\}~.
\]
It follows that for all $n\ge\max\{n_0,k_0\}$,
\begin{align*}
P(B_n)&\le P(B_n\cap (C_n^c\cap D^c))+P(B_n\cap (C_n^c\cap D^c)^c)\\
&\le P(B_n\cap (C_n^c\cap D^c))+P(C_n)+P(D)\le 3\epsilon~,
\end{align*}
by definition of $k(x,\epsilon)$, results \eqref{Eqn: Extended CMT, aux4} and \eqref{Eqn: Extended CMT, aux5}, and we are done since $\epsilon$ is arbitrary.\qed

\section{Results for Examples \ref{Ex: mean squared} - \ref{Ex: Overidentification Test}} \label{Sec: Results for Exs}

\subsection*{Example \ref{Ex: moment inequalities}: Moment Inequalities}

In this example, it is a simple exercise to show that
\begin{align}\label{Eqn: derivative, MI}
\phi_{\theta}'(h)=\begin{cases}
2\theta h &\text{ if }\theta>0\\
0 & \text{ if }\theta\le 0
\end{cases}~,\,\phi_\theta''(h)=\begin{cases}
h^2 & \text{ if }\theta>0\\
(\max\{h,0\})^2 &\text{ if }\theta=0\\
0 &\text{ if }\theta<0
\end{cases}~.
\end{align}
Thus, $\phi$ is Hadamard differentiable with the derivative $\phi_\theta'$ degenerate at $\theta\le 0$. Moreover, $\phi$ is second order Hadamard directionally differentiable. The derivative $\phi_{\theta}''$ is nondegenerate at 0, though degenerate whenever $\theta<0$. Exploiting the structure in \eqref{Eqn: derivative, MI}, we may easily estimate the derivative by
\begin{align}
\hat\phi_n''(h)=\begin{cases}
h^2 & \text{ if }\obar X_n>\kappa_n\\
(\max\{h,0\})^2 &\text{ if }|\obar X_n|\le \kappa_n\\
0 & \text{ if }\obar X_n<\kappa_n
\end{cases}~,
\end{align}
where $\kappa_n\darrow 0$ satisfies $\sqrt{n} \kappa_n\uarrow\infty$, and $\obar X_n\equiv\frac{1}{n}\sum_{i=1}^n X_i$. Interestingly, construction of $\hat\phi_n''$ as above amounts to the generalized moment selection procedure as in \citet{AndrewsandSoares2010} for conducting inference in moment inequalities models.

\subsection*{Example \ref{Ex: Cramer-von Mises}: Cramer-von Mises Functionals}

Cramer-von Mises functionals can be viewed as generalized Wald functionals. It is straightforward to show that $\phi$ is first and second Hadamard differentiable at any $\theta\in\ell^\infty(\mathbf R^{d_x})$ with derivatives satisfying:
\[
\phi_\theta'(h)=2\int(\theta-F_0)h\,dF_0~,\, \phi_\theta''(h)=\int h^2\,dF_0~,
\]
for all $h\in\ell^\infty(\mathbf R^{d_x})$. Note that first order derivative $\phi_\theta'$ is degenerate when $\theta=F_0$, while second order derivative $\phi_\theta''$ is nowhere degenerate. The corresponding bilinear map $\Phi_\theta'': \ell^\infty(\mathbf R^{d_x})\times \ell^\infty(\mathbf R^{d_x})\to\mathbf  R$ is given by $\Phi_\theta''(h,g)=\int h g\, dF_0$. In this example, there is no need for derivative estimation because $\phi_{\theta_0}''$ is a known map.

\subsection*{Example \ref{Ex: stochastic dominance}: Stochastic Dominance}

\begin{lem}\label{Lem: stochastic dominance, HDD}
Let $w:\mathbf R \rightarrow \mathbf R^+$ satisfy $\int_{\mathbf R} w(u)du < \infty$ and $\phi:\ell^\infty(\mathbf R)\times \ell^\infty(\mathbf R) \rightarrow \mathbf R$ be given by $\phi(\theta) = \int_{\mathbf R}\max\{\theta^{(1)}(u) - \theta^{(2)}(u),0\}^2w(u)du$ for any $\theta = (\theta^{(1)},\theta^{(2)}) \in \ell^\infty(\mathbf R)\times \ell^\infty(\mathbf R)$. Then it follows that
\begin{enumerate}
\item[(i)]$\phi$ is first order Hadamard differentiable at any $\theta \in \ell^\infty(\mathbf R)\times \ell^\infty(\mathbf R)$ with $\phi_\theta^\prime:\ell^\infty(\mathbf R)\times \ell^\infty(\mathbf R) \rightarrow \mathbf R$ satisfying for any $h = (h^{(1)},h^{(2)})\in \ell^\infty(\mathbf R)\times \ell^\infty(\mathbf R)$
\[
\phi^\prime_\theta(h) = 2\int_{B_+(\theta)}[\theta^{(1)}(u)-\theta^{(2)}(u)][h^{(1)}(u)-h^{(2)}(u)]w(u)du~,
\]
where $B_+(\theta) \equiv \{u \in \mathbf R: \theta^{(1)}(u) > \theta^{(2)}(u)\}$.
\item[(ii)] $\phi$ is second order Hadamard directionally differentiable at any $\theta \in \ell^\infty(\mathbf R)\times \ell^\infty(\mathbf R)$ and the derivative $\phi_\theta'':\ell^\infty(\mathbf R)\times \ell^\infty(\mathbf R) \rightarrow \mathbf R$ is given by: for any $h = (h^{(1)},h^{(2)})\in \ell^\infty(\mathbf R)\times \ell^\infty(\mathbf R)$
\[
\phi''_\theta(h) = \int_{B_0(\theta)} \max\{h^{(1)}(u)-h^{(2)}(u),0\}^2w(u)du+\int_{B_+(\theta)} [h^{(1)}(u)-h^{(2)}(u)]^2w(u)du~,
\]
where $B_0(\theta) \equiv \{u \in \mathbf R: \theta^{(1)}(u) = \theta^{(2)}(u)\}$.
\end{enumerate}
\end{lem}
\noindent{\sc Proof:} Fix $\theta\in\ell^\infty(\mathbf R)\times \ell^\infty(\mathbf R)$. Further, let $t_n\darrow 0$, $\{h_n\} = \{(h_{n}^{(1)},h_{n}^{(2)})\}$ be a sequence in $\ell^\infty(\mathbf R) \times \ell^\infty(\mathbf R)$ satisfying $\|h_{n}^{(1)} - h^{(1)}\|_\infty  \vee \|h_{n}^{(2)} - h^{(2)}\|_\infty = o(1)$ for some $h = (h^{(1)},h^{(2)}) \in\ell^\infty(\mathbf R)\times \ell^\infty(\mathbf R)$, and
\begin{equation*}
B_-(\theta) \equiv \{u \in \mathbf R : \theta^{(1)}(u) < \theta^{(2)}(u)\} ~.
\end{equation*}
Observe that since $\theta^{(1)}(u) - \theta^{(2)}(u) < 0$ for all $u \in B_-(\theta)$, and $\|h_{n}^{(1)} - h_{n}^{(2)}\|_\infty = O(1)$ due to $\|h^{(1)}- h^{(2)}\|_\infty < \infty$, the dominated convergence theorem yields that:
\begin{gather}\label{Eqn: stochastic dominance, aux1}
\lim_{n\to\infty}\frac{1}{t_n}\int_{B_-(\theta)}\max\{(\theta^{(1)}(u) - \theta^{(2)}(u)) + t_n(h_{n}^{(1)}(u) - h_{n}^{(2)}(u)),0\}^2w(u)du=0~,\\
\lim_{n\to\infty}\frac{1}{t_n}\int_{B_0(\theta)}\max\{(\theta^{(1)}(u) - \theta^{(2)}(u)) + t_n(h_{n}^{(1)}(u) - h_{n}^{(2)}(u)),0\}^2w(u)du=0~,
\end{gather}
and
\begin{align}\label{Eqn: stochastic dominance, aux3}
\lim_{n\to\infty}&\frac{1}{t_n}\big[\int_{B_+(\theta)}\max\{(\theta^{(1)}(u) - \theta^{(2)}(u)) + t_n(h_{n}^{(1)}(u) - h_{n}^{(2)}(u)),0\}^2w(u)du\notag\\
&\hspace{2.5in}-\int_{B_+(\theta)}(\theta^{(1)}(u) - \theta^{(2)}(u))^2w(u)du\big]\notag\\
&=\lim_{n\to\infty}\int_{B_+(\theta)}\frac{1}{t_n}\big[\max\{(\theta^{(1)}(u) - \theta^{(2)}(u)) + t_n(h_{n}^{(1)}(u) - h_{n}^{(2)}(u)),0\}^2\notag\\
&\hspace{2.85in}-(\theta^{(1)}(u) - \theta^{(2)}(u))^2\big]w(u)du\notag\\
&=2\int_{B_+(\theta)}[\theta^{(1)}(u)-\theta^{(2)}(u)][h^{(1)}(u)-h^{(2)}(u)]w(u)du~.
\end{align}
Combining results \eqref{Eqn: stochastic dominance, aux1} - \eqref{Eqn: stochastic dominance, aux3} yields
\begin{align*}
\phi_\theta'(h)\equiv\lim_{n\to\infty}\frac{\phi(\theta + t_nh_n) - \phi(\theta)}{t_n}= 2\int_{B_+(\theta)}[\theta^{(1)}(u)-\theta^{(2)}(u)][h^{(1)}(u)-h^{(2)}(u)]w(u)du~,
\end{align*}
which establishes the first claim of the lemma.

Next fix $\theta\in\ell^\infty(\mathbf R)\times \ell^\infty(\mathbf R)$ and let $\{h_n\}$ and $\{t_n\}$ be as before. Therefore, by the dominated convergence theorem we have
\begin{align}\label{Eqn: stochastic dominance, aux4}
\lim_{n\to\infty}&\int_{B_-(\theta)}\frac{1}{t_n^2}\max\{(\theta^{(1)}(u) - \theta^{(2)}(u)) + t_n(h_{n}^{(1)}(u) - h_{n}^{(2)}(u)),0\}^2w(u)du=0~,\\
\lim_{n\to\infty}&\int_{B_0(\theta)}\frac{1}{t_n^2}\max\{(\theta^{(1)}(u) - \theta^{(2)}(u)) + t_n(h_{n}^{(1)}(u) - h_{n}^{(2)}(u)),0\}^2w(u)du\notag\\
&=\int_{B_0(\theta)} \max\{h^{(1)}(u)-h^{(2)}(u),0\}^2w(u)du~,
\end{align}
and
\begin{align}\label{Eqn: stochastic dominance, aux5}
&\lim_{n\to\infty}\frac{1}{t_n^2}\Big[\int_{B_+(\theta)}\max\{(\theta^{(1)}(u) - \theta^{(2)}(u)) + t_n(h_{n}^{(1)}(u) - h_{n}^{(2)}(u)),0\}^2w(u)du\notag\\
&-\int_{B_+(\theta)}(\theta^{(1)}(u) - \theta^{(2)}(u))^2w(u)du-t_n 2 \int_{B_+(\theta)}[\theta^{(1)}(u)-\theta^{(2)}(u)][h_n^{(1)}(u)-h_n^{(2)}(u)]w(u)du\Big]\notag\\
&\to \int_{B_+(\theta)}[h^{(1)}(u)-h^{(2)}(u)]^2w(u)du~.
\end{align}
It follows from results \eqref{Eqn: stochastic dominance, aux4}-\eqref{Eqn: stochastic dominance, aux5} that
\begin{align*}
\phi_\theta''(h)&\equiv\lim_{n\to\infty}\frac{\phi(\theta + t_nh_n) -\phi(\theta)- t_n\phi_\theta'(h_n)}{t_n^2}\\
&=\int_{B_0(\theta)} \max\{h^{(1)}(u)-h^{(2)}(u),0\}^2w(u)du+\int_{B_+(\theta)}[h^{(1)}(u)-h^{(2)}(u)]^2w(u)du~.
\end{align*}
This competes the proof of the second claim and we are done.\qed

Note that if $B_+(\theta)$ has Lebesgue measure zero, i.e., $\theta^{(1)}\le\theta^{(2)}$ almost everywhere, then $\phi^\prime_\theta(h)=0$ and $\phi''_\theta(h)$ simplifies to $\phi''_\theta(h) = \int_{B_0(\theta)} \max\{h^{(1)}(u)-h^{(2)}(u)\}^2w(u)du$. If in addition the contact set $B_0(\theta)$ has Lebesgue measure zero, then $\phi''_\theta$ in turn is degenerate, corresponding to the degenerate limits obtained in Theorem 1 of \citet{Linton2010}. Let $\hat B_0(\theta_0)$ be an estimator of $B_0(\theta_0)$. Then we may estimate $\phi_{\theta_0}''$ by
\begin{align}
\hat\phi_n''(h)= \int_{\hat B_0(\theta)} \max\{h^{(1)}(u)-h^{(2)}(u),0\}^2w(u)du~.
\end{align}
It is a simple exercise to verify that Assumption \ref{Ass: derivative estimation} is satisfied provided
\begin{align}
\int_{\mathbf R} 1\{u\in \hat B_0(\theta_0)\triangle B_0(\theta_0)\}w(u)du\convp 0~,
\end{align}
where $A\triangle B$ denotes the set difference between sets $A$ and $B$. Such a construction corresponds to the bootstrap procedure studied in \citet{Linton2010}.

\subsection*{Example \ref{Ex: conditional moment inequalities, KS}: Conditional Moment Inequalities}

\begin{lem}\label{Lem: KS, HDD 2nd}
Let $\mathcal F$ be compact under some metric $d$ and $\phi: \ell^\infty(\mathcal F)\times \ell^\infty(\mathcal F)\to\mathbf R$ be given by $\phi(\theta)=\sup_{f\in\mathcal F}\{[\max(\theta^{(1)}(f),0)]^2+[\theta^{(2)}(f)]^2\}$. Then it follows that:
\begin{itemize}
\item[(i)] $\phi$ is Hadamard differentiable at any $\theta\in\ell^\infty(\mathcal F)\times \ell^\infty(\mathcal F)$ satisfying $\theta^{(1)}\le 0$ and $\theta^{(2)}=0$, and its derivative $\phi_{\theta}'(h)=0$ for any $h\in\ell^\infty(\mathcal F)\times \ell^\infty(\mathcal F)$
\item[(ii)] $\phi$ is second order Hadamard directionally differentiable at any $\theta\in C(\mathcal F)\times C(\mathcal F)$ satisfying $\theta^{(1)}\le 0$ and $\theta^{(2)}=0$ tangentially to $C(\mathcal F)\times C(\mathcal F)$, and the derivative is given by: for any $h\in C(\mathcal F)\times C(\mathcal F)$,
     \[
     \phi_{\theta}''(h)=\max\{\sup_{f\in\mathcal F_0}\{\max(h^{(1)}(f),0)^2+[h^{(2)}(f)]^2\},\sup_{f\in\mathcal F\setminus\mathcal F^0}[h^{(2)}(f)]^2\}~,
     \]
     where $\mathcal F_0\equiv\{f\in\mathcal F: \theta^{(1)}(f)=0\}$, and $\sup \emptyset\equiv0$.
\end{itemize}
\end{lem}
\begin{rem}
Note that if $\mathcal F_0=\emptyset$, then $\phi_{\theta}''$ simplifies to $\phi_{\theta}''(h)=\sup_{f\in\mathcal F}[h^{(2)}(f)]^2$.\qed
\end{rem}
\noindent{\sc Proof:} Let $\theta\in\ell^\infty(\mathcal F)\times \ell^\infty(\mathcal F)$ satisfying $\theta^{(1)}\le 0$ and $\theta^{(2)}=0$, $\{h_n\}\subset \ell^\infty(\mathcal F)\times \ell^\infty(\mathcal F)$ such that $h_n\to h\in \ell^\infty(\mathcal F)\times \ell^\infty(\mathcal F)$, and $t_n\darrow 0$. Combining $\theta^{(1)}\le 0$, $\theta^{(2)}=0$ so that $\phi(\theta)=0$ and the triangle inequality, we have
\begin{align}
|\phi(\theta+t_nh_n)-\phi(\theta)|&=\big|\sup_{f\in\mathcal F}\{[\max(\theta^{(1)}(f)+t_nh_n^{(1)}(f),0)]^2+[\theta^{(2)}(f)+t_nh_n^{(2)}(f)]^2\}\big|\notag\\
&\le \sup_{f\in\mathcal F}[\max(\theta^{(1)}(f)+t_nh_n^{(1)}(f),0)]^2+t_n^2\sup_{f\in\mathcal F}[h_n^{(2)}(f)]^2\notag\\
&\le \sup_{f\in\mathcal F}[\max(t_nh_n^{(1)}(f),0)]^2+t_n^2\sup_{f\in\mathcal F}[h_n^{(2)}(f)]^2=o(t_n)~,
\end{align}
as desired in part (i), where in the last step we used the fact that $h_n^{(1)}=h_n^{(2)}=O(1)$.

As for the second claim, let $\theta\in C(\mathcal F)\times C(\mathcal F)$ satisfying $\theta^{(1)}\le 0$ and $\theta^{(2)}=0$, $\{h_n\}\subset \ell^\infty(\mathcal F)\times \ell^\infty(\mathcal F)$ such that $h_n\to h\in C(\mathcal F)\times C(\mathcal F)$, and $t_n\darrow 0$. By $\theta^{(1)}\le 0$ and $\theta^{(2)}=0$, Lipschtiz continuity of the sup operator and the triangle inequality we have
\begin{align}\label{Eqn: KS, HDD 2nd, aux1}
|\phi(\theta+t_nh_n)-&\phi(\theta+t_nh)|\notag\\
&=|\sup_{f\in\mathcal F}\{\max(\theta^{(1)}(f)+t_nh_n^{(1)}(f),0)^2+[t_nh_n^{(2)}(f)]^2\}\notag\\
&\qquad-\sup_{f\in\mathcal F}\{\max(\theta^{(1)}(f)+t_nh^{(1)}(f),0)^2+[t_nh^{(2)}(f)]^2\}|\notag\\
&\le \sup_{f\in\mathcal F}|\max(\theta^{(1)}(f)+t_nh_n^{(1)}(f),0)^2-\max(\theta^{(1)}(f)+t_nh^{(1)}(f),0)^2|\notag\\
&\qquad+  \sup_{f\in\mathcal F}|[t_nh_n^{(2)}(f)]^2-[t_nh^{(2)}(f)]^2|~.
\end{align}
Since $\|h_n-h\|_\infty=o(1)$ and $\theta^{(1)}\le 0$, it follows that
\begin{align}\label{Eqn: KS, HDD 2nd, aux2}
\sup_{f\in\mathcal F}&|\max(\theta^{(1)}(f)+t_nh_n^{(1)}(f),0)^2-\max(\theta^{(1)}(f)+t_nh^{(1)}(f),0)^2|\notag\\
&\le \sup_{f\in\mathcal F}|\max(\theta^{(1)}(f)+t_nh_n^{(1)}(f),0)-\max(\theta^{(1)}(f)+t_nh^{(1)}(f),0)|\notag\\
&\qquad\times \sup_{f\in\mathcal F}|\max(\theta^{(1)}(f)+t_nh_n^{(1)}(f),0)+\max(\theta^{(1)}(f)+t_nh^{(1)}(f),0)|\notag\\
&\le \sup_{f\in\mathcal F}|t_nh_n^{(1)}(f)-t_nh^{(1)}(f)|\sup_{f\in\mathcal F}\{\max(t_nh_n^{(1)}(f),0)+\max(t_nh^{(1)}(f),0)\}\notag\\
&=o(t_n)O(t_n)=o(t_n^2)~,
\end{align}
and that
\begin{align}\label{Eqn: KS, HDD 2nd, aux3}
\sup_{f\in\mathcal F}|[t_nh_n^{(2)}(f)]^2-[t_nh^{(2)}(f)]^2|=o(t_n^2)~.
\end{align}
Combination of results \eqref{Eqn: KS, HDD 2nd, aux1}, \eqref{Eqn: KS, HDD 2nd, aux2} and \eqref{Eqn: KS, HDD 2nd, aux3} leads to
\begin{align}\label{Eqn: KS, HDD 2nd, aux4}
|\phi(\theta+t_nh_n)-\phi(\theta+t_nh)|=o(t_n^2)~.
\end{align}
Next, fix $\delta>0$. By definition of $\mathcal F_0^\delta$, compactness of $\mathcal F$ and continuity of $\theta^{(1)}$, we see that $\sup_{f\in\mathcal F\setminus\mathcal F_0^\delta}\theta^{(1)}(f)< 0$. Since also $t_nh^{(1)}=o(1)$ and $h^{(1)}\in C(\mathcal F)$, it follows that $\theta^{(1)}(f)+t_nh^{(1)}(f)<0$ for all $f\in f\in\mathcal F\setminus \mathcal F_0^\delta$ and for all $n$ large. In turn we have
\begin{align}\label{Eqn: KS, HDD 2nd, aux5}
\lim_{\delta\darrow 0}\lim_{n\to\infty}t_n^{-2}\sup_{f\in\mathcal F\setminus \mathcal F_0^\delta}\{\max(\theta^{(1)}(f)+&t_nh^{(1)}(f),0)^2+[t_nh^{(2)}(f)]^2\}\notag\\
&=\lim_{\delta\darrow 0}\sup_{f\in\mathcal F\setminus \mathcal F_0^\delta}[h^{(2)}(f)]^2=\sup_{f\in\mathcal F\setminus \mathcal F_0}[h^{(2)}(f)]^2~,
\end{align}
where the last step is due to $h^{(2)}\in C(\mathcal F)$. On the other hand, we have,
\begin{align}\label{Eqn: KS, HDD 2nd, aux6}
&\lim_{\delta\darrow 0}\limsup_{n\to\infty}t_n^{-2}\Big|\sup_{f\in\mathcal F_0^\delta}\{\max(\theta^{(1)}(f)+t_nh^{(1)}(f),0)^2+[t_nh^{(2)}(f)]^2\}\notag\\
&\hspace{1.5in}-t_n^2\sup_{f\in\mathcal F_0}\{\max(h^{(1)}(f),0)^2+[h^{(2)}(f)]^2\}\Big|\notag\\
&\le \lim_{\delta\darrow 0}\limsup_{n\to\infty}t_n^{-2}\sup_{f\in\mathcal F_0^\delta}\{\max(t_nh^{(1)}(f),0)^2+[t_nh^{(2)}(f)]^2\}\notag\\
&\hspace{1.5in}-\sup_{f\in\mathcal F_0}\{\max(t_nh^{(1)}(f),0)^2+[t_nh^{(2)}(f)]^2\}\notag\\
&\le \lim_{\delta\darrow 0}\limsup_{n\to\infty}t_n^{-2}\sup_{f,g\in\mathcal{F}: d(f,g)\le\delta}|\max(t_nh^{(1)}(f),0)^2-\max(t_nh^{(1)}(g),0)^2|\notag\\
&\le\lim_{\delta\darrow 0}\sup_{f,g\in\mathcal{F}: d(f,g)\le\delta}|\max(h^{(1)}(f),0)^2-\max(h^{(1)}(g),0)^2|=0~,
\end{align}
where the first inequality is due to $\theta(f)=0$ for all $f\in\mathcal F_0$ and $\theta^{(1)}\le 0$, the second inequality exploits the definition and compactness of $\mathcal F_0^\delta$, and the equality is due to uniform continuity of $h^{(1)}$ on $\mathcal F$ since $h^{(1)}\in C(\mathcal F)$ and $\mathcal F$ is compact.

Finally, combining results \eqref{Eqn: KS, HDD 2nd, aux5}, \eqref{Eqn: KS, HDD 2nd, aux6}, and $\phi(\theta)=0$ we have:
\begin{align}\label{Eqn: KS, HDD 2nd, aux7}
\limsup_{n\to\infty}t_n^{-2}&\{\phi(\theta+t_nh)-\phi(\theta)\}=\limsup_{n\to\infty}t_n^{-2}\phi(\theta+t_nh)\notag\\
&=\limsup_{n\to\infty}t_n^{-2}\sup_{f\in\mathcal F}\{\max(\theta^{(1)}(f)+t_nh^{(1)}(f),0)^2+[t_nh^{(2)}(f)]^2\}\notag\\
&=\lim_{\delta\darrow 0}\limsup_{n\to\infty}t_n^{-2}\max\Big\{\sup_{f\in\mathcal F_0^\delta}\{\max(\theta^{(1)}(f)+t_nh^{(1)}(f),0)^2+[t_nh^{(2)}(f)]^2\},\notag\\
&\hspace{1in} \sup_{f\in\mathcal F\setminus\mathcal F_0^\delta}\{\max(\theta^{(1)}(f)+t_nh^{(1)}(f),0)^2+[t_nh^{(2)}(f)]^2\}\Big\}\notag\\
&=\max\Big\{\sup_{f\in\mathcal F_0}\{\max(h^{(1)}(f),0)^2+[h^{(2)}(f)]^2\},\sup_{f\in\mathcal F\setminus \mathcal F_0}[h^{(2)}(f)]^2\Big\}~.
\end{align}\label{Eqn: KS, HDD 2nd, aux8}
It follows from $\phi_\theta'=0$, \eqref{Eqn: KS, HDD 2nd, aux4} and \eqref{Eqn: KS, HDD 2nd, aux7} that
\begin{align}
\lim_{n\to\infty}&\frac{\phi(\theta+t_nh_n)-\phi(\theta)-t_n\phi_\theta'(h_n)}{t_n^2}\notag\\
&\hspace{0.5in}= \max\Big\{\sup_{f\in\mathcal F\setminus \mathcal F_0}[h^{(2)}(f)]^2, \sup_{f\in\mathcal F_0}\{\max(h^{(1)}(f),0)^2+[h^{(2)}(f)]^2\}\Big\}~,
\end{align}
as desired for the second claim of the lemma. \qed

Suppose that $\hat{\mathcal F}_0$ and $\hat{\mathcal F}_{0,c}$ are respectively estimators of $\mathcal F_0\equiv\{f\in\mathcal F: \theta_0^{(1)}(f)=0\}$ and $\mathcal F\setminus \mathcal F_0$ that satisfy\footnote{We note that for two generic sets $A$ and $B$ in a metric space, neither $d_H(A,B)$ controls $d_H(A^c,B^c)$ nor $d_H(A^c,B^c)$ controls $d_H(A,B)$ \citep{Lemenant_Milakis_Spinolo2014}.}
\begin{align}
d_H(\hat{\mathcal F}_0,\mathcal F_0; L^2(W))=o_p(1)\text{ and }d_H(\hat{\mathcal F}_{0,c},\mathcal F\setminus\mathcal F_0; L^2(W))=o_P(1)~.
\end{align}
Based on $\hat{\mathcal F}_0$ and $\hat{\mathcal F}_{0,c}$ and in view of Lemma B.3 in \citet{Fang_Santos2014HDD}, we may estimate the derivative as follows:
\begin{align}
\hat\phi_n''(h)=\max\{\sup_{f\in\hat{\mathcal F}_0}\{\max(h^{(1)}(f),0)^2+[h^{(2)}(f)]^2\},\sup_{f\in\hat{\mathcal F}_{0,c}}[h^{(2)}(f)]^2\}~.
\end{align}
The estimation of $\mathcal F_0$ and $\mathcal F\setminus \mathcal F_0$ is in accordance with the generalized moment selection in \citet{Andrews_Shi2013CMI}; see also \citet{Kaido_Santos2013}.

\subsection*{Example \ref{Ex: Overidentification Test}: Overidentification Test}

\begin{lem}\label{Lem: over ID, HDD}
Let $\Gamma\subset\mathbf R^k$ be a compact set, and $\phi: \prod_{j=1}^m \ell^\infty(\Gamma)\to\mathbf R$ be given by $\phi(\theta)=\inf_{\gamma\in\Gamma}\theta(\gamma)^\intercal W\theta(\gamma)$ where $\theta \in \prod_{j=1}^m \ell^\infty(\Gamma)$ and $W$ is a $m\times m$ symmetric positive definite matrix. Then we have
\begin{itemize}
\item[(i)] $\phi$ is Hadamard differentiable at any $\theta\in \prod_{j=1}^m \ell^\infty(\Gamma)$ satisfying $\theta(\gamma)=0$ for some $\gamma\in\Gamma$ with the derivative given by $\phi_{\theta}'(h)=0$ for all $h\in \prod_{j=1}^m \ell^\infty(\Gamma)$.
    \item[(ii)] If $\Gamma_0(\theta)\equiv\{\gamma_0\in\Gamma: \theta(\gamma_0)=0\}$ is in the interior of $\Gamma$, $\theta\in \prod_{j=1}^m C^1(\Gamma)$ satisfies $\phi(\theta)=0$, and for all small $\epsilon>0$, $\inf_{\gamma\in\Gamma\setminus\Gamma_0(\theta)^\epsilon}\|\theta(\gamma)\|\ge C\epsilon^\kappa$ for some $\kappa\in(0,1]$ and some $C>0$, then $\phi$ is second order Hadamard directionally differentiable at $\theta$ tangentially to $\prod_{j=1}^m C(\Gamma)$ with the derivative given by: for any $h\in \prod_{j=1}^m C(\Gamma)$
        \[
        \phi_{\theta}''(h)=\min_{\gamma_0\in\Gamma_0(\theta)}\min_{v\in \mathbf R^k}\{h(\gamma_0)-J(\gamma_0)v\}^\intercal W\{h(\gamma_0)-J(\gamma_0)v\}~,
        \]
        where $J: \Gamma_0(\theta)\to\mathbf M^{m\times k}$ is the Jacobian matrix defined by $J(\gamma_0)\equiv\frac{d\theta(\gamma)}{d\gamma^\intercal}\big|_{\gamma=\gamma_0}$.
\end{itemize}
\end{lem}
\noindent{\sc Proof:} Fix $\theta\in \prod_{j=1}^m \ell^\infty(\Gamma)$ and let $t_n\darrow 0$ and $\{h_n,h\}\subset \prod_{j=1}^m \ell^\infty(\Gamma)$ such that $h_n\to h$. For a vector $a\in\mathbf R^m$, define the norm $\|a\|_W=\sqrt{a^\intercal Wa}$. It follows that
\begin{align}
|\phi(\theta&+t_nh_n)-\phi(\theta)|=\inf_{\gamma\in\Gamma}\|\theta(\gamma)+t_nh_n(\gamma)\|_W^2\notag\\
&\le\inf_{\gamma_0\in\Gamma_0(\theta)}\|\theta(\gamma_0)+t_nh_n(\gamma_0)\|_W^2\le t_n^2\inf_{\gamma_0\in\Gamma_0(\theta)}\|h_n(\gamma_0)\|_W^2=o(t_n)~,
\end{align}
where the second inequality is because $\theta(\gamma_0)=0$ for all $\gamma_0\in\Gamma_0(\theta)$ and the last step is due to $h_n=O(1)$ by assumption. This establishes part (i).

For part (ii), fix $\theta\in \prod_{j=1}^m C^1(\Gamma)$ with $\phi(\theta)=0$ and let $t_n\darrow 0$ and $\{h_n\}\subset \prod_{j=1}^m \ell^\infty(\Gamma)$ such that $h_n\to h\in \prod_{j=1}^m C(\Gamma)$. First of all, note that for $\gamma_0\in\Gamma_0(\theta)$,
\begin{align}\label{Eqn: over ID, aux1}
|\phi(\theta+t_nh_n)-\phi(\theta+t_nh)|&=\big|\inf_{\gamma\in\Gamma}\|\theta(\gamma)+t_nh_n(\gamma)\|_W^2-\inf_{\gamma\in\Gamma}\|\theta(\gamma)+t_nh(\gamma)\|_W^2\big|\notag\\
&=\big|\inf_{\gamma\in\Gamma}\|\theta(\gamma)+t_nh_n(\gamma)\|_W-\inf_{\gamma\in\Gamma}\|\theta(\gamma)+t_nh(\gamma)\|_W\big|\notag\\
&\qquad\times\big|\inf_{\gamma\in\Gamma}\|\theta(\gamma)+t_nh_n(\gamma)\|_W+\inf_{\gamma\in\Gamma}\|\theta(\gamma)+t_nh(\gamma)\|_W\big|\notag\\
&\le t_n\|h_n-h\|_\infty \{\|\theta(\gamma_0)+t_nh_n(\gamma_0)\|_W+\|\theta(\gamma_0)+t_nh(\gamma_0)\|_W\}\notag\\
&\le t_n^2\|h_n-h\|_\infty \{\|h_n(\gamma_0)\|_W+\|h(\gamma_0)\|_W\}=o(t_n^2)~,
\end{align}
where the first inequality is by Lipschitz continuity of the $\inf$ operator and the triangle inequality, and the last inequality follows from $h_n\to h$ and $\theta(\gamma_0)=0$ for $\gamma_0\in\Gamma_0(\theta)$.

Next, for each fixed $a\ge (3\lambda_0^{-1/2}C^{-1}\max_{\gamma\in\Gamma}\|h(\gamma)\|_W)^{1/\kappa}$ with $h\neq 0$ and $\lambda_0>0$ the smallest eigenvalue of $W$, by assumption and the triangle inequality we have: for all $n$ sufficiently large so that $t_n^\kappa\ge t_n$,
\begin{align}
\inf_{\gamma\in\Gamma\setminus\Gamma_0(\theta)^{at_n}}&\|\theta(\gamma)+t_nh(\gamma)\|_W\ge \inf_{\gamma\in\Gamma\setminus\Gamma_0(\theta)^{at_n}}\|\theta(\gamma)\|_W-t_n\sup_{\gamma\in\Gamma\setminus\Gamma_0(\theta)^{at_n}}\|h(\gamma)\|_W\notag\\
&\ge \lambda_0^{1/2} \inf_{\gamma\in\Gamma\setminus\Gamma_0(\theta)^{at_n}}\|\theta(\gamma)\|-t_n\sup_{\gamma\in\Gamma\setminus\Gamma_0(\theta)^{at_n}}\|h(\gamma)\|_W\notag\\
&\ge \lambda_0^{1/2} C(at_n)^\kappa-t_n\max_{\gamma\in\Gamma}\|h(\gamma)\|_W\ge 3t_n^\kappa\max_{\gamma\in\Gamma}\|h(\gamma)\|_W-t_n\max_{\gamma\in\Gamma}\|h(\gamma)\|_W\notag\\
&\ge 3t_n\max_{\gamma\in\Gamma}\|h(\gamma)\|_W-t_n\max_{\gamma\in\Gamma}\|h(\gamma)\|_W> t_n\min_{\gamma_0\in\Gamma_0(\theta)}\|h(\gamma_0)\|_W\notag\\
&=\min_{\gamma_0\in\Gamma_0(\theta)}\|\theta(\gamma_0)+t_nh(\gamma_0)\|_W\ge\sqrt{\phi(\theta+t_nh)}~,
\end{align}
where the strict inequality is due to $h\neq 0$. This in turn implies that for all $n$ large,
\begin{align}\label{Eqn: over ID, aux2}
\phi(\theta+t_nh)=\min_{\gamma\in\Gamma_0(\theta)^{at_n}}\|\theta(\gamma)+t_nh(\gamma)\|_W^2~.
\end{align}

Now for $\gamma_0\in\Gamma_0(\theta)$, set $V_{n,\gamma_0}(a)\equiv\{v\in\mathbf R^k: \gamma_0+t_nv\in\Gamma,\|v\|\le a\}$ and $V(a)\equiv\{v\in\mathbf R^k: \|v\|\le a\}$. Note that $\bigcup_{\gamma_0\in\Gamma_0(\theta)}V_{n,\gamma_0}(a)=\Gamma_0(\theta)^{at_n}$. Since $\theta$ and $h$ are continuous, it then follows that
\begin{align}\label{Eqn: over ID, aux3}
\phi(\theta+t_nh)=\min_{\gamma_0\in\Gamma_0(\theta)}\min_{v\in V_{n,\gamma_0}(a)}\|\theta(\gamma_0+t_nv)+t_nh(\gamma_0+t_nv)\|_W^2~.
\end{align}
In turn, notice that
\begin{align}\label{Eqn: over ID, aux4}
\big|\phi(\theta+t_nh)-&\min_{\gamma_0\in\Gamma_0(\theta)}\min_{v\in V_{n,\gamma_0}(a)}\|\theta(\gamma_0+t_nv)+t_nh(\gamma_0)\|_W^2\big|\notag\\
&\le 2t_n\|h(\gamma_0)\|_W\cdot t_n\max_{\gamma_0\in\Gamma_0(\theta)}\max_{v\in V_{n,\gamma_0}(a)}\|h(\gamma_0+t_nv)-h(\gamma_0)\|_W\notag\\
&\le 2t_n^2\max_{\gamma_1,\gamma_2\in\Gamma: \|\gamma_1-\gamma_2\|\le at_n}\|h(\gamma_1)-h(\gamma_2)\|_W=o(t_n^2)~,
\end{align}
where the first inequality follows from the formula $|b^2-c^2|\le |b+c||b-c|$ and that $\gamma_0$ is any fixed element in $\Gamma_0(\theta)$, and the last step follows from uniform continuity of $h$ on $\Gamma$ because $h$ is continuous on $\Gamma$ and $\Gamma$ is compact.

Since $\theta\in\prod_{j=1}^m C^1(\Gamma)$, we further have,
\begin{align}\label{Eqn: over ID, aux5}
&\big|\min_{\gamma_0\in\Gamma_0(\theta)}\min_{v\in V_{n,\gamma_0}(a)}\|\theta(\gamma_0+t_nv)+t_nh(\gamma_0)\|_W^2&\notag\\
&\hspace{1in}-\min_{\gamma_0\in\Gamma_0(\theta)}\min_{v\in V_{n,\gamma_0}(a)}\|\theta(\gamma_0)+J(\gamma_0)t_nv+t_nh(\gamma_0)\|_W^2\big|\notag\\
&\le 2t_n\max_{\gamma_0\in\Gamma_0(\theta)}\|h(\gamma_0)\|_W\cdot \max_{\gamma_0\in\Gamma_0(\theta)}\max_{v\in V_{n,\gamma_0}(a)}\|\theta(\gamma_0+t_nv)-\theta(\gamma_0)-J(\gamma_0)t_nv\|_W~.
\end{align}
By the mean value theorem applied entry-wise to $\theta(\gamma_0+t_nv)-\theta(\gamma_0)$, there exist $\tilde\gamma_n^{(1)}(\gamma_0,v),\ldots,\tilde\gamma_n^{(m)}(\gamma_0,v)$ all between $\theta_0$ and $\theta_0+t_nv$ such that
\begin{align}\label{Eqn: over ID, aux6}
\|\theta(\gamma_0+t_nv)-\theta(\gamma_0)-J(\gamma_0)t_nv\|=\|J(\tilde\gamma_n)t_n v-J(\gamma_0)t_nv\|~,
\end{align}
where by abuse of notation we write
\[
J(\tilde\gamma_n)\equiv\begin{bmatrix}
\frac{d\theta^{(1)}}{d\gamma^\intercal}\big|_{\gamma=\tilde\gamma_n^{(1)}(\gamma_0,v)}\\
\vdots\\
\frac{d\theta^{(m)}}{d\gamma^\intercal}\big|_{\gamma=\tilde\gamma_n^{(m)}(\gamma_0,v)}\\
\end{bmatrix}~.
\]
Since $\theta\in \prod_{j=1}^m C^1(\Gamma)$ and $\Gamma$ is compact, $J(\cdot)$ is uniformly continuous on $\Gamma$ and hence
\begin{align}\label{Eqn: over ID, aux7}
\max_{\gamma_0\in\Gamma_0(\theta)}\max_{v\in V_{n,\gamma_0}(a)}\|&J(\tilde\gamma_n)t_n v-J(\gamma_0)t_nv\|\notag\\
&\le t_n \max_{\gamma_0\in\Gamma_0(\theta)}\max_{v\in V_{n,\gamma_0}(a)}\{\|J(\tilde\gamma_n)-J(\gamma_0)\|\|v\|\}=o(t_n)~.
\end{align}
Since all norms in finite dimensional spaces are equivalent, it follows from results \eqref{Eqn: over ID, aux4}, \eqref{Eqn: over ID, aux5}, \eqref{Eqn: over ID, aux6}, \eqref{Eqn: over ID, aux7} and $\theta(\gamma_0)=0$ for all $\gamma_0\in\Gamma_0(\theta)$ that
\begin{align}\label{Eqn: over ID, aux8}
\big|\phi(\theta+t_nh)-\min_{\gamma_0\in\Gamma_0(\theta)}\min_{v\in V_{n,\gamma_0}(a)}\|J(\gamma_0)t_nv+t_nh(\gamma_0)\|_W^2\big|=o(t_n^2)~.
\end{align}

By assumption, $\Gamma_0(\theta)$ is in the interior of $\Gamma$ and so $V_{n,\gamma_0}(a)=V(a)$ for all $n$ sufficiently large. It follows that
\begin{align}\label{Eqn: over ID, aux9}
\min_{\gamma_0\in\Gamma_0(\theta)}\min_{v\in V_{n,\gamma_0}(a)}\|J(\gamma_0)t_nv&+t_nh(\gamma_0)\|_W^2=t_n^2\min_{\gamma_0\in\Gamma_0(\theta)}\min_{v\in V(a)}\|h(\gamma_0)+J(\gamma_0)v\|_W^2\notag\\
&=t_n^2\min_{\gamma_0\in\Gamma_0(\theta)}\min_{v\in V(a)}\|h(\gamma_0)-J(\gamma_0)v\|_W^2~,
\end{align}
where the second equality exploits the fact that $V(a)$ is symmetric. For each $\gamma_0\in\Gamma_0(\theta)$, by the projection theorem there is some $v^*\in\mathbf R^k$ such that
\begin{align}\label{Eqn: over ID, aux10}
\min_{v\in \mathbf R^k}\|h(\gamma_0)-J(\gamma_0)v\|_W^2=\|h(\gamma_0)-J(\gamma_0)v^*\|_W^2~.
\end{align}
Thus, by choosing $a$ large if necessary so that $v^*\in V(a)$, we have from results \eqref{Eqn: over ID, aux8}, \eqref{Eqn: over ID, aux9} and \eqref{Eqn: over ID, aux10} that
\begin{align}\label{Eqn: over ID, aux11}
\big|\phi(\theta+t_nh)-t_n^2\min_{\gamma_0\in\Gamma_0(\theta)}\min_{v\in \mathbf R^k}\|h(\gamma_0)-J(\gamma_0)v\|_W^2\big|=o(t_n^2)~.
\end{align}
Combining \eqref{Eqn: over ID, aux11}, $\phi(\theta)=0$ and part (i), we then arrive at part (ii).\qed

\begin{rem}\label{Rem: over ID, HDD}
The condition that ``{\it for all small $\epsilon>0$, $\inf_{\gamma\in\Gamma\setminus\Gamma_0(\theta)^\epsilon}\|\theta(\gamma)\|\ge C\epsilon^\kappa$ for some $\kappa\in(0,1]$ and some $C>0$}'' in Lemma \ref{Lem: over ID, HDD} effectively imposes restrictions on the Jacobian matrix that prevent one directly applying Lemma \ref{Lem: over ID, HDD} to the setup of \citet{Dovonon_Renault2013testing} where $\Gamma_0(\theta)=\{\gamma_0\}$ is a singleton. To see this, let $\theta$ be the moment function $\rho$ in \citet{Dovonon_Renault2013testing}. Then, for any $\gamma\in\Gamma\setminus\Gamma_0(\theta)^\epsilon$ with $\|\gamma-\gamma_0\|=a\epsilon$ for $a>1$, we have by \citet[p.2570]{Dovonon_Renault2013testing},
\begin{align}
\|\theta(\gamma)\|&=\|\theta(\gamma_0)+\frac{d\theta(\gamma)}{d\gamma^\intercal}\Big|_{\gamma=\gamma_0}\{\gamma-\gamma_0\}+\frac{1}{2}\Big(\{\gamma-\gamma_0\}^\intercal \frac{d^2\theta^{(j)}(\gamma)}{d\gamma d\gamma^\intercal}\Big|_{\gamma=\gamma_0}\{\gamma-\gamma_0\}\Big)_{j=1,\ldots,m}\notag\|\\
&=\|\frac{1}{2}\Big(\{\gamma-\gamma_0\}^\intercal \frac{d^2\theta^{(j)}(\gamma)}{d\gamma d\gamma^\intercal}\Big|_{\gamma=\gamma_0}\{\gamma-\gamma_0\}\Big)_{j=1,\ldots,m}\|\le  C'\epsilon^2~,
\end{align}
for some constant $C'>0$ depending on the eigenvalues of the Hessian matrices (evaluated at $\gamma_0$) of the maps $\gamma\mapsto \theta^{(j)}(\gamma)$, where for the second equality we exploited the facts that (i) $\theta(\gamma_0)=0$, (ii) the Jacobian matrix is degenerate, and (iii) $\|\gamma-\gamma_0\|=a\epsilon$. But by assumption, for the same $\gamma$,
\begin{align}
\|\theta(\gamma)\|\ge C\epsilon^\kappa>C'\epsilon^2~,
\end{align}
for all $\epsilon>0$ sufficiently small since $\kappa\in(0,1]$, a contradiction. The conclusion holds more generally: the condition in fact excludes Jacobian matrices of deficient rank, regardless of whether $\gamma_0$ is point or partially identified. To see this, let $J(\gamma_0)a=0$ for some nonzero $a\in\mathbf R^k$. Then we may choose $\gamma=\gamma_0+\lambda a\in \Gamma\setminus\Gamma_0(\theta)^\epsilon$ for some suitable $\lambda\in\mathbf R$ and for all small $\epsilon>0$ -- this is possible since $\Gamma_0(\theta)$ is required to be in the interior of $\Gamma$. Then the previous arguments apply with such a choice of $\gamma$ and any $\gamma_0\in\Gamma_0(\theta)$. \qed
\end{rem}

\section{Proofs for Section \ref{Sec: Application}}

\begin{lem}\label{Lem: Common CH, HDD2}
Let $\phi: \prod_{j=1}^m\ell^\infty(\mathbb S^k)\to\mathbf R$ be given by $\phi(\theta)=\inf_{\gamma\in\mathbb S^k}\Vert\theta(\gamma)\Vert^{2}$. Then
\begin{itemize}
\item[(i)] $\phi $ is Hadamard differentiable at any $\theta \in \prod_{j=1}^m\ell^\infty(\mathbb S^k)$ satisfying $\theta(\gamma_0)=0$ for some $\gamma_0\in\mathbb S^k$ and the derivative satisfies $\phi _{\theta }^{\prime }( h) =0$ for all $h\in \prod_{j=1}^m\ell^\infty(\mathbb S^k)$.
\item[(ii)] $\phi $ is second order Hadamard directionally differentiable at any $\theta_0(\gamma)\equiv E[Z_{t}\{( \gamma ^\intercal Y_{t+1}) ^{2}-c( \gamma )\}]$ under Assumption \ref{Ass: Common CH, setup} tangentially to $\prod_{j=1}^m C(\mathbb S^k)$ with the derivative given by: for all $h\in\prod_{j=1}^m C(\mathbb S^k)$,
\begin{equation}\label{SOC: application}
\phi _{\theta_0}^{\prime \prime }(h) =\min_{\gamma_0 \in \Gamma_0}\min_{v\in \mathbf{R}^k}\Vert h(\gamma_0) + G \vecop(vv^\intercal)\Vert ^{2}~,
\end{equation}
where $\Gamma_0=\{\gamma_0\in\mathbb S^k: \theta_0(\gamma_0)=0\}$ is the (nonempty) identified set of $\gamma_0$, and $G\in\mathbf M^{m\times k^2}$ with the $j$th row given by $\vecop(\Delta_j)^\intercal$ and
\[
\Delta_j=E[Z_t^{(j)}(Y_{t+1}Y_{t+1}^\intercal-E[Y_{t+1}Y_{t+1}^\intercal])]~.
\]
\end{itemize}
\end{lem}
\noindent{\sc Proof:} Fix $\theta \in \prod_{j=1}^m\ell^\infty(\mathbb S^k) $ satisfying $\theta(\gamma_0)=0$ for some $\gamma_0\in\mathbb S^k$, $\{ h_{n}\} \subset \prod_{j=1}^m\ell^\infty(\mathbb S^k) $ such that $h_{n}\rightarrow h\in \prod_{j=1}^m\ell^\infty(\mathbb S^k)$, and $t_{n}\downarrow0$. It follows that
\begin{align*}
\vert\phi(\theta +t_{n}h_{n})-\phi(\theta )\vert &=\inf_{\gamma \in \mathbb S^k}\Vert \theta ( \gamma) +t_{n}h_{n}( \gamma )\Vert ^{2}\\
&\le \Vert \theta ( \gamma_0 )+t_{n}h_{n}( \gamma_0)\Vert ^{2}=t_{n}^{2}\Vert h_{n}(\gamma_0)\Vert^{2}=o( t_{n}) .
\end{align*}
where in the last step we used the fact that $\sup_{\gamma\in\mathbb S^k}\|h_n(\gamma)\|=O(1)$. So $\phi _{\theta }^{\prime }( h)=0$ for any $h\in \prod_{j=1}^m\ell^\infty(\mathbb S^k)$, as desired for the first claim of the lemma.

Now consider $\theta_0(\gamma)\equiv E[Z_{t}\{( \gamma ^\intercal Y_{t+1}) ^{2}-c( \gamma )\}]$ and suppose that Assumption \ref{Ass: Common CH, setup} holds. Pick $\{h_{n}\}\subset \prod_{j=1}^m\ell^\infty(\mathbb S^k)$ such
that $h_{n}\rightarrow h\in \prod_{j=1}^m C(\mathbb S^k)$, and $t_{n}\downarrow 0$. Note that $\phi(\theta_0) =0$ under Assumption \ref{Ass: Common CH, setup}. Then first, we have
\begin{eqnarray}
\vert \phi ( \theta_0 +t_{n}h_{n}) -\phi ( \theta_0+t_{n}h)\vert &=&\vert \inf_{\gamma \in \mathbb S^k}\Vert \theta_0 ( \gamma ) +t_{n}h_{n}( \gamma )\Vert ^{2}-\inf_{\gamma \in \mathbb S^k}\Vert \theta_0 (\gamma ) +t_{n}h( \gamma ) \Vert ^{2}\vert\notag \\
&\leq &\vert \inf_{\gamma \in \mathbb S^k}\Vert \theta_0 (
\gamma ) +t_{n}h_{n}( \gamma ) \Vert +\inf_{\gamma
\in \mathbb S^k}\Vert \theta_0 ( \gamma ) +t_{n}h( \gamma
) \Vert\vert \notag\\&& \cdot\, t_{n}\sup_{\gamma \in \mathbb S^k}\Vert
h_{n}( \gamma ) -h( \gamma ) \Vert  \notag \\
&\leq &t_{n}\vert \inf_{\gamma_0 \in \Gamma_0}\Vert \theta_0 (
\gamma_0 ) +t_{n}h_{n}( \gamma_0 ) \Vert +\inf_{\gamma_0
\in \Gamma_0}\Vert \theta_0 ( \gamma_0 ) +t_{n}h( \gamma_0
) \Vert\vert \notag\\&& \cdot\, \sup_{\gamma \in \mathbb S^k}\Vert
h_{n}( \gamma ) -h( \gamma ) \Vert  \notag \\
&=&o( t_{n}^{2})~.\label{eqn: apdc1}
\end{eqnarray}%
Next, let $\Gamma_0^{\epsilon }\equiv \{\gamma \in \mathbb S^k:\min_{s\in \Gamma_0}\Vert s-\gamma \Vert \leq \epsilon\}$ and $\Gamma_1^{\epsilon }\equiv \{\gamma \in \mathbb S^k:\min_{s\in \Gamma_0}\Vert s-\gamma\Vert \geq \epsilon\}$. By Equation (7) in \citet{Dovonon_Renault2013testing}, $\theta_0(\gamma ) =$ Cov$(Z_{t},\sigma_t^2)$Diag$(\Lambda ^\intercal\gamma \gamma^\intercal \Lambda)$), where for a $p\times p$ matrix $A$, $\text{Diag}(A)$ denotes the $p\times 1$ vector consisting of diagonal entries. Also, let $\lambda _{\min}(\cdot)$ and $\lambda _{\min }^{+}(\cdot)$ denote the smallest and the smallest positive singular values, respectively. We then have for $C\equiv p^{-1/2}\lambda_{\min}^{+}(\Lambda^\intercal)\lambda _{\min }( \text{Cov}(Z_{t}, \sigma_t^2))/2$,
\begin{eqnarray*}
\min_{\gamma \in \Gamma_1^{\epsilon }}\Vert \theta_0 (\gamma )\Vert &\geq& \min_{\gamma \in \Gamma_1^{\epsilon }}\Vert {\text{Diag}}( \Lambda ^\intercal\gamma
\gamma ^\intercal\Lambda ) \Vert \lambda _{\min }(\text{Cov}(Z_{t},\sigma_t^2)) \\
&\geq&\min_{\gamma \in \Gamma_1^{\epsilon }}\Vert \Lambda ^\intercal\gamma \Vert ^{2}p^{-1/2}\lambda _{\min }( \text{Cov}(
Z_{t},\sigma_t^2 )) \geq C\epsilon^{2}~,
\end{eqnarray*}%
where the first inequality follows from a simple application of the singular value decomposition of $\text{Cov}(Z_{t},\sigma_t^2 )$, the second inequality exploits the generalized mean inequality, and last inequality is by Lemma \ref{Lem: Common CH, HDD2, aux}. Note that $\lambda _{\min }( \text{Cov}(Z_{t},\sigma_t^2 ))>0$ by Assumption \ref{Ass: Common CH, setup}(v). Let $\Delta \equiv \lbrack 3C^{-1}\max_{\gamma \in \mathbb S^k
}\Vert h( \gamma ) \Vert ]^{1/2}>0$ for the nontrivial case $\max_{\gamma\in\mathbb S^k}\|h(\gamma)\|>0$. Then it follows by the triangle inequality that for $n$ sufficiently large such that $t_n\le \sqrt{t_n}$,
\begin{align*}
\min_{\gamma \in \Gamma_1^{\sqrt{t_{n}}\Delta }}\Vert \theta_0 (\gamma )&+t_{n}h(\gamma )\Vert \geq \min_{\gamma\in \Gamma_1^{\sqrt{t_{n}}\Delta }}\Vert \theta_0 ( \gamma
)\Vert -t_{n}\max_{\gamma \in \mathbb S^k }\Vert h( \gamma) \Vert \\
&\geq 3t_{n}\max_{\gamma \in \mathbb S^k }\Vert h( \gamma )\Vert -t_{n}\max_{\gamma \in \mathbb S^k }\Vert h( \gamma )\Vert > t_{n}\min_{\gamma_0 \in \Gamma_0}\Vert h( \gamma_0 )\Vert \geq \sqrt{\phi ( \theta_0 +t_{n}h)}~,
\end{align*}
and therefore
\begin{equation*}
\phi(\theta_0 +t_{n}h) =\min_{\gamma \in \Gamma_0^{\sqrt{t_{n}}\Delta }}\Vert \theta_0 ( \gamma ) +t_{n}h( \gamma)\Vert ^{2}~.
\end{equation*}%
For $\gamma_0 \in \Gamma_0$, let $V_{n,\gamma_0 }^{\Delta }\equiv \{v\in\mathbf{R}^k:\gamma_0 +\sqrt{t_{n}}v\in \mathbb S^k$ and $\Vert v\Vert \leq \Delta \}$ and $V_{\gamma_0 }^{\Delta }\equiv \{v\in \mathbf{R}^k: \gamma_0^\intercal v=0$ and $\Vert v\Vert\leq \Delta \}$. Then we have
\begin{eqnarray}
\phi( \theta_0 +t_{n}h) &=&\min_{\gamma_0 \in \Gamma_0}\min_{v\in V_{n,\gamma_0 }^{\Delta }}\Vert \theta_0 ( \gamma_0 +
\sqrt{t_{n}}v) +t_{n}h( \gamma_0 +\sqrt{t_{n}}v) \Vert^{2}  \notag \\
&=&\min_{\gamma_0 \in \Gamma_0}\min_{v\in V_{n,\gamma_0 }^{\Delta}}\Vert \theta_0 ( \gamma_0 +\sqrt{t_{n}}v) +t_{n}h( \gamma_0
)\Vert ^{2}+o(t_{n}^{2})~,  \label{eqn: apdc2}
\end{eqnarray}%
where the first equality is due to the definition of $\Gamma_0^{\sqrt{t_{n}}\Delta }$ and the second follows by
\begin{align*}
&\vert \min_{\gamma_0 \in \Gamma_0}\min_{v\in V_{n,\gamma_0}^{\Delta }}\Vert \theta_0( \gamma_0 +\sqrt{t_{n}}v)+t_{n}h(\gamma_0+\sqrt{t_{n}}v)\Vert^2 -\min_{\gamma_0 \in
\Gamma_0}\min_{v\in V_{n,\gamma_0 }^{\Delta }}\Vert \theta_0(\gamma_0 +\sqrt{t_{n}}v) +t_{n}h(\gamma_0)\Vert^2\vert \\
&\leq 2t_{n}\|h(\gamma_0)\|\cdot t_n\max_{\gamma_0 \in \Gamma_0}\max_{v\in V_{n,\gamma_0}^{\Delta }}\Vert h( \gamma_0 +\sqrt{t_{n}}v) -h( \gamma_0)\Vert\\
&\leq 2t_{n}^2\|h(\gamma_0)\|\max_{\gamma _{1},\gamma _{2}\in \mathbb S^k ,\Vert\gamma _{1}-\gamma _{2}\Vert \leq \sqrt{t_{n}}\Delta }\Vert
h( \gamma _{1}) -h( \gamma _{2}) \Vert =o(t_{n}^2)~,
\end{align*}%
where $\gamma_0$ in the first inequality is any fixed element in $\Gamma_0$, the last equality follows by the uniform continuity of $h$ over $\mathbb S^k$. Noting that $( \gamma ^\intercal Y_{t+1}) ^{2}=\gamma^\transpose Y_{t+1}Y_{t+1}^\intercal\gamma$ and so $c(\gamma)=\gamma^\transpose E[Y_{t+1}Y_{t+1}^\intercal]\gamma$, we may write
\begin{multline}\label{Eqn: vec}
         \theta_0(\gamma)= \begin{bmatrix}
                                          E[Z_{t}^{(1)}\{( \gamma ^\intercal Y_{t+1}) ^{2}-c( \gamma )\}] \\
                                          \vdots \\
                                          E[Z_{t}^{(m)}\{( \gamma ^\intercal Y_{t+1}) ^{2}-c( \gamma )\}] \\
                                        \end{bmatrix} \\
                                       =  \begin{bmatrix}
                                          \gamma^\intercal\Delta_1\gamma \\
                                          \vdots \\
                                          \gamma^\intercal\Delta_m\gamma \\
                                        \end{bmatrix} = \begin{bmatrix}
                                          \vecop(\Delta_1)^\intercal\mathrm{vec}(\gamma\gamma^\intercal) \\
                                          \vdots \\
                                          \vecop(\Delta_m)^\intercal\mathrm{vec}(\gamma\gamma^\intercal)\\
                                        \end{bmatrix}= G\mathrm{vec}(\gamma\gamma^\intercal)~,
      \end{multline}
where we made use of some facts on the vec operator \citep[p.282]{AbadirandMagnus}. In turn, by \eqref{Eqn: vec} and the definition of $\Gamma_0$, we have
\begin{align}
\min_{\gamma_0 \in \Gamma_0}\min_{v\in V_{n,\gamma_0 }^{\Delta}}\Vert \theta_0( \gamma_0 +\sqrt{t_{n}}v) + & t_{n}h( \gamma_0)\Vert ^{2}
=t_{n}^{2}\min_{\gamma_0 \in \Gamma_0}\min_{v\in V_{n,\gamma_0 }^{\Delta }}\Vert G \vecop(vv^{\intercal})+h(\gamma_0 )\Vert ^{2} \notag \\
&=t_{n}^{2}\min_{\gamma_0 \in \Gamma_0}\min_{v\in V_{\gamma_0 }^{\Delta}}\Vert G \vecop(vv^{\intercal})+h( \gamma_0)\Vert ^{2}+o( t_{n}^{2}) \notag \\
&=t_{n}^{2}\min_{\gamma_0 \in \Gamma_0}\min_{v\in \mathbf{R}^k}\Vert G \vecop(vv^{\intercal})+h(\gamma_0)\Vert ^{2}+o( t_{n}^{2})~,  \label{eqn: apdc3}
\end{align}%
where the second equality follows by the fact that $V_{n,\gamma_0 }^{\Delta }$ converges to $V_{\gamma_0 }^{\Delta}$ uniformly in $\gamma_0\in\Gamma_0$ with respect to the Hausdorff metric by Lemma \ref{Lem: Common CH, set conv} and Lemma B.3 in \citet{Fang_Santos2014HDD}, and the third equality by the facts that $ G \vecop(vu^{\intercal})=0$ for all $v\in\Gamma_0$ and all $u\in\mathbf R^k$ (to be proved shortly) and that the inside minimum can be attained in $V_{\gamma_0 }^{\Delta}$ for all $\Delta$ large enough. Combining \eqref{eqn: apdc1}, \eqref{eqn: apdc2} and \eqref{eqn: apdc3} yields
\begin{equation*}
\phi _{\theta_0 }''( h) =\lim_{n\rightarrow \infty }\frac{\phi( \theta_0 +t_{n}h_{n}) }{t_{n}^{2}}=\min_{\gamma_0 \in\Gamma_0}\min_{v\in \mathbf{R}^k}\Vert h(\gamma_0)+G \vecop(vv^{\intercal})\Vert ^{2},
\end{equation*}%
as desired. It remains to show $ G \vecop(vu^{\intercal})=0$ for all $v\in\Gamma_0$ and all $u\in\mathbf R^k$. Fix $v\in\Gamma_0$ and $u\in\mathbb R^k$. By similar arguments (in reverse order) that led to \eqref{Eqn: vec}, we obtain
      \begin{align}\label{Eqn: vec1}
         G\mathrm{vec}(vu^\intercal)& =  \begin{bmatrix}
                                          v^\intercal\Delta_1u \\
                                          \vdots \\
                                          v^\intercal\Delta_m u \\
                                        \end{bmatrix}~.
      \end{align}
      Next, note that, by the law of iterated expectations, we have
      \begin{align}\label{Eqn: vec1a}
      \Delta_j&=E[Z_t^{(j)}(Y_{t+1}Y_{t+1}^\intercal-E[Y_{t+1}Y_{t+1}^\intercal])] \notag\\
       &= E\big[Z_t^{(j)}E[(Y_{t+1}Y_{t+1}^\intercal-E[Y_{t+1}Y_{t+1}^\intercal])|\mathcal{F}_t]\big] \notag\\
       &=  \Lambda E[Z_t^{(j)}(D_t-E[D_t])]\Lambda^\intercal~,
      \end{align}
      where the third inequality follows by the model specified in display \eqref{Eqn: Common CH, defn} and Assumption \ref{Ass: Common CH, setup}(ii). Result \eqref{Eqn: vec1a} in turn implies that, for all $j=1,\ldots,m$,
        \begin{align}\label{Eqn: vec2}
        v^\intercal\Delta_ju = v^\intercal\Lambda E[Z_t^{(j)}(D_t-E[D_t])]\Lambda^\intercal u = 0~,
      \end{align}
  where $v^\intercal\Lambda = 0$ because $v\in\Gamma_0 = \{\gamma_0\in\mathbb S^k: \theta_0(\gamma_0)=0\}$ which is equal to the intersection of $\mathbb S^k$ and the null space of $\Lambda^{\intercal}$ -- see our discussions below Assumption \ref{Ass: Common CH, setup}. The claim now follows by combining \eqref{Eqn: vec1} and \eqref{Eqn: vec2}. \qed

\begin{rem}\label{Rem: Common CH, derivative est}
The derivative \eqref{Eqn: Common CH, Derivative} can be rewritten as:
\begin{align}\label{Eqn: Common CH, Derivative22}
\phi_{\theta_0}''(h)=\min_{\gamma_0\in\Gamma_0}\min_{v\in\Gamma_0^\perp}\|h(\gamma_0)+G\vecop(vv^\intercal)\|^2~,
\end{align}
where $\Gamma_0^\perp\equiv\{\lambda\in\mathbf R^k: \lambda^\intercal \gamma_0=0~,\,\forall\,\gamma_0\in\Gamma_0\}$ denotes the orthogonal complement of $\Gamma_0$. Then for $\hat\Gamma_{T,\perp}=\{\gamma\in\mathbf{R}^k: \sup_{\lambda\in\hat\Gamma_T}|\gamma^\intercal\lambda|\leq\kappa_{T}^{1/2}\}$ and $B_T\equiv\{v\in\mathbf R^k: \|v\|\le \kappa_T^{-1/2}\}$, we may estimate $\phi_{\theta_0}''(h)$ by
\[
\hat{\phi} _{T }^{\prime \prime }(h) =\inf_{\gamma \in \hat\Gamma_T}\min_{v\in \hat\Gamma_{T,\perp}\cap B_T}\Vert h(\gamma) + \hat{G} \vecop(vv^\intercal)\Vert ^{2}~.\tag*{\qed}
\]
\end{rem}

\begin{lem}\label{Lem: Common CH, weak limit for theta}
Under Assumptions \ref{Ass: Common CH, setup} and \ref{Ass: Common CH, CLT}, we have
\[
\sqrt{T}\{\hat{\theta}_T-\theta_0\}\overset{L}{\rightarrow}\mathbb{G} \text{ in } \prod_{j=1}^m\ell^\infty(\mathbb S^k)~,
\]
where $\mathbb{G}$ is a zero mean Gaussian process with the covariance functional satisfying: for any $\gamma_{1}$, $\gamma_{2}\in\Gamma_0$ and $\mu_z=E[Z_t]$,
\begin{align*}
E[\mathbb{G}(\gamma_{1})\mathbb{G}(\gamma_{2})]=E[(Z_{t}-\mu_{z})(Z_{t}-\mu_{z})^\intercal\{(\gamma_{1}^\intercal Y_{t+1})^{2}-c(\gamma_{1})\}\{(\gamma_{2}^\intercal Y_{t+1})^{2}-c(\gamma_{2})\}] ~.
\end{align*}
\end{lem}
\noindent{\sc Proof:} By elementary rearrangements we have
\begin{align*}
\sqrt{T}\{\hat{\theta}_T(\gamma)-\theta_0(\gamma)\}=\sqrt{T}G_{T}(\gamma)- \sqrt{T}(\hat{\mu}_{z}-\mu_{z})\{\hat{c}(\gamma)-c(\gamma)\}~,
\end{align*}
where $\hat{\mu}_{z}=\frac{1}{T}\sum_{t=1}^{T}Z_{t}$, $\hat c(\gamma)=\frac{1}{T}\sum_{t=1}^T(\gamma^\intercal Y_{t+1})^2$, and
\begin{align*}
G_{T}(\gamma)=\frac{1}{T}\sum_{t=1}^{T}(Z_{t}-\mu_{z})\{(\gamma^{\intercal}Y_{t+1})^{2}-c(\gamma)\}-E[(Z_{t}-\mu_{z})\{(\gamma^{\intercal}Y_{t+1})^{2}-c(\gamma)\}]~.
\end{align*}
By Assumptions \ref{Ass: Common CH, setup}(vi) and \ref{Ass: Common CH, CLT}, and the law of large numbers for stationary and ergodic sequences and the compactness of $\mathbb S^k$, we have
\[
\sqrt{T}(\hat{\mu}_{z}-\mu_{z})(\hat{c}-c)=o_{p}(1) \text{ in } \prod_{j=1}^m\ell^{\infty}(\mathbb S^k)~.
\]
Once again by Assumptions \ref{Ass: Common CH, setup}(vi) and \ref{Ass: Common CH, CLT}, together with $\sqrt{T}G_{T}(\gamma)=\sqrt{T}\tilde{G}\vecop({\gamma\gamma^{\intercal}})$ where $\tilde{G}\in\mathbf{M}^{m\times k^{2}}$ having its $j$th row given by $(\vecop(\tilde{\Delta}_{j}))^{\intercal}$ for
\begin{align*}
\tilde{\Delta}_{j}=\frac{1}{T}\sum_{t=1}^{T}(Z_t^{(j)}&-\mu_{z}^{(j)})\{Y_{t+1}Y_{t+1}^{\intercal}-E(Y_{t+1}Y_{t+1}^{\intercal})\}\\
&\hspace{1cm}-E[(Z_t^{(j)}-\mu_{z}^{(j)})\{Y_{t+1}Y_{t+1}^{\intercal}-E(Y_{t+1}Y_{t+1}^{\intercal})\}] ~,
\end{align*}
we have by the compactness of $\mathbb S^k$ that
\[
\sqrt{T}G_{T}\overset{L}{\to}\mathbb{G} \text{ in } \prod_{j=1}^m\ell^{\infty}(\mathbb S^k)
\]
for some Gaussian process $\mathbb{G}(\gamma)$. In particular, for $\gamma\in\Gamma_0$ the summand in $G_{T}(\gamma)$ is a martingale difference sequence, so for any $\gamma_{1}$, $\gamma_{2}\in\Gamma_0$, the covariance functional satisfies
\begin{align*}
E[\mathbb{G}(\gamma_{1})\mathbb{G}(\gamma_{2})]=E[(Z_{t}-\mu_{z})(Z_{t}-\mu_{z})^\intercal\{(\gamma_{1}^\intercal Y_{t+1})^{2}-c(\gamma_{1})\}\{(\gamma_{2}^\intercal Y_{t+1})^{2}-c(\gamma_{2})\}]~.
\end{align*}
This completes the proof of the lemma. \qed

\begin{lem}\label{Lemma: Common CH, derivative convergence}
Suppose Assumptions \ref{Ass: Common CH, setup}, \ref{Ass: Common CH, CLT} and \ref{Ass: Common CH, tuning parameter} hold. Let $\hat{\phi}_{T }''$ be constructed as in \eqref{Eqn: Common CH, derivative estimate}. Then we have: whenever $h_{T}\to h$ as $T\rightarrow\infty$ for a sequence $\{h_{T}\}\subset\prod_{j=1}^m \ell^\infty(\mathbb S^k)$ and $h\in\prod_{j=1}^m C(\mathbb S^k)$, it follows that
\[
\hat{\phi} _{T }''(h_{T})\overset{p}{\rightarrow}{\phi} _{\theta_0}^{\prime \prime }(h)~.
\]
\end{lem}
\noindent{\sc Proof:} Pick a sequence $\{h_{T}\}\subset\prod_{j=1}^m \ell^\infty(\mathbb S^k)$ and $h\in\prod_{j=1}^m C(\mathbb S^k)$ such that $h_T\to h$ as $T\to\infty$. Define
\begin{equation*}
\tilde{\phi} _{T }''(h) =\min_{\gamma \in \hat\Gamma_T}\min_{v\in B_T}\Vert h(\gamma) + G\vecop(vv^\intercal)\Vert ^{2}~.
\end{equation*}
Then we have
\begin{align}\label{Eqn: Common CH, derivative conv, aux1}
|\hat{\phi} _{T }''(h_T)&-\tilde{\phi} _{T }''(h)|\notag\\
&\le \big|\inf_{\gamma \in \hat\Gamma_T}\min_{v\in B_T}\|h_T(\gamma)+\hat G\vecop(vv^\intercal)\|+\inf_{\gamma \in \hat\Gamma_T}\min_{v\in B_T}\|h(\gamma)+ G\vecop(vv^\intercal)\|\big|\notag\\
&\quad \cdot \big|\inf_{\gamma \in \hat\Gamma_T}\min_{v\in B_T}\|h_T(\gamma)+\hat G\vecop(vv^\intercal)\|-\inf_{\gamma \in \hat\Gamma_T}\min_{v\in B_T}\|h(\gamma)+G\vecop(vv^\intercal)\|\big|\notag\\
&\le (\sup_{\gamma\in \mathbb S^k}\|h_T(\gamma)\|+\sup_{\gamma\in \mathbb S^k}\|h(\gamma)\|)\sup_{\gamma\in\mathbb S^k}\|h_T(\gamma)-h(\gamma)\| \sup_{v\in B_T}\|\vecop(vv^\intercal)\|\|\hat G-G\|\notag\\
&\lesssim\sup_{v\in B_T}T^{-1/2}\|v\|^2\|\sqrt{T}\{\hat G-G\}\|\le T^{-1/2}\kappa_T^{-1}\|\sqrt{T}\{\hat G-G\}\|=o_p(1)~,
\end{align}
where ``$\lesssim$'' follows from $h_T\to h$, and the last step is by Assumptions \ref{Ass: Common CH, CLT} and \ref{Ass: Common CH, tuning parameter}.

Next, under Assumptions \ref{Ass: Common CH, setup}, \ref{Ass: Common CH, CLT} and \ref{Ass: Common CH, tuning parameter}, we have by Theorem 3.1 in \citet{CHT2007} that $d_H(\hat\Gamma_T,\Gamma_0)\convp 0$ as $T\to\infty$, with $a_T=T$, $b_T=\sqrt{T}$, and $\hat c=T\kappa_T^2$. Let
\[
\bar\phi_{T}''(h)=\min_{\gamma\in\Gamma_0}\min_{v\in B_T}\|h(\gamma)+G\vecop(vv^\intercal)\|^2~.
\]
Since $h\in\prod_{j=1}^m C(\mathbb S^k)$ and $\mathbb S^k$ is compact, together with $d_H(\hat\Gamma_T,\Gamma_0)\convp 0$, it follows that
\begin{align}\label{Eqn: Common CH, derivative conv, aux2}
|&\tilde{\phi} _{T }''(h)-\bar\phi_{T}''(h)|\notag\\
&\le \sup_{\|\gamma_1-\gamma_2\|_{\mathbb D}\le d_H(\hat\Gamma_T,\Gamma_0)}\big|\min_{v\in B_T}\|h(\gamma_1)+G\vecop(vv^\intercal)\|^2-\min_{v\in B_T}\|h(\gamma_2)+G\vecop(vv^\intercal)\|^2\big|\notag\\
&\le \sup_{\|\gamma_1-\gamma_2\|_{\mathbb D}\le d_H(\hat\Gamma_T,\Gamma_0)}\|h(\gamma_1)-h(\gamma_2)\|=o_p(1)~.
\end{align}
Since $\bar\phi_{T}''(h)$ is monotonically decreasing as $T\uarrow\infty$, we further have
\begin{align}\label{Eqn: Common CH, derivative conv, aux3}
\bar\phi_{T}''(h)\to \min_{\gamma\in\Gamma_0}\min_{v\in\mathbf R^k}\|h(\gamma)+G\vecop(vv^\intercal)\|^2=\phi_{\theta_0}''(h)~.
\end{align}
The lemma then follows from results \eqref{Eqn: Common CH, derivative conv, aux1}, \eqref{Eqn: Common CH, derivative conv, aux2} and \eqref{Eqn: Common CH, derivative conv, aux3}. \qed

\noindent{\sc Proof of Proposition \ref{Pro: Common CH, test validity}:} By Lemmas \ref{Lem: Common CH, weak limit for theta} and \ref{Lemma: Common CH, derivative convergence}, Assumptions \ref{Ass: bootstrap for theta} and \ref{Ass: measurability}, and the cdf of the weak limit being strictly increasing at $c_{1-\alpha}$, we have $\hat{c}_{1-\alpha}\overset{p}{\to}c_{1-\alpha}$ following exactly the same proof of Corollary 3.2 in \citet{Fang_Santos2014HDD}.\footnote{Note $\phi_{\theta_0}''$ trivially admits a continuous extension on $\prod_{j=1}^m\ell^\infty(\mathbb S^k)$ with the first min replaced by $\inf$.} Then under $\mathrm H_0$, the conclusion follows from combining Proposition \ref{Pro: Common CH, weak limit}, Slutsky thoerem, ${c}_{1-\alpha}$ being a continuity point of the weak limit and the portmanteau theorem. \qed

\begin{lem}\label{Lem: Common CH, HDD2, aux}
Let $\Lambda$ and $\Gamma_1^\epsilon$ be given as in the proof of Lemma \ref{Lem: Common CH, HDD2}. Then under Assumption \ref{Ass: Common CH, setup} and $\mathrm H_0$, for all sufficiently small $\epsilon>0$, we have
\[
\min_{\gamma\in\Gamma_1^\epsilon}\|\Lambda^\intercal \gamma\|\ge \frac{\epsilon}{\sqrt{2}}\sigma_{\min}^+(\Lambda^\intercal)~,
\]
where $\sigma_{\min}^+(\Lambda^\intercal)$ denotes the smallest positive singular value of $\Lambda^\intercal$.
\end{lem}
\noindent{\sc Proof:} To begin with, note that i) $\Gamma_0=\amin_{\gamma\in\mathbb S^k}\|\Lambda^\intercal\gamma\|$ by Assumption \ref{Ass: Common CH, setup}, ii) $\Gamma_0\neq\emptyset$ under the null, iii) $\sigma_{\min}^+(\Lambda^\intercal)$ is well-defined by Assumption \ref{Ass: Common CH, setup}(i) so that $\Gamma_0\subsetneqq\mathbb S^k$. Let $\Lambda^\intercal=P\Sigma Q^\intercal$ be the singular value decomposition of $\Lambda^\intercal$, where $P\in\mathbf M^{p\times p}$ and $Q\in\mathbf M^{k\times k}$ are orthonormal, and $\Sigma\in\mathbf M^{p\times k}$ is a diagonal matrix with diagonal entries in descending order. Since $\Lambda$ is of full column rank, $\sigma_{\min}^+(\Lambda^\intercal)$ is equal to the $p$th diagonal entry of $\Sigma$ with $p<k$.

Fix $\gamma\in\Gamma_1^\epsilon$. Let $a_\gamma\equiv Q^\intercal\gamma$ and write $a_\gamma=[a_\gamma^{(1)\intercal}, a_\gamma^{(2)\intercal}]^\intercal$ for $a_\gamma^{(1)}\in\mathbf R^{p}$ and $a_\gamma^{(2)}\in\mathbf R^{k-p}$. Suppose first that $\|a_\gamma^{(2)}\|\neq 0$. Then we have
\begin{align}\label{Eqn: Common CH, HDD2, aux1}
\|[0,a_\gamma^{(2)\intercal}]^{\intercal}/\|a_\gamma^{(2)}\|-a_\gamma\|=\|Q[0,a_\gamma^{(2)\intercal}]^{\intercal}/\|a_\gamma^{(2)}\|-\gamma\|\geq\min_{s\in\Gamma_{0}}\|s-\gamma\|\geq \epsilon~,
\end{align}
since $Q[0,a_\gamma^{(2)\intercal}]^{\intercal}/\|a_\gamma^{(2)}\|\in\Gamma_{0}$ by direct calculations. In turn, result \eqref{Eqn: Common CH, HDD2, aux1} implies
\begin{align}\label{Eqn: Common CH, HDD2, aux2}
\|a_\gamma^{(1)}\|^{2} + (1-\|a_\gamma^{(2)}\|)^2\geq \epsilon^{2}  ~.
\end{align}
Moreover, we know from $Q\in\mathbf M^{k\times k}$ being orthonormal and $\gamma\in\mathbb S^k$ that
\begin{align}\label{Eqn: Common CH, HDD2, aux3}
\|a_\gamma^{(1)}\|^{2}+\|a_\gamma^{(2)}\|^{2}=1~.
\end{align}
Combining results \eqref{Eqn: Common CH, HDD2, aux1} and \eqref{Eqn: Common CH, HDD2, aux2} we may thus conclude that
\begin{align}\label{Eqn: Common CH, HDD2, aux4}
2\|a_\gamma^{(1)}\|^{2}= \|a_\gamma^{(1)}\|^{2} + 1- \|a_\gamma^{(2)}\|^2\geq \|a_\gamma^{(1)}\|^{2} + (1-\|a_\gamma^{(2)}\|)^2\geq \epsilon^{2}~,
\end{align}
implying that $\|a_\gamma^{(1)}\|\geq \frac{\epsilon}{\sqrt{2}}$. This also holds for all sufficiently small $\epsilon>0$ when $\|a_\gamma^{(2)}\|=0$ in which case $\|a_\gamma^{(1)}\|=1$ in view of \eqref{Eqn: Common CH, HDD2, aux3}. Consequently, we have
\begin{align}\label{Eqn: phi, diff lemma, ID, aux4}
\min_{\gamma\in \Gamma_{1}^{\epsilon}}\|\Lambda^{\intercal} \gamma\|&=\min_{\gamma\in \Gamma_{1}^{\epsilon}}\|P\Sigma Q^{\intercal} \gamma\|=\min_{\gamma\in \Gamma_{1}^{\epsilon}}\|\Sigma a_\gamma\|\notag\\
&\ge \lambda^{+}_{\min}(\Lambda^{\intercal})\min_{\gamma\in \Gamma_{1}^{\epsilon}}\|a_\gamma^{(1)}\|\geq \lambda^{+}_{\min}(\Lambda^{\intercal})\frac{\epsilon}{\sqrt{2}}~,
\end{align}
for all sufficiently small $\epsilon>0$. This completes the proof of the lemma. \qed

\begin{lem}\label{Lem: Common CH, set conv}
Let $V_{n,\gamma_0}^\Delta$ and $V_{\gamma_0}^\Delta$ be defined as in the proof of Lemma \ref{Lem: Common CH, HDD2}. Then $d_H(V_{n,\gamma_0}^\Delta,V_{\gamma_0}^\Delta)\to 0$ uniformly in $\gamma_0\in\Gamma_0$ as $n\to\infty$.
\end{lem}
\noindent{\sc Proof:} First, note that $V_{n,\gamma_0}^\Delta=\{v\in\mathbf R^k: \gamma_0+\sqrt{t_n}v\in\mathbb S^k\text{ and }\|v\|\le \Delta\}$. For $u\in V_{n,\gamma_0}^\Delta$, set $u^*\equiv u-(\gamma_0^\intercal u)\gamma_0$. It is a simple exercise to verify that $u^*\in V_{\gamma_0}^\Delta$. It follows that
\begin{align}\label{Eqn: Common CH, set conv, aux1}
\min_{v\in V_{\gamma_0}^\Delta}\|u-v\|\le \|u-u^*\|\le \frac{1}{2}\sqrt{t_n}\Delta^2~.
\end{align}
In turn, result \eqref{Eqn: Common CH, set conv, aux1} implies that: for all $\gamma_0\in\Gamma_0$,
\begin{align}\label{Eqn: Common CH, set conv, aux2}
\max_{u\in V_{n,\gamma_0}^\Delta}\min_{v\in V_{\gamma_0}^\Delta}\|u-v\|\le \frac{1}{2}\sqrt{t_n}\Delta^2~.
\end{align}
On the other hand, for $v\in V_{\gamma_0}^\Delta$, set $v^*=v- b_n \gamma_0$ for $b_n=(1-\sqrt{1-t_n\|v\|})/\sqrt{t_n}$ if $\|v\|< \Delta$, and $v^*=a_n v- b_n \gamma_0$ for $a_n=1-\sqrt{t_n}$ and $b_n=(1-\sqrt{1-t_n(1-\sqrt{t_n})^2\|v\|^2})/\sqrt{t_n}$ if $\|v\|=\Delta$. In any case, $v^*\in V_{n,\gamma_0}^\Delta$ by direct calculations. Therefore,
\begin{align}\label{Eqn: Common CH, set conv, aux3}
\min_{v\in V_{\gamma_0}^\Delta}\max_{u\in V_{n,\gamma_0}^\Delta}\|u-v\|\le \min_{v\in V_{\gamma_0}^\Delta}\|v-v^*\|=O(\sqrt{t_n})~,
\end{align}
uniformly in $\gamma_0\in\Gamma_0$, where we exploited the facts that $b_n=O(\sqrt{t_n})$ uniformly in $\gamma_0\in\Gamma_0$ and that $V_{\gamma_0}^\Delta$ is bounded. The lemma then follows from \eqref{Eqn: Common CH, set conv, aux2} and \eqref{Eqn: Common CH, set conv, aux3}. \qed

Our final lemma shows the work in Section 4 is consistent with \citet{Dovonon_Renault2013testing} in the case they studied when the weighting matrix is the identity matrix. We note that the essential difference between $G$ and $\bar G$ in \eqref{Eqn: Common CH, DR comparison} and \eqref{Eqn: Common CH, DR comparison1} is: the former consists of the second order derivatives of the moment function with respect to all $k$ entries of $\gamma$, whereas the latter the second order derivatives of the moment function with the $k$-th entry $\gamma^{(k)}$ of $\gamma$ substituted by $\gamma^{(k)}=1-\sum_{j=1}^{k-1}\gamma^{(j)}$.
\begin{lem}\label{Lem: Common CH, compare with DR}
The limit $J^W$ with $W=I_m$ in Theorem 3.1 of \citet{Dovonon_Renault2013testing} can be represented as: for $\mathbb G$ and $G$ defined in Section \ref{Sec: Application},
\begin{align}
\min_{v\in\mathbf R^k} \{\mathbb G(\gamma_{0})^\intercal \mathbb G(\gamma_{0})+\mathbb G(\gamma_0)^\intercal G\mathrm{vec}(vv^\intercal)+\frac{1}{4}(\mathrm{vec}(vv^\intercal))^\intercal G^\intercal G\mathrm{vec}(vv^\intercal)\}~.
\end{align}
\end{lem}
\noindent{\sc Proof:} First, note that by \citet[p.2576]{Dovonon_Renault2013testing}, $J^W$ with $W=I_m$ can be represented as in \eqref{Eqn: Common CH, DR comparison1} where $\mathbb G(\gamma_0)$ is centered Gaussian with variance $E[(Z_t-E[Z_t])(Z_t-E[Z_t])^\intercal\{(\gamma_0^\intercal Y_{t+1})^2-E[(\gamma_0^\intercal Y_{t+1})^2]\}]$. Next, simple algebra shows that
\begin{align}
\bar G\mathrm{vec}(uu^\intercal)=u^\intercal A \Delta_j A^{\intercal} u=G\mathrm{vec}(A^{\intercal} u (A^{\intercal}u)^\intercal)~,
\end{align}
where $A\equiv [I_{k-1},-\jmath_{k-1}]$, where $\jmath_{k-1}$ is the $(k-1)\times 1$ vector of ones. It follows that
\begin{align}\label{Eqn: Common CH, DR comparison2}
J^{I_m}&=\min_{u\in\mathbf R^{k-1}} \{\mathbb G(\gamma_0)^\intercal \mathbb G(\gamma_0)+\mathbb G(\gamma_0)^\intercal G\mathrm{vec}(A^{\intercal} u (A^{\intercal}u)^\intercal)\notag\\
&\hspace{0.6in}+\frac{1}{4}(\mathrm{vec}(A^{\intercal} u (A^{\intercal}u)^\intercal))^\intercal G^\intercal G\mathrm{vec}(A^{\intercal} u (A^{\intercal}u)^\intercal)\}\notag\\
&= \min_{a\in\mathbf{R}}\min_{u\in\mathbf R^{k-1}} \{\mathbb G(\gamma_0)^\intercal \mathbb G(\gamma_0)+\mathbb G(\gamma_0)^\intercal G\mathrm{vec}((A^{\intercal} u+a \gamma_{0}) (A^{\intercal}u+a \gamma_{0})^\intercal)\notag\\
&\hspace{0.6in}+\frac{1}{4}(\mathrm{vec}((A^{\intercal} u+a \gamma_{0}) (A^{\intercal}u+a \gamma_{0})^\intercal))^\intercal G^\intercal G\mathrm{vec}((A^{\intercal} u+a \gamma_{0}) (A^{\intercal}u+a \gamma_{0})^\intercal)\}\notag\\
&=\min_{v\in\mathbf R^k} \{\mathbb G(\gamma_{0})^\intercal \mathbb G(\gamma_{0})+\mathbb G(\gamma_0)^\intercal G\mathrm{vec}(vv^\intercal)+\frac{1}{4}(\mathrm{vec}(vv^\intercal))^\intercal G^\intercal G\mathrm{vec}(vv^\intercal)\}~,
\end{align}
as desired, where the second equality exploited the facts that $\theta_0(\gamma_0)=0$ and that $G\mathrm{vec}(v\gamma_{0}^{\intercal})=0$ for any $v\in\mathbf{R}^{k}$, and the third equality follows from the fact that the $(k-1)$ columns in $A^{\intercal}\in\mathbf M^{k\times(k-1)}$ and $\gamma_0$ form a basis for $\mathbf{R}^{k}$. To see this last fact, note first that the columns of $A^\intercal$ are clearly linearly independent; moreover, if $\gamma_0=A^{\intercal} c^{\ast}$ for some nonzero $c^{\ast}\in\mathbf{R}^{k-1}$, then $\gamma_0^{\intercal} \jmath_{k}=0$ by simple algebra, contradicting the linear normalization that $\sum_{j=1}^{k}\gamma_0^{(j)}\neq 0$. \qed

\end{appendices}

\addcontentsline{toc}{section}{References}
\putbib
\end{bibunit}

\end{document}